\documentclass[aps,pra,twocolumn,superscriptaddress,longbibliography,10pt]{revtex4-2}
\usepackage{amsmath}
\usepackage{hyperref}
\hypersetup{
	colorlinks=true,
	linkcolor=blue,
	filecolor=blue,
	citecolor = black,      
	urlcolor=cyan,
}
\usepackage{amssymb}
\usepackage{amsmath,amsthm}
\usepackage{graphicx}
\usepackage{siunitx}

\usepackage{array}

\usepackage{xargs}                      
\usepackage[pdftex,dvipsnames]{xcolor}
\usepackage{color}

\usepackage[papersize={8.5in,11in}, margin=1in]{geometry}

\usepackage{placeins}

\newcommand{\magn}[1]{\left|#1\right|}

\newcommand{\bra}[1]{\left\langle#1\right|}
\newcommand{\ket}[1]{\left|#1\right\rangle}
\newcommand{\bracket}[2]{\left\langle#1|#2\right\rangle}
\newcommand{\adj}[1]{#1^{\dagger}}

\newcommand{\Exp}{\mathop{\mathbb{E}}}
\newcommand{\Var}{\mathop{\text{Var}}}
\newcommand{\XEB}{\textsf{XEB}}
\newcommand{\lXEB}{\log \textsf{XEB}}
\newcommand{\Heff}{H^{(k)}_{\mathrm{eff}}}
\newcommand{\heis}[2]{\vec{\sigma}_{#1} \cdot \vec{\sigma}_{#2}}

\newcommand{\bx}{b_{x}}

\begin{document}

\title{Sample-efficient benchmarking of shallow all-to-all random quantum circuits}

\author{Gregory Bentsen}
\affiliation{Department of Physics, The College of William \& Mary}
\author{Bill Fefferman}
\author{Soumik Ghosh}
\affiliation{Department of Computer Science, The University of Chicago}
\author{Michael J. Gullans}
\affiliation{Joint Center for Quantum Information and Computer Science,
University of Maryland and NIST}
\author{Yinchen Liu}
\affiliation{Department of Combinatorics and Optimization, University of Waterloo}
\affiliation{Institute for Quantum Computing, University of Waterloo}

\begin{abstract}
Random circuit sampling (RCS) remains one of the most competitive frameworks for demonstrating quantum advantage in near-term noisy intermediate-scale quantum (NISQ) hardware. Unfortunately, absent error-correction, existing benchmarks to characterize these experiments, like linear cross-entropy, have been classically spoofed due to noise. Because of this, there are interesting regimes, like shallow-depth random quantum circuits, where sampling is plausibly classically intractable, but no existing benchmark can distinguish between a noisy quantum computer and an adversarial classical spoofer. In this paper, we demonstrate that the \emph{nonlinear} cross-entropy provides a sample-efficient benchmark for shallow-depth all-to-all random quantum circuits whose score cleanly separates noisy quantum computers from state-of-the-art classical spoofers, even in the presence of depolarizing noise. Further, we develop a binary classifier based on the notion of heavy output generation that features \emph{logarithmic} sample complexity at short depth. Our evidence comes from exact analytic expressions for all-to-all Brownian circuit ensembles derived using replica tricks, and numerical simulations that corroborate these results for discrete Haar-random unitary circuits.

\end{abstract}

\maketitle

\section{Introduction}
\label{sec:intro}

A major goal in the near-term quantum era is to achieve quantum advantage — the first unequivocal experimental demonstration of a quantum computational speedup.  Many quantum advantage experiments involve implementations of random quantum circuits in which many layers of independently-sampled random quantum gates are applied to a simple initial state followed by measurement of all qubits in the computational basis \cite{boixo2018characterizing,Arute2019-zu, zhu2021quantumcomputationaladvantage60qubit, morvan2023phasetransitionrandomcircuit}.

Why are random quantum circuits hard to simulate classically? There are two perspectives on this question. The traditional perspective is that the hard problem is sampling from the output distribution $q(\mathbf{x})$ of a random quantum circuit, where $q(\mathbf{x}) := \magn{\bra{\mathbf{x}} U \ket{\mathbf{0}}}^2$ is the probability of measuring the bitstring $\mathbf{x} = x_1 x_2 \ldots x_n$ at the end of the random circuit $U$.  There is rigorous complexity-theoretic evidence that such sampling problems cannot be solved by any efficient classical algorithm (see e.g.,
\cite{aaronson2016complexity, Bouland_2018, movassagh2020quantumsupremacyrandomcircuits, Bouland_2022, bouland2025exponentialimprovementsaveragecasehardness}). However, verifying
that the experimental output distribution is close to the ideal
quantum output distribution is a daunting computational task. Another caveat is that the hardness evidence so far only pertains to the noiseless ideal output distributions of random quantum circuits.

A second perspective is that the hard problem solved by random quantum circuit sampling experiments is not necessarily to faithfully sample from a hard distribution but merely to score sufficiently well on a benchmark \cite{aaronson2016complexity, aaronson2020classicalhardnessspoofinglinear}. In a similar spirit to a Bell inequality violation, the hope would be that scoring above some threshold, with respect to a benchmark, is a hard problem for any efficient classical algorithm, and so can be used to demonstrate quantum advantage directly. If such a hope could be proven true for a benchmark that has a large gap between the noiseless quantum score and the best possible classical score, it would be able to certify quantum advantage even for a noisy quantum experiment achieving an intermediate score.

Ideally, in order to ensure that the task can be efficiently solved by a quantum computer, we would want the benchmark to be sample-efficient. There are existing sample-efficient benchmarks for deep random circuits, like linear cross-entropy (LXEB) \cite{Arute2019-zu, aaronson2020classicalhardnessspoofinglinear}. However, recently developed classical algorithms have been able to spoof LXEB without sampling from the output distribution of the ideal circuit \cite{barak2020spoofing, gao2024limitations}. 
But this leaves open whether there are any sample-efficient benchmarks for \emph{shallow} random quantum circuits. This is an interesting regime to study because there is evidence that beyond a certain threshold depth, sampling from shallow random circuits is potentially classically intractable for many architectures \cite{NappShallow2022,mcginley2025measurement,bene2025quantum}.

In this paper we argue that the \emph{nonlinear} cross-entropy
\begin{equation}
    \lXEB := - \langle \log q \rangle
\end{equation}
is an example of a sample-efficient benchmark for sublogarithmic-depth all-to-all random circuits, even in the presence of depolarizing noise. This metric was considered in the study of random circuit sampling before, for e.g. see \cite{boixo2018characterizing}. However, it was only studied for deep circuits, under the assumption that the output distribution looks like a Porter-Thomas distribution, which is \emph{not} the case at sublogarithmic depths \cite{Dalzell_2022, Deshpande_2022}. Our analysis of this metric involves new analytical methods of independent interest.

Furthermore, our analytical tools also facilitate the development of a new binary classification protocol that can be used to distinguish between genuine quantum hardware and state-of-the-art classical spoofers using exponentially fewer samples than is required for the nonlinear cross-entropy. In particular, this binary classifier features \emph{logarithmic} sample complexity, requiring only $m \sim \log n$ samples to guarantee a success probability of at least $1 - 1/\mathrm{poly}(n)$.
This protocol is based on the notion of heavy output generation (HOG) which captures the propensity of a random quantum circuit to favor the production of certain special bitstrings that serve as `fingerprints' of the random circuit dynamics. Our HOG classifier obtains this favorable sample complexity because it is based on a decision boundary separating the quantum sampler from the classical spoofer. Whereas estimating the means of these two distributions still requires a polynomial number of samples, if we are armed with a decision boundary the probability of guessing the sampler incorrectly is given by the tail of the distribution, which decays exponentially with the number of samples.

~\\~\\
\textbf{Motivating a quantum advantage experiment:~} Our results motivate a quantum advantage experiment using sublogarithmic-depth random circuits with all-to-all connectivity. Our two new benchmarks are secure against known state-of-the-art classical spoofers; however a more rigorous analysis against spoofing is left for future work. Additionally, one feature of shallow-depth random circuits is that they do not anticoncentrate \cite{Dalzell_2022, Deshpande_2022}. This means that existing Pauli path based techniques, like \cite{Aharonov_2023}, do not work to sample from our system in the presence of noise. However, this also means that classical hardness-of-sampling proofs break down in the noiseless case. While there is evidence that there is a phase transition in classical simulation complexity beyond a certain critical constant depth \cite{NappShallow2022,mcginley2025measurement,bene2025quantum}, a rigorous proof of classical hardness remains open.
~\\~\\

\begin{figure}
    \centering
    \includegraphics[width=0.9\linewidth]{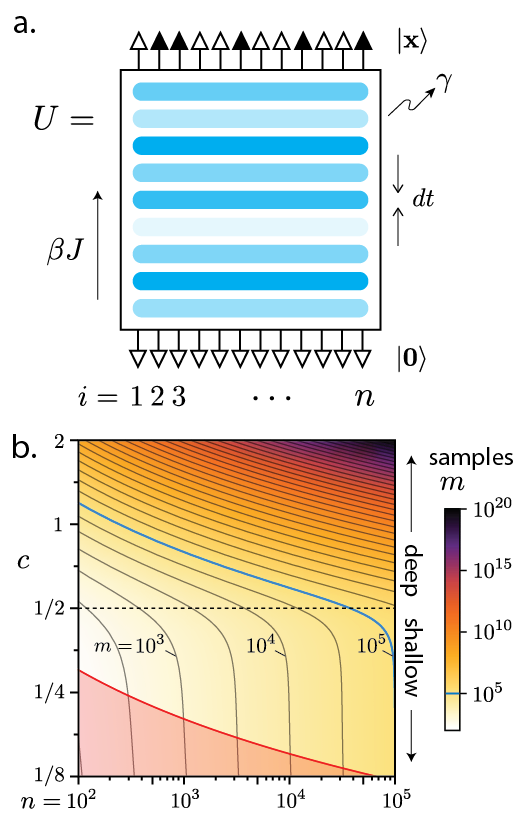}
    \caption{\textbf{Brownian random circuit models} generate analytically tractable matrix ensembles with tunable circuit depth $\beta J$ and noise rate $\gamma$. A system of $n$ qubits initialized in $\ket{\mathbf{0}}$ evolves under the Brownian circuit $U$ (a, blue stripes) and produces an output bitstring $\ket{\mathbf{x}}$ with probability $q({\mathbf{x}}) = \magn{\bra{\mathbf{x}}U \ket{\mathbf{0}}}^2$. Using the analytic expressions in Eq. \eqref{eq:noisyxeb} and \eqref{eq:noisyxebvar} we plot the number $m$ of bitstring samples (b) required to reliably estimate the nonlinear cross-entropy as a function 
    of system size $n$ and circuit depth $12 \beta J = c \log n$, with noise rate $r = n \gamma / 12 J = 1$. The requisite number of samples $m$ is linear in system size at shallow circuit depths $c < 1/2$ and grows as a polynomial in $n$ at larger depths $c > 1/2$, with a crossover at $c = 1/2$ (dashed black). Our methods become unreliable at extremely short depths $12 \beta J \leq \log 4$ (shaded red). We mark $m = 10^5$ samples in blue as a reference.}
    \label{fig:PTBrownianSetup}
\end{figure}

\textbf{Our techniques:~} Our evidence comes in two forms: first, we model all-to-all random circuits using a Brownian random circuit ensemble. This ensemble, while not formally equivalent to conventional random quantum circuits models constructed from discrete Haar-random gates, has often been used as a close proxy
in the physics literature (e.g., \cite{Lashkari2013towards, bentsen2021measurement,
jian2023linear}). In the confines of this Brownian model, we are able to derive exact results using tools from quantum many-body theory. In particular, performing the ensemble average over Brownian circuits yields an exact mapping to a quantum statistical mechanics problem at inverse temperature $\beta$:
\begin{equation}
    \mathbb{E}_U \left[ U \otimes U_{\mathcal{T}} \otimes \cdots \right] = e^{-\beta H^{(k)}_{\mathrm{eff}}}
\end{equation}
where $H^{(k)}_{\mathrm{eff}}$ is a many-body Hamiltonian that can be studied using tools from quantum many-body theory such as mean-field (large-$n$) methods. This mapping from random circuit dynamics to a statistical mechanics problem is analogous to the statistical mechanical mappings that have been widely used to argue results about conventional discrete Haar-random circuits; for e.g. see \cite{NappShallow2022, Dalzell_2022,mcginley2025measurement}. The reason to study a Brownian circuit ensemble is that it allows us to analytically compute arbitrarily high moments of the output probability distribution. Replica tricks can then be employed to obtain exact expressions for nonlinear quantities such as the mean $\lXEB$ and fluctuations around the mean. This is a notoriously difficult problem for discrete unitary circuits for which higher moment quantities are currently unavailable.

Second, we present numerical evidence that our results for the Brownian circuit ensemble agree with discrete Haar-random circuit ensembles with all-to-all connectivity. In particular, based on numerical simulations of up to $40$ qubits, we observe qualitative and quantitative behaviors matching our Brownian analytical predictions in discrete all-to-all random quantum circuits composed of Haar-random $2$-qubit gates. See Appendix \ref{app:numerics} for full details.

In the course of our calculations, we introduce new connections between random circuit sampling and tools from condensed matter and high-energy physics, like large-$n$ (mean-field) methods which could be of independent interest. 
~\\~\\
\textbf{Our model:~} Our circuit model in this work is a system of $n$ qubits initialized in the all-zero state $\ket{\mathbf{0}} \equiv \ket{00 \ldots 0}$ and subjected to random circuit dynamics described by a unitary matrix $U \in \mathcal{B}$ sampled from the ensemble $\mathcal{B}$. Our analytic results are derived for circuits sampled from the all-to-all Brownian circuit ensemble indexed by the circuit depth $\beta J$ as illustrated in Fig. \ref{fig:PTBrownianSetup}, where $J$ is a coupling strength and $\beta$ is the amount of time the circuit is allowed to run \footnote{$\beta J$ is the depth of the Brownian circuit and is a good proxy for the depth of analogous discrete all-to-all circuits.}. At the end of the circuit evolution we projectively measure each qubit in the computational basis and obtain an output bitstring $\ket{\mathbf{x}} \equiv \ket{x_1 x_2\ldots x_{n}}$. 

~\\~\\
\textbf{Our results:~} Our primary results are a pair of analytic expressions for the ensemble-averaged nonlinear cross-entropy and its fluctuations as a function of circuit depth $\beta J$, system size $n$, and single-qubit depolarizing noise at a rate $\gamma$. Throughout, it is convenient to parameterize circuit depth in terms of the quantity $a = e^{-12 \beta J}$, where $a \rightarrow 0$ corresponds to large circuit depth.

\vspace{0.5in}

Our first result is an exact expression for the ensemble-averaged nonlinear cross-entropy:
\begin{align}
    \lXEB = nH\left(\frac{1+a}{2}\right) + \gamma_e - e^{-n \beta \gamma}
    \label{eq:noisyxeb}
\end{align}
where $H(p) = - p \log p - (1-p) \log (1-p)$ is the binary entropy function and $\gamma_e$ is the Euler-Mascheroni constant (which is distinct from the noise rate $\gamma$). Throughout, $\log(x)$ refers to the natural logarithm. Eq. \eqref{eq:noisyxeb} generalizes the well-known Haar-random expression to finite circuit depth and in the presence of single-qubit depolarizing noise \cite{boixo2018characterizing}. The final noise-dependent term $e^{-n \beta \gamma}$ should be regarded in this scenario as a `signal' that serves both as a certification of quantum advantage and as a measure of the average noise in the system. In particular, this signal establishes a strict hierarchy between three classes of physical samplers: perfectly clean samplers with $\gamma = 0$, noisy samplers with average noise rate $\gamma > 0$, and the `Harvard' spoofing algorithm corresponding in our setup to $\gamma \rightarrow \infty$ (see Appendix \ref{app:harvard}). While our results do not rule out other spoofing algorithms, our benchmarks at the very least reliably distinguish the Harvard spoofer from genuine noisy quantum hardware.

Our second result is an analytic expression for the ensemble variance:
\begin{align}
    &\mathrm{var}(\lXEB) = \nonumber \\
    & \quad \quad \quad \frac{\pi^2}{6} - e^{-2 n \beta \gamma} + n \frac{(1-a^2)}{4} \mathrm{logit}^2 \left( \frac{1+a}{2} \right)
    \label{eq:noisyxebvar}
\end{align}
where $\mathrm{logit}(p) := \log(p / (1-p) ) = -dH(p)/dp$ is the log-odds function.
Together, these expressions establish the nonlinear cross-entropy as a sample-efficient benchmark at shallow circuit depths $\log n \gtrsim \beta J \gg 1$. In particular, parameterizing the circuit depth as $a = n^{-c}$ for an $O(1)$ constant $c$ and the noise rate as $r = n \gamma / 12 J \sim O(1)$, the key noise-dependent `signal' term $e^{-n \beta \gamma} = a^r = n^{-rc}$ scales as an inverse polynomial in system size. Meanwhile, the variance exhibits two different scalings depending on the circuit depth. At shallow depths $c < 1/2$ the final term
\begin{equation}
    n \frac{(1-a^2)}{4} \mathrm{logit}^2 \left( \frac{1+a}{2} \right) \approx n a^2 = n^{1-2c}
\end{equation}
dominates whereas at larger depths $c > 1/2$ the leading constant term $\pi^2 / 6$ dominates, with a crossover between these two regimes near $c = 1/2$. In either case the signal-to-noise ratio $\mathrm{SNR} \approx a^r / \sqrt{\pi^2 / 6 - a^{2r} + n a^2}$ scales like an inverse polynomial, meaning that the nonlinear cross-entropy is sample-efficient to estimate in this shallow-depth regime.

Practically speaking, Eqs. \eqref{eq:noisyxeb} and \eqref{eq:noisyxebvar} allow us to estimate the number of samples $m$ required to reliably estimate the nonlinear cross-entropy (see Fig. \ref{fig:PTBrownianSetup}.b). The standard error in our estimate of the nonlinear cross-entropy is
\begin{equation}
    \sigma_{\mathrm{se}} = \sigma / \sqrt{m}
\end{equation}
where $\sigma^2 = \mathrm{var}(\log \mathrm{XEB})$ is the variance. To reliably estimate the signal $X = e^{-n \beta \gamma}$ of interest we require that the standard error is smaller than the mean $\overline{X}$, such that we need at least
\begin{equation}
    m \gtrsim \sigma^2 / \ \overline{X}^2 \sim \begin{cases}
    n^{1+2c(r-1)} & c < 1/2 \\
    \frac{\pi^2}{6} n^{2cr} & c > 1/2
\end{cases}
\end{equation}
samples to ensure that our estimate of the nonlinear cross-entropy is not dominated by statistical fluctuations. We plot the requisite number of samples $m$ as a function of system size $n$ and circuit depth $c$ in Fig. \ref{fig:PTBrownianSetup}.b. which clearly shows the two regimes of behavior at different circuit depths.

Our results are facilitated by analytically tractable all-to-all Brownian circuits (Fig. \ref{fig:PTBrownianSetup}.a.) that yield an exact mapping between the Brownian random circuit ensemble and a thermal partition function governed by Boltzmann weights $e^{-\beta \Heff}$ where $\Heff$ is a quantum many-body Hamiltonian and the inverse temperature $\beta$ is proportional to the circuit depth. This mapping between random circuit sampling and quantum statistical mechanics is an especially powerful -- and quite general -- result because it allows us to bring the tools of quantum many-body theory and quantum statistical mechanics to bear on the sampling problem at hand. In particular, shallow (deep) circuits correspond to partition functions with higher (lower) temperature. Practically speaking, this means that the infinite depth behavior is controlled by the ground states of $\Heff$ and the timescale required to reach this large-depth regime is controlled by the many-body energy gap $\Delta$.

This dual description of Brownian random circuits in terms of a statistical partition function is similar to the statistical mechanics mappings usually applied to the theoretical study of discrete Haar-random circuits \cite{zabalo2020critical,bao2020theory,choi2020quantum}, where the ensemble average over discrete Haar-random gates yields a classical spin model that can be analyzed using domain-wall arguments \cite{li2019measurement,skinner2019measurement}. In contrast to those methods, which are typically only capable of computing low-order moments $m^{(k)}$ of the output probability distribution, our Brownian circuit tools allow us to compute arbitrary moments $m^{(k)}$ for $k = 1,2,3,\ldots$, thereby facilitating both the complete characterization of the resulting distribution $P_a(q)$ as well as the exact calculation of means and shot-to-shot fluctuations in various benchmark scores.

In the following sections we elaborate on our Brownian circuit methods, use them to derive the XEB results advertised above, and develop additional theoretical tools relevant to benchmarking of quantum hardware. In Section \ref{sec:browniancircuits} we introduce the all-to-all Brownian circuit ensemble, which serves as the central theoretical tool in this work. In Section \ref{sec:distributions} we develop this tool to study random circuit sampling and showcase its capabilities by obtaining closed-form expressions for the distribution $P(q)$ of bitstring probabilities at arbitrary, continuously-tunable circuit depth. This distribution converges asymptotically to the Porter-Thomas distribution as circuit depth is increased. In Section \ref{sec:logXEB} we combine these results with replica tricks to derive the exact expressions Eq. \eqref{eq:noisyxeb} and \eqref{eq:noisyxebvar} for the ensemble-averaged nonlinear cross-entropy and its variance at arbitrary circuit depth and in the presence of single-qubit depolarizing noise. These results allow us to conclude that the nonlinear cross-entropy is sample-efficient to estimate in the shallow-depth regime. In Section \ref{sec:sshog} we propose a new binary classification protocol based on the notion of heavy output generation (HOG) for distinguishing noisy quantum samplers from a class of classical spoofers using only a single bitstring sample. We conclude with a discussion of experimental and theoretical outlook in Section \ref{sec:discussion}.

\section{All-to-all Brownian Circuits}
\label{sec:browniancircuits}

Our results are facilitated by an analytically tractable ensemble of all-to-all Brownian quantum circuits $\mathcal{B}$ \cite{bentsen2021measurement,sahu2022entanglement,jian2023linear}, which serve as an especially powerful theoretical `laboratory' for conducting RCS \textit{gedanken}-experiments on systems of $n$ qubits. These Brownian circuits are highly chaotic and rapidly scramble quantum information into delocalized patterns of entanglement, becoming $k$-designs (i.e. closely resembling the Haar ensemble $\mathcal{H}$) in a circuit depth scaling linearly with $k$ \cite{jian2023linear}. These properties make Brownian circuits an ideal candidate for RCS experiments, which are designed to take advantage of the classical intractability of reproducing the random `speckle' pattern obtained from transition amplitudes $\alpha(\mathbf{x}) = \bra{\mathbf{x}} U \ket{\mathbf{0}}$ when $U$ is drawn from a highly chaotic quantum system.

Each circuit $U \in \mathcal{B}$ in the ensemble is composed of a series of short unitary steps $U_t$ (see Fig. \ref{fig:PTBrownianSetup}.a) that evolve the system for a short time $dt$ under random all-to-all 2-qubit interactions:
\begin{equation}
    U = \prod_t U_t = \prod_t \exp{\left[-i \sum_{\substack{j < k \\ \alpha, \beta}} J_{jk}^{\alpha \beta}(t) \sigma_j^{\alpha} \sigma_k^{\beta} dt \right]},
    \label{eq:brownianunitary}
\end{equation}
where $\sigma_j^{\alpha} = (\sigma_j^{x},\sigma_j^{y},\sigma_j^{z})$ are the Pauli matrices acting on qubit $j$. At each timestep $t$ the 2-qubit interactions are governed by randomly-chosen couplings $J_{jk}^{\alpha \beta}(t)$, which are Gaussian white-noise random variables with zero mean and variance
\begin{equation}
    \mathbb{E} \left[ J_{jk}^{\alpha \beta}(t) J_{j'k'}^{\alpha' \beta'}(t') \right] = \frac{J}{n dt} \delta_{jj'} \delta_{kk'} \delta^{\alpha \alpha'} \delta^{\beta \beta'} \delta_{tt'}
    \label{eq:couplingvariance}
\end{equation}
where $J$ is a constant that sets the overall timescale of the model. The total depth of the circuit is given by $\beta = K dt$ where $K$ is the number of unitary steps $U_t$ comprising $U$. When we speak of ``Brownian'' circuits it means that we are working in the limit of many infinitesimally small timesteps $dt \rightarrow 0$ (with $K \rightarrow \infty$ and total circuit depth $\beta$ fixed).

Taking the ensemble average over the disordered couplings $J_{jk}^{\alpha \beta}(t)$ we obtain an exact mapping between the moments of $2k$ unitary Brownian matrices and a quantum statistical mechanics problem at inverse temperature $\beta$:
\begin{equation}
    \mathbb{E}_U \left[ U \otimes U_{\mathcal{T}} \otimes \cdots \right] = e^{-\beta H^{(k)}_{\mathrm{eff}}}
\end{equation}
where $H^{(k)}_{\mathrm{eff}}$ is a many-body Hamiltonian whose ground states govern the large-depth dynamics. This mapping between the moments of unitary matrices and a stat-mech problem is analogous to known mappings between discrete Haar-random quantum circuits and stat-mech Ising models that have been widely used to study entanglement transitions \cite{li2019measurement,skinner2019measurement,bao2020theory,zabalo2020critical}, randomized benchmarking \cite{heinrich2022randomized}, the approach to $k$-designs \cite{brandao2021models}, and linear XEB \cite{ware2023sharp,gao2024limitations}. Whereas these discrete Haar-random circuits are limited to finite moments $k \leq 2$, our Brownian circuit stat-mech mapping enables us to compute arbitrary moments $k$ analytically, opening the door to replica tricks that can be employed to compute information-theoretic quantities of interest such as the nonlinear cross-entropy.

Brownian circuits are continuous analogues of all-to-all random quantum circuit composed of Haar-random 2-qubit gates, and they have been used to model quantum information scrambling \cite{Lashkari2013towards}, entanglement transitions \cite{bentsen2021measurement}, and the convergence to $k$-designs \cite{jian2023linear}. The reason to study the Brownian circuit ensemble, in place of other ensembles, is because it shares many similarities with discrete random quantum circuit ensembles, while also allowing for direct calculation of higher moments at arbitrary circuit depth and in the presence of decoherence. From a technical perspective, the all-to-all interactions in our model allow us to make use of large-$n$ (mean-field) methods in which the effective dynamics becomes semi-classical and readily solvable. The large-$n$ `saddle-point' methods we employ build on a long history of mean-field-theory techniques in studies of classical spin glasses \cite{edwards1975theory,sherrington1975solvable,thouless1977solution,almeida1978stability,parisi1979toward,gross1984simplest,mezard1986spin,young1997spin}, quantum spin glasses \cite{bray1980replica,sachdev1993gapless,miller1993zerotemperature,read1995landau,kopec1995continuous,georges2000mean,georges2001quantum}, condensed matter physics \cite{parcollet1999nonfermi,fitzpatrick2014nonfermi,sachdev2015bekenstein,werman2017nonquasiparticle,davison2017thermoelectric,gu2020notes,chowdhury2022sachdev}, high-energy physics \cite{brezin1978planar,witten1980expansion,coleman1985aspects,thooft1993planar,moshe2003quantum,kitaev2015simple,maldacena2016remarks,fu2016numerical,gu2017local,berkooz2017higher,fu2017supersymmetric,kitaev2018soft,hartnoll2018holographic,sarosi2018ads2,saad2019semiclassicalrampsykgravity,rosenhaus2019introduction,berkooz2021complex}, and, more recently, quantum information science \cite{bentsen2021measurement,sahu2022entanglement,jian2023linear}.

\section{Output Distributions at Arbitrary Depth}
\label{sec:distributions}

In this section we apply Brownian circuit technology to compute the output sampling distributions obtained from finite-depth random quantum circuits. This affords an opportunity to both showcase Brownian circuit methods and to establish technical results that will be essential to the calculations we discuss in the following sections. Our large-$n$ methods allow us to find a family of distributions $P_a(q)$ indexed by the circuit depth $a = e^{- 12 \beta J}$ that smoothly converges to the Porter-Thomas (PT) distribution at large depth $a \rightarrow 0$.

Because every distribution (on a bounded domain) is uniquely determined by its moments
\begin{equation}
    m^{(k)} := \mathbb{E}[q^k] \equiv \int_0^1 dq \ q^k P(q)
    \label{eq:momentexpr}
\end{equation}
we can compute these moments and use them to obtain the distribution $P_a(q)$ itself. For example, the Porter-Thomas distribution $Q(q)$ has moments $m^{(k)} = k! / d^k$, which allows us to construct the moment-generating function
\begin{equation}
    M(t) = \sum_{k=0}^{\infty} \frac{m^{(k)}}{k!} t^k = \frac{d}{d-t}
    \label{eq:momentgenfunc}
\end{equation}
and apply the inverse Laplace transform to obtain the distribution
\begin{equation}
    Q(q) = \mathcal{L}^{-1}\left[M(-t)\right] = d e^{- d q}.
    \label{eq:porterthomas}
\end{equation}
In the following we will use the same strategy to obtain an expression for the family of distributions $P_a(q)$ at arbitrary circuit depth $a$.

\begin{figure}
    \centering
    \includegraphics[width=0.95\linewidth]{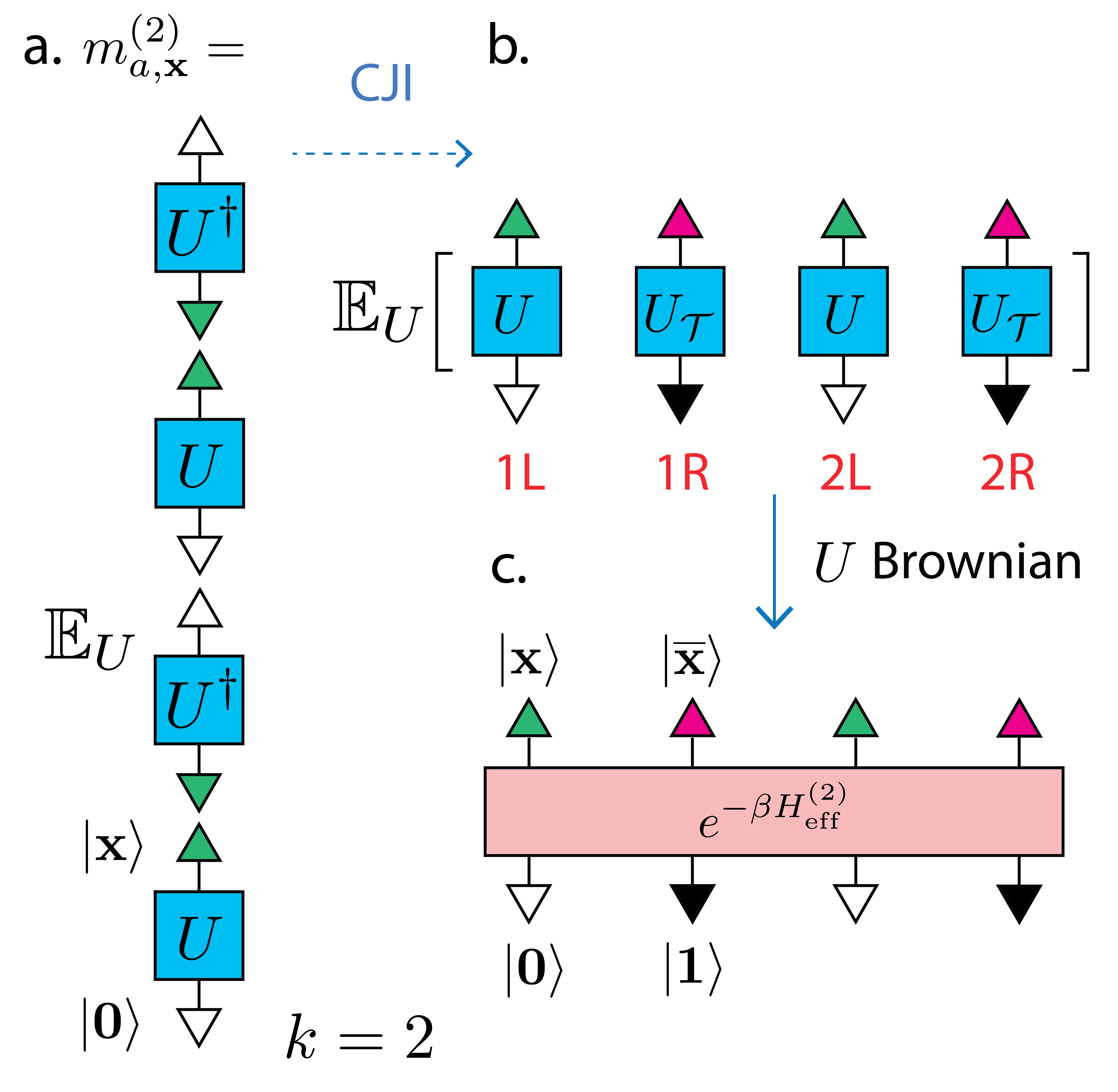}
    \caption{\textbf{The Choi-Jamio\l kowski isomorphism (CJI) and disorder averaging} allow us to map the Brownian circuit sampling problem onto a quantum statistical mechanics problem at inverse temperature $\beta$. We first apply CJI to convert quantities with multiple forward $U$ and inverse $U^{\dagger}$ evolutions (a, for $k=2$) into an expectation value on multiple replicas (b) labeled 1L,1R,2L,2R. We then take the ensemble average over the disordered couplings $J_{ij}^{\alpha \beta}(t)$ which yields a Gibbs weight $\exp(-\beta \Heff )$ governed by an effective Hamiltonian $\Heff$ acting on $2kn$ qubits.}
    \label{fig:cji}
\end{figure}

Consider an ensemble of Brownian circuits $U \in \mathcal{B}_a$ at depth $12\beta J = - \log a$ that produces bitstrings $\mathbf{x}$ with probability $q_{\mathbf{x},U}$. We are interested in computing the moments
\begin{equation}
    m_{a,\mathbf{x}}^{(k)} := \mathbb{E}_{U \in \mathcal{B}_a} \left[ \magn{\bra{\mathbf{x}} U \ket{\mathbf{0}}}^{2k} \right]
    \label{eq:brownianmoments}
\end{equation}
averaged over the circuit ensemble at a fixed depth $a$ and for a particular output bitstring $\mathbf{x}$. Using the Choi-Jamio\l kowski isomorphism (channel-state duality) \cite{choi1975completely,jamiolkowski1972linear}, we can equivalently express these moments as a tensor product over $2k$ copies or \emph{replicas} (see Fig. \ref{fig:cji})
\begin{equation}
    m_{a,\mathbf{x}}^{(k)} = (-1)^{xk}\bra{\mathbf{x} \overline{\mathbf{x}} \ldots } \mathbb{E}_U \left[ U \otimes U_{\mathcal{T}} \otimes \cdots \right] \ket{\mathbf{0} \mathbf{1} \ldots}
\end{equation}
where $\overline{\mathbf{x}}$ is the inverse of the bitstring $\mathbf{x}$, with $\overline{0} = 1$ and $\overline{1} = 0$ and $x := \magn{\mathbf{x}}$ is the Hamming weight of $\mathbf{x}$. Time-reversal $\mathcal{T}$ plays a crucial role in this expression: because we are dealing with a model composed of spin-1/2 particles, we must define the time-reversal operator $\mathcal{T}$ as:
\begin{equation}
	\mathcal{T}(U) \equiv U_{\mathcal{T}} := (i Y)^{\dagger} U^* (i Y)
\end{equation}
where
\begin{equation}
	iY = \prod_i \left(i \sigma_i^y\right) = \bigotimes_i \begin{bmatrix}
0 & 1 \\
-1 & 0 \\
\end{bmatrix}_i
\end{equation}
and $(i Y)^{\dagger} (i Y) = (i Y) (i Y)^{\dagger} = 1$. Note that the Pauli operators uniformly reverse sign under the time-reversal operation: $\mathcal{T}(\sigma_i^{\alpha}) = - \sigma_i^{\alpha}$, whereas they do not under complex conjugation alone.

The `forward' (L) replicas $U$ are labeled by indices $rL$ for $r = 1,2,\ldots,k$, while the time-reversed (R) replicas $U_{\mathcal{T}}$ are labeled by indices $rR$. Taking the expectation value over the disordered Brownian couplings $J_{jk}^{\alpha \beta}(t)$ according to Eq. \eqref{eq:couplingvariance}, we can write the moments in terms of a thermal partition function
\begin{equation}
	m_{a,\mathbf{x}}^{(k)} = (-1)^{xk}\bra{\mathbf{x} \overline{\mathbf{x}} \ldots} e^{- \beta H^{(k)}_{\mathrm{eff}}} \ket{\mathbf{01} \ldots}
    \label{eq:brownianmomentstatmechmap}
\end{equation}
governed by an effective Hamiltonian
\begin{equation}
    H^{(k)}_{\mathrm{eff}} = \frac{J}{2n} \sum_{\substack{i<j \\ \alpha \beta}} \left( \sum_{r = 1}^k \sigma_{irL}^{\alpha} \sigma_{jrL}^{\beta} - \sigma_{irR}^{\alpha} \sigma_{jrR}^{\beta} \right)^2
    \label{eq:arbitrarykeffham}
\end{equation}
where the circuit depth $\beta$ plays the role of inverse temperature. It is useful at this point to disregard the Brownian origins of this Hamiltonian and treat it as an interesting strongly-interacting quantum many-body system; this change in perspective is at the heart of the methods we bring to bear in this paper.

Crucially, the Hamiltonian Eq. \eqref{eq:arbitrarykeffham} is invariant under global $\mathrm{SU}(2)$ rotations, which substantially simplifies the analysis. In addition, $\Heff$ is invariant under arbitrary permutations $\sigma \in S_k$ of the forward (L) replicas among themselves and under arbitrary permutations $\sigma' \in S_k$ of the backward (R) replicas among themselves, where $S_k$ is the symmetric group on $k$ items. In addition, the Hamiltonian is invariant under exchange of L with R, corresponding to a global time-reversal operation $\mathcal{T}$. Together, these discrete symmetry operations generate a wreath product
\begin{equation}
    G = (S_k \times S_k) \rtimes \mathbb{Z}_2  \equiv S_k \wr \mathbb{Z}_2.
\end{equation}
This discrete symmetry group, in addition to global $\mathrm{SU}(2)$ invariance, will be helpful in constructing the spectrum of $H^{(k)}_{\mathrm{eff}}$.

Our analysis is substantially simplified by the mean-field nature of the system, allowing us to bring large-$n$ (mean-field) methods to bear on the problem. In particular, introducing the mean fields 
\begin{equation}
    G_{rs}^{ab} := \frac{1}{n} \sum_{i=1}^n \vec{\sigma}_{ira} \cdot \vec{\sigma}_{isb}
\end{equation}
we may rewrite the effective Hamiltonian Eq. \eqref{eq:arbitrarykeffham} as
\begin{align}
    \Heff = \frac{Jn}{2} \sum_{ra < sb} (-1)^{a+b} \left( G_{rs}^{ab} \right)^2 + \frac{9}{2} Jkn + O(1)
    \label{eq:heff}
\end{align}
in which the global $\mathrm{SU}(2)$ invariance is clearly manifested. The final term indicates that we have dropped terms of $O(1)$ that are subleading at large $n$.

\begin{figure}
    \centering
    \includegraphics[width=0.95\linewidth]{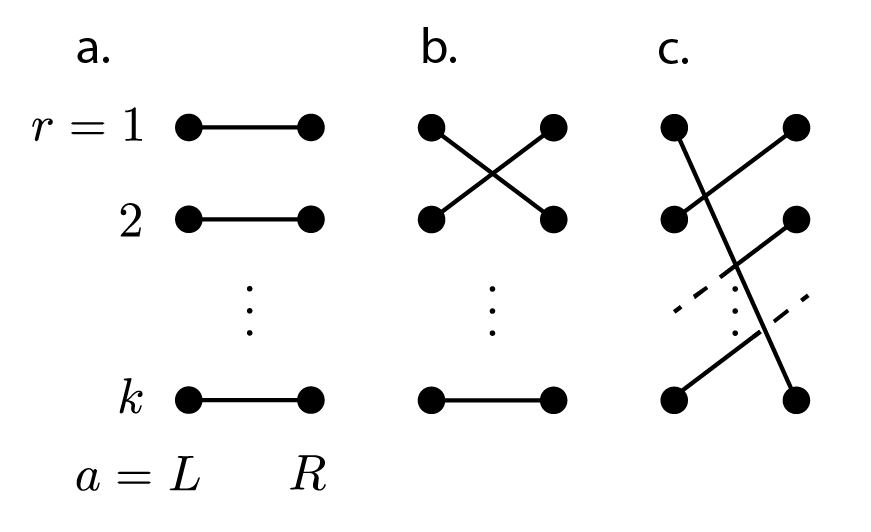}
    \caption{\textbf{Ground states of the effective Hamiltonian $\Heff$} where each solid line represents a collection of spin singlets $\bigotimes_{i=1}^n (\ket{01} - \ket{10})_{irs,LR}/\sqrt{2}$ connecting the forward (L) and time-reversed (R) replicas. Equivalently, each solid line represents a non-vanishing mean field $G_{rs}^{LR} = -3$, with all other mean fields vanishing. Whereas the `ladder' state (a) is the simplest pairing, we may use the invariance of $\Heff$ under permutations $\sigma \in S_k$ of the time-reversed replicas (R) among themselves to obtain all other ground states (b,c), yielding $k! = \magn{S_k}$ ground states in total.}
    \label{fig:groundstatediagrams}
\end{figure}

In the large-$n$ limit, the mean fields $G_{rs}^{ab}$ behave like semiclassical c-numbers with vanishingly small quantum fluctuations as $n \rightarrow \infty$ \cite{bentsen2021measurement,jian2023linear}. This allows us to study the spectrum of $\Heff$ analytically. We first study the ground states. The `ladder' state (Fig. \ref{fig:groundstatediagrams}.a.)
\begin{equation}
    \ket{\Omega} := \bigotimes_{\substack{i=1,\ldots,n \\ r = 1, \ldots, k}} \frac{1}{\sqrt{2}} \left( \ket{01} - \ket{10} \right)_{ir,LR}
    \label{eq:ladderground}
\end{equation}
consists of pairwise spin singlets connecting forward (L) and backward (R) replicas for each $i = 1,\ldots,n$ and each $r = 1,\ldots,k$. Equivalently, this configuration corresponds to mean fields $G_{rr}^{LR} = -3$ with all other mean fields vanishing. Exploiting the invariance of $\Heff$ under arbitrary permutations $\sigma \in S_k$ of the time-reversed replicas, we can immediately obtain all other ground states (Fig. \ref{fig:groundstatediagrams}.b,c). Running through all possible permutations $\sigma$ yields $k! = \magn{S_k}$ ground states in total, each corresponding to a different pairing between $L,R$.

The excited states simply build on top of each ground state. For each ground state pairing, we can either leave the pair $\vec{\sigma}_{irL} \cdot \vec{\sigma}_{isR}$ as a singlet with eigenvalue $-3$, or we can flip it to a triplet state with eigenvalue $+1$. The singlet state is the ground state with Gibbs weight 1, whereas the triplet state is an excited state with energy $12 J$ and comes with a Gibbs weight $(-1)^{x_i} e^{-12 \beta J}$. The sign $(-1)^{x_i}$ depends on the value of the sampled output bit $x_i$ and originates from the overlap between the output bra $\bra{\mathbf{x}}$ and the ground states $\ket{\Omega}$ \cite{bentsen2024complexity}. 

Putting everything together, these calculations yield a relatively simple final expression for the moments of the transition probability:

\begin{equation}
    m_{a,\mathbf{x}}^{(k)} = \frac{k!}{\left(d b_{x} \right)^k}
    \label{eq:momkxfinal2}
\end{equation}
where
\begin{align}
    b_{x}^{-1} &:= \prod_{i=1}^n \left( 1 + (-1)^{x_i} e^{-12 \beta J} \right) \nonumber \\
    &= (1-a)^{x} (1+a)^{n-x}.
\end{align}
Here $a = e^{-12 \beta J}$ parameterizes the circuit depth and $x = \magn{\mathbf{x}}$ is the Hamming weight of the bitstring $\mathbf{x}$. Crucially, our results are reliable only when $a \ll 1$ is sufficiently small.
Notice that these moments depend only on the Hamming weight $x$ of the bitstring. Eq. \eqref{eq:momkxfinal2} is our first primary technical result, and will serve as the starting point for our analytic results in this section and the following sections. It passes some basic sanity checks: for infinite depth $\beta \rightarrow \infty$ ($a \rightarrow 0$) we obtain $m_{a,\mathbf{x}}^{(k)} \rightarrow k!/d^k$, which is precisely the large-$n$ expression for the moments of the Porter-Thomas distribution. In fact, the factor $\bx$ appears to simply scale the overall Hilbert space dimension $d \bx$. Notice that if we blindly take the limit $\beta \rightarrow 0$ ($a \rightarrow 1$) we obtain $m_{a,\mathbf{x}}^{(k)} \rightarrow \delta_{x0} k! / d^k$, but this is clearly unreliable since our results rely on the assumption $\beta J \gg 1$. This extremely short-depth regime deserves further study, but for the current discussion we limit ourselves to $0 \leq a \ll 1/4$.

With the moments $m_{a,\mathbf{x}}^{(k)}$ in hand, we may reconstruct the probability distribution $P_{a,x}(q)$ for the return probability $q$ similar to Eqs. \eqref{eq:momentexpr} - \eqref{eq:porterthomas}. As above, any such distribution $P_{a,x}(q)$ (on a bounded interval $0 \leq q \leq 1$) is uniquely identified by its moments $m_{a,\mathbf{x}}^{(k)} = \int_0^1 P_{a,x}(q) q^k dq$; conversely, given the moments $m_k$ we may reconstruct the distribution $P(q)$. Starting from Eq. \eqref{eq:momkxfinal2}, we construct the moment-generating function
\begin{equation}
    M_{a,x}(t) = \sum_k \frac{m_{a,\mathbf{x}}^{(k)}}{k!} t^k = \frac{d \bx}{d \bx - t}
\end{equation}
and apply the inverse Laplace transform to obtain
\begin{equation}
    P_{a,x}(q) = \mathcal{L}^{-1}\left[M_{a,x}(-t)\right] = d \bx e^{- d \bx q}.
    \label{eq:singlestringporterthomas}
\end{equation}
Here we have retained the subscript $\mathbf{x}$ to remind the reader that these expressions are for a particular choice of output bitstring $\mathbf{x}$ (the distribution in Eq. \eqref{eq:singlestringporterthomas} is over different choices of Brownian circuit, \emph{not} over different bitstrings). The resulting distribution $P_{a,x}(q)$ is identical to the Porter-Thomas distribution, except that the Hilbert space dimension $d$ is replaced by the scaled dimension $d \bx$, which approaches the Porter-Thomas value $d \bx \rightarrow d$ as $a \rightarrow 0$ (infinite circuit depth).

\begin{figure}
    \includegraphics[width=0.95\linewidth]{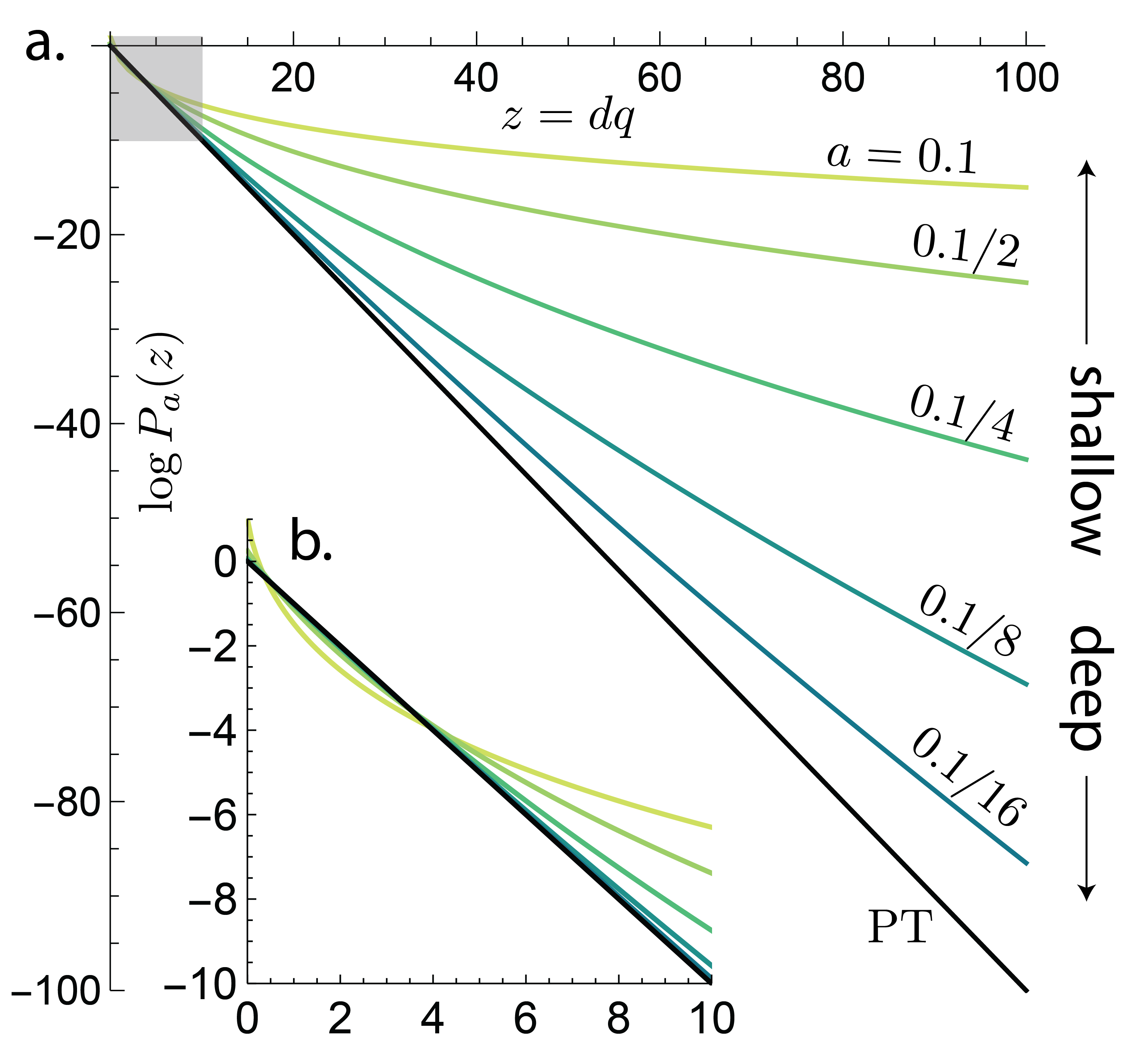}
    \caption{\textbf{Brownian circuit ensembles converge to the Haar random matrix ensemble} for increasing circuit depth $12\beta J = - \log a$. The finite-depth Brownian sampling distributions $P_a(z)$ for $n = 100$ qubits, plotted as a function of $z = dq$ for depths $a = 0.1,0.1/2,\ldots,0.1/16$ (yellow to dark green) converge to the infinite-depth Porter-Thomas (PT) distribution $\lim_{a \rightarrow 0} P_a(z) \equiv Q(z) = e^{-z}$ (black). The inset (b) shows detail of the upper-left region (grey box).}
    \label{fig:ApproachToPT}
\end{figure}

Ultimately, the distributions $P_{a,x}(q)$ are somewhat fictitious, as they can only be experimentally measured using post-selection on the desired bitstrings $\mathbf{x}$. The more physically relevant quantity is the distribution averaged over bitstrings $\mathbf{x}$
\begin{equation}
    P_a(q) = \frac{1}{d} \sum_{\mathbf{x}} P_{a,x}(q) = \frac{1}{d} \sum_{x=0}^n \binom{n}{x} P_{a,x}(q)
    \label{eq:porterthomasdeptha}
\end{equation}
where in going from the second to the third term we have reduced from a sum over all bitstrings $\mathbf{x}$ to a sum over all Hamming weights $x$. While we are currently unaware of an analytic method to evaluate this sum, it is straightforward to perform numerically and can be done efficiently because we must only sum over the Hamming weights $x$, not over every individual bitstring $\mathbf{x}$. We plot the resulting distribution $P_a(z)$ as a function of $z = d q$ for $n=100$ qubits in Fig. \ref{fig:ApproachToPT} for various choices of circuit depth $a \ll 1/4$. The Brownian distribution clearly approaches the ideal Porter-Thomas distribution as $a \rightarrow 0$ (large circuit depth). 

\section{Sample-Efficiency of Nonlinear Cross-Entropy at Shallow Depths}
\label{sec:logXEB}

The previous section demonstrated the capabilities of Brownian circuit ensembles to provide detailed information about random circuit dynamics at arbitrary tunable circuit depths. In particular, the analytic expressions derived in the previous section allowed us to completely characterize the convergence of Brownian circuit ensembles to the Haar-random ensemble as a function of tunable circuit depth $\beta J$. In this section we build on these results to study the nonlinear cross-entropy ($\lXEB$), which has played a prominent role in ongoing work on random circuit sampling, quantum advantage experiments, and randomized benchmarking. In particular, the analytic results in this section allow us to establish the nonlinear cross-entropy as a sample-efficient benchmark at short circuit depths $\log n \gtrsim \beta J \gg 1$ that reliably distinguishes between a perfectly clean sampler ($\gamma = 0$), a noisy sampler ($\gamma > 0$), and the `Harvard' spoofing algorithm ($\gamma \rightarrow \infty$), where $\gamma$ parameterizes the average noise rate in the system.

The cross-entropy generally serves as a standard measure of similarity between two probability distributions $p,q$ and is therefore a natural choice for a benchmark score where $q$ is the ideal distribution and $p$ is the true distribution. Applied to RCS tasks, the cross-entropy serves as a precise benchmark capable of measuring how close a noisy quantum sampler comes to imitating the `real thing' i.e. a sampler built from the $n$-qubit Haar ensemble. This realization has led to intense interest over the past decade in using the cross-entropy as a quantitative benchmark (often referred to as the cross-entropy benchmark or XEB for short) to evaluate and characterize the performance of noisy intermediate-scale quantum hardware \cite{boixo2018characterizing,Arute2019-zu,heinrich2022randomized,gao2024limitations}. 

Unfortunately it is well-known that the nonlinear cross entropy in the presence of noise is sample-inefficient at polynomial depth -- meaning that $O(\exp(n))$ samples are necessary to accurately estimate it -- so this particular measure is not especially useful in practice, at least at large circuit depths. To circumvent this issue, many groups over the past decade have instead resorted to the \emph{linear} cross-entropy benchmark, which provides some information about the closeness of the physical distribution to the ideal distribution while also remaining sample-efficient \cite{boixo2018characterizing,Arute2019-zu,gao2024limitations}. Sufficiently high benchmark scores using the linear XEB have recently been used to make bold claims about demonstrations of quantum advantage, meaning that quantum hardware has efficiently performed a computing task that would be inefficient to perform on a classical machine. More recent work has thrown such claims in dispute, partly due to improvements in classical algorithms to simulate many-body physics \cite{pednault2019leveraging}, and partly due to fundamental weaknesses in the linear XEB. In particular, recent work has demonstrated classically efficient algorithms that can ``spoof'' the linear XEB by achieving high benchmark scores comparable to quantum hardware \cite{gao2024limitations}.

We tackle these issues head-on in this section by using the technology of all-to-all Brownian circuits to demonstrate that the nonlinear cross-entropy benchmark serves as a sample-efficient benchmark for shallow-depth random circuits. Moreover, we show that this benchmark establishes a strict hierarchy between clean quantum hardware, noisy quantum hardware, and classical spoofing algorithms.

We frame the discussion in terms of a concrete physical setup. Imagine that we are in possession of a physical piece of hardware that we call the \emph{sampler}. The sampler takes as input a target circuit ensemble $\mathcal{B}$ (for example an all-to-all Brownian ensemble at depth $\beta J$) and produces a series of length-$n$ bitstrings $[\mathbf{x}_{1}, \mathbf{x}_{2}, \ldots, \mathbf{x}_{m}]$, ostensibly sampled according to the \emph{ideal} quantum transition probabilities
$q(\mathbf{x}) = \magn{\bra{\mathbf{x}} U \ket{\mathbf{0}}}^2$
where the unitaries $U$ are sampled uniformly at random from the ensemble $\mathcal{B}$. In reality however, this black box produces bitstrings with \emph{true} probabilities $p(\mathbf{x})$ that are apriori unknown and depend sensitively on the internal workings of the sampler.
Based on the bitstrings that we collect from the sampler, we are interested in characterizing whether this black box is a bonafide noisy quantum system attempting to sample bitstrings from the true quantum distribution, or whether it is a classical machine running a spoofing algorithm (or perhaps something else entirely).

To perform this task, suppose we are also in possession of a classical supercomputer capable of simulating arbitrary $n$-qubit dynamics. We make no assumptions about the efficiency of this classical simulation, and in fact we expect it to be exponentially inefficient (otherwise our efforts here are clearly pointless). We use this supercomputer to analyze the bitstrings $\mathbf{x}_s$ and compute the nonlinear cross-entropy score by taking each sampled bitstring $\mathbf{x}_s$ and simulating the many-body unitary dynamics $U$ to calculate the ideal transition probability $q(\mathbf{x}_s)$. We repeat this process for all $m$ bitstrings, take the logarithm, and compute the mean to obtain an estimate of the nonlinear cross-entropy \cite{boixo2018characterizing}:
\begin{equation}
    \mathrm{log \ XEB}^* := - \frac{1}{m} \sum_{s=1}^{m} \log q(\mathbf{x}_s).
\end{equation}
We label this quantity with a star to remind us that it is just a finite-sample estimate of the mean $\lXEB$ score
\begin{align}
    \log \mathrm{XEB} &:= - \sum_{\mathbf{x}} \mathbb{E}_U \left[ p(\mathbf{x}) \log q(\mathbf{x}) \right] \nonumber \\
    &\equiv -\langle \log q \rangle_{p,U}
    \label{eq:xeb}
\end{align}
where in the last expression we use angled brackets to represent the combined average over circuits $U$ and bitstrings $\mathbf{x}$ sampled with probability $p(\mathbf{x})$. Our primary claim is that the nonlinear cross-entropy can \emph{sample-efficiently} and \emph{reliably} distinguish between different kinds of samplers at shallow circuit depth, including ideal clean quantum hardware, realistic noisy quantum hardware, classical spoofing algorithms, and other classical distributions.

To demonstrate this, we leverage our prior results in Sec. \ref{sec:distributions} and make use of replica tricks, which have a long history in physics especially in studying spin glass physics \cite{edwards1975theory,sherrington1975solvable}. The key issue is that the nonlinear cross-entropy in Eq. \eqref{eq:xeb} involves taking expectation values over a logarithm. To circumvent this issue, recall that the logarithm can always be viewed as a limiting process:
\begin{equation}
    \langle \log q \rangle = \lim_{k \rightarrow 0} \frac{\partial}{\partial k} \langle q^k \rangle.
\end{equation}
So to compute the mean of the logarithm, the replica trick instructs us to compute the moments $\langle q^k \rangle$ for all positive integers $k$, analytically continue $k$ to the complex plane, take a derivative with respect to $k$, and then take a formal limit $k \rightarrow 0$. Due to the necessity of analytic continuation in $k$, the replica trick requires an analytic expression for \emph{all} moments $\langle q^k \rangle$, so it is often challenging to apply this technique in practice. In the present case, however, our mean-field (large-$n$) techniques for Brownian circuits enables us to compute just such an analytic expression.

Our task then is to compute the moments
\begin{equation}
    \langle q^k \rangle = \sum_{\mathbf{x}} \mathbb{E}_U \left[ p(\mathbf{x}) q^k(\mathbf{x}) \right]
    \label{eq:qkmomentsdef}
\end{equation}
where $p(\mathbf{x})$ is the true probability of the physical black-box sampler returning the string $\mathbf{x}$ and $q(\mathbf{x})$ is the ideal transition probability computed by the classical supercomputer. Let us first consider the case where our black-box sampler is actually a perfectly clean quantum machine -- meaning that $p = q$ -- with the circuit $U$ sampled from the all-to-all Brownian circuit ensemble $\mathcal{B}$. In this case we already have the answer from the previous sections (see Eqs. \eqref{eq:brownianmoments} and \eqref{eq:momkxfinal2}):
\begin{align}
    \langle q^k \rangle &\equiv \sum_{\mathbf{x}} m_{a,\mathbf{x}}^{(k+1)} \nonumber \\
    &= \frac{(k+1)!}{2^{n(k+1)}} \left[ (1-a)^{k+1} + (1+a)^{k+1} \right]^n
    \label{eq:qkmoments}
\end{align}
which is an analytic expression in the variables $k,a,n$. Analytically continuing $k$ to the complex plane, taking a single derivative $\partial / \partial k$, and taking the limit $k \rightarrow 0$ we find the simple result
\begin{equation}
    \log \mathrm{XEB} = -\langle \log q \rangle = nH\left(\frac{1+a}{2}\right) + \gamma_e - 1
    \label{eq:exactcleanxeb}
\end{equation}
where $\gamma_e$ is the Euler-Mascheroni constant and $H$ is the binary entropy function. This generalizes the well-known formula \cite{boixo2018characterizing}, which we recover in the limit $a \rightarrow 0$.

We can use the same technology to compute the variance of this score:
\begin{align}
    \mathrm{var}(\mathrm{XEB}) &:= \langle \log^2 q \rangle - \langle \log q \rangle^2 \nonumber \\
    &= \frac{\pi^2}{6} - 1 + n \frac{(1-a^2)}{4} \mathrm{logit}^2 \left( \frac{1+a}{2} \right) \nonumber \\
    &\approx \frac{\pi^2}{6} - 1 + n a^2
    \label{eq:exactcleanxebvar}
\end{align}
where $\mathrm{logit}(p) := \log [p /(1-p)]$ is the log-odds function and in the final line we have expanded in a Taylor series in the small parameter $a$. This is a linear function of system size $n$ with a slope that depends on the circuit depth $a$.
To compute the second moment of the logarithm we used the more general form of the replica trick:
\begin{equation}
    \langle \log^{\ell} Z \rangle = \lim_{k \rightarrow 0} \frac{\partial^{\ell}}{\partial k^{\ell}} \langle Z^k \rangle
\end{equation}
for positive integer $\ell$. The exact expressions in Eqs. \eqref{eq:exactcleanxeb} and \eqref{eq:exactcleanxebvar} illustrate a key point for circuits in the constant-depth regime: whereas the `signal' in the nonlinear cross-entropy is $\mathcal{O}(1)$, the variance scales linearly in $n$. As a result, we only require a polynomial number of samples to accurately estimate the nonlinear cross-entropy. In other words, we see that the $\lXEB$ is sample efficient for shallow depth circuits, at least for the perfectly clean case. Of course, while these calculations are useful for the perfectly idealized noiseless case, we know that noise and dissipation play a critical role in real experiments.

Similar techniques can be used to analyze the scores one expects to see in noisy quantum hardware, and even for certain classical spoofers. For example, if our black box sampler is a noisy quantum system implementing the ideal Brownian circuit $U \in \mathcal{B}$ while also experiencing single-qubit depolarizing noise at a rate $\gamma \ll J$, we find the following expression for the moments (see Appendix \ref{app:noise} for technical details):
\begin{align}
    \langle q^k \rangle_{\gamma} &= \sum_{\mathbf{x}} \frac{k!}{(d b_{\mathbf{x}})^{k+1} } \left[ 1 + k \ e^{- n \beta \gamma} \right] \nonumber \\
    &= \frac{k!}{2^{n(k+1)}} \left[ (1-a)^{k+1} + (1+a)^{k+1} \right]^n \nonumber \\
    & \quad \quad \quad \quad \times \left[ 1 + k e^{- n \beta \gamma} \right]
    \label{eq:qkmomentsnoisy}
\end{align}
which clearly reduces to Eq. \eqref{eq:qkmoments} as $\gamma \rightarrow 0$. Note that the above expression is only valid for very small dissipation rates $\gamma < J / n$; higher noise rates lead to a phase transition in the XEB \cite{ware2023sharp}, which we leave for future study. Although it is not obvious, the moments \eqref{eq:qkmomentsnoisy} also naturally contain the physics of the Harvard spoofing algorithm, which corresponds to taking the formal limit $\gamma \rightarrow \infty$. While this limit is actually unphysical from the perspective of the noisy Brownian circuit model (since it is prohibited by the requirement $n \gamma < J$) our analysis demonstrates that the physics of the Harvard spoofer in the Brownian circuit ensemble is nevertheless correctly captured by this limit, see Appendix \ref{app:harvard}.

Performing the replica trick as before, we find our primary results Eqs. \eqref{eq:noisyxeb} and \eqref{eq:noisyxebvar},
which clearly reduce to Eqs. \eqref{eq:exactcleanxeb} and \eqref{eq:exactcleanxebvar} in the limit $\gamma \rightarrow 0$. Equations \eqref{eq:noisyxeb} and \eqref{eq:noisyxebvar} represent our core technical results, and can be immediately employed to understand various regimes of circuit depth. At shallow circuit depths $\log n \gtrsim \beta J \gg 1$ both the nonlinear cross-entropy and its variance scale as polynomials in $n$ as discussed in Sec. \ref{sec:intro}, indicating the the nonlinear cross-entropy is sample-efficient at both constant depth $\beta J \sim O(1)$ and log depth $\beta J \sim O(\log n)$. By contrast, at linear depths $\beta J \sim O(n)$ the nonlinear cross-entropy is no longer sample efficient to estimate. Assuming the noise is small $n \gamma \sim J$, we find $n \beta \gamma \sim \beta J \sim O(n)$, so the signal $e^{- n \beta \gamma} \sim O(e^{-n})$ is exponentially small in $n$. Meanwhile, the variance at these depths is dominated by the leading constant $c = \pi^2/6$ since everything else in the expression, including $a$, is exponentially small; as a result, one requires an exponential number of samples to accurately estimate the nonlinear cross-entropy in the linear-depth regime. This result essentially demonstrates the oft-repeated lore that the nonlinear cross-entropy is not sample-efficient at polynomial circuit depths.

As an interesting and direct secondary outcome of our Brownian circuit tools, we can also compute the mean and variance of the \emph{linear} XEB in different depth regimes. These calculations demonstrate that the fluctuations are exponentially large compared to the signal for shallow constant-depth circuits \cite{bentsen2024complexity}. By contrast, at log-depth and deeper, with $n a^2 \ll 1$, we find that the signal and fluctuations are comparable, so we expect the linear cross-entropy to be sample-efficient at these depths. We present these results in Appendix \ref{app:linearxeb}.

\section{Heavy Output Generation Classifier}
\label{sec:sshog}

While our results above already provide us with a sample-efficient method for distinguishing perfectly clean quantum hardware from noisy quantum hardware from classical spoofers at shallow depth, we can use the same theoretical tools to construct a binary classifier capable of performing this distinction using far fewer bitstring samples than the nonlinear cross-entropy. In particular, the classifier we construct here allows us to reliably guess which sampler we have even if we have access to only a \emph{single bitstring} $\mathbf{x}$ at a time. In the adversarial setting for example, the sampler could be switched out every time we sample, and our binary classifier is still capable of distinguishing between different kinds of quantum and classical samplers. Furthermore, given multiple samples from the same sampler our binary classifier features \emph{logarithmic} sample complexity, where merely $m \sim \log n$ samples are sufficient to guarantee a success probability of at least $1 - 1/\mathrm{poly(n)}$.

To motivate this new benchmark, we appeal to the notion of `heavy output generation' (HOG) \cite{aaronson2016complexity,boixo2018characterizing,Arute2019-zu}: the propensity for random quantum circuit samplers to produce a few special `heavy' bitstrings with relatively high probability, while the remaining bitstrings are produced with relatively low probability. These heavy bitstrings serve as `fingerprints' of the chaotic quantum circuit $U$ that can be used to certify faithful implementation on the hardware under test. This key property of output distributions of shallow-depth random quantum circuits stands in contrast to trivial samplers like the uniform distribution, which lacks HOG because it produces bitstrings without fear or favor.

To capture this notion quantitatively, we consider fixing a circuit $U$, sampling a single bitstring $\mathbf{x}$ from our sampler and assigning it a score $z := d q(\mathbf{x}) = d\magn{\bra{\mathbf{x}}U\ket{\mathbf{0}}}^2$ proportional to the bitstring's ideal output probability as computed by a classical supercomputer. If our sampler is in fact a high-quality quantum sampler possessing the HOG property, then it is likely that the sampled bitstring $\mathbf{x}$ is one of these preferred `heavy' strings, meaning that it has a relatively large probability $q(\mathbf{x})$ and therefore is likely to receive a high score $z$. On the other hand, for a sampler that does not possess HOG, the bitstring $\mathbf{x}$ will be chosen more or less at random, and the corresponding score $z$ is likely to be much lower.

In the following, we will show how to construct a binary classifier capable of distinguishing with high probability a noisy quantum sampler from a classical spoofer running the Harvard algorithm using only a \emph{single} sample $\mathbf{x}$. More generally, this heavy output generation classifier can also characterize the rate of decoherence in the sampler; as we show below, we generally expect noisy quantum samplers to smoothly interpolate between the ideal clean quantum sampler and the Harvard spoofer, similar to what we saw in previous sections.

\begin{figure*}
    \centering
    \includegraphics[width=0.85\linewidth]{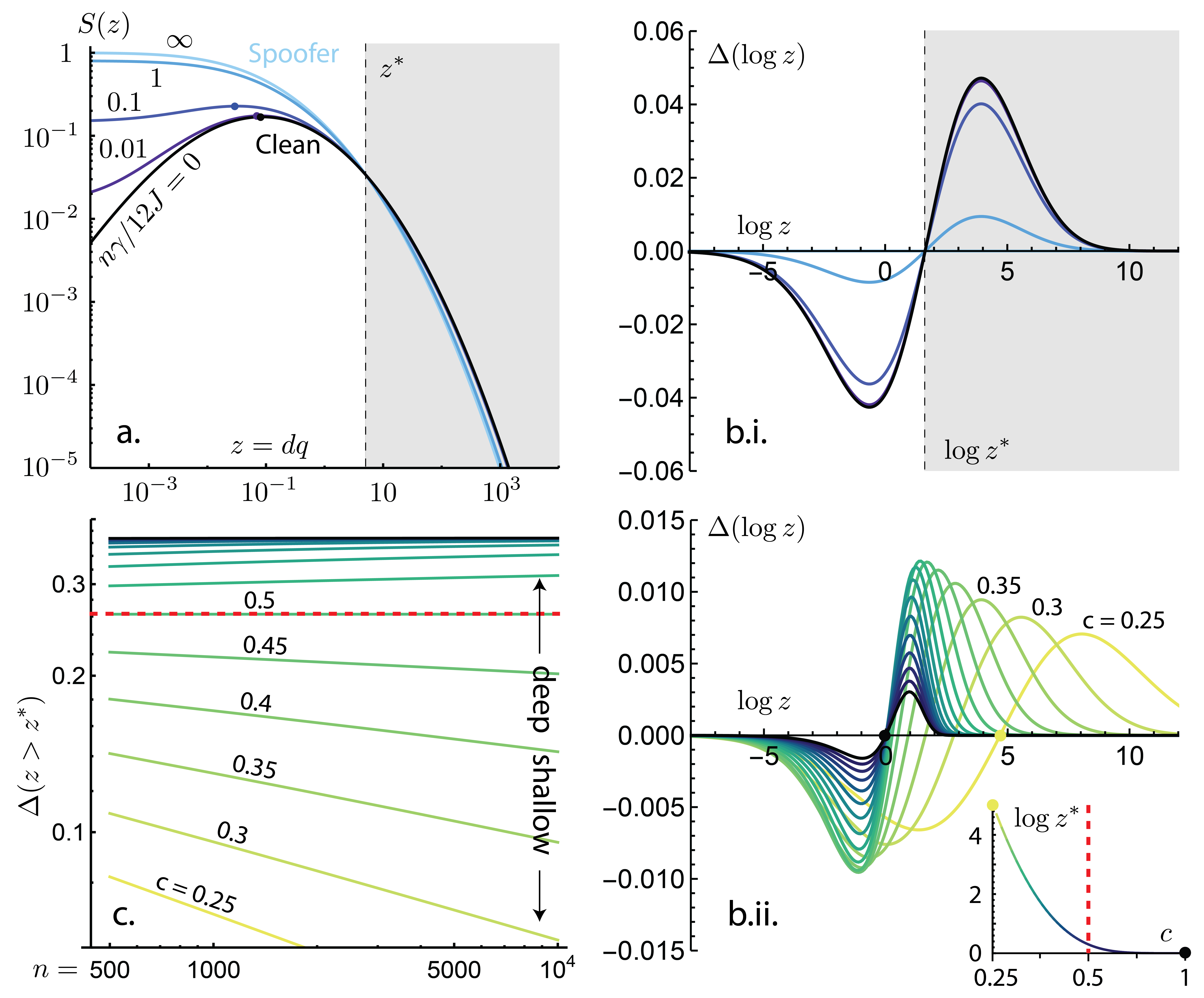}
    \caption{\textbf{The heavy output generation classifier reliably distinguishes} between a clean Brownian circuit, a noisy circuit, and the Harvard spoofing algorithm using only a single bitstring sample. The distribution $S(z)$ over scores $z = d q$ (a) shifts its weight about the decision boundary $z^*$ as we tune the noise rate $n \gamma / 12 J = 0, 0.01, 0.1, 1, \infty$ (black to light blue), where $n = 100$ and $a = 1/5$. The clean system ($\gamma = 0$, black) exhibits `score repulsion' with heavier weight at large scores $z > z^*$ (shaded gray) relative to the classical spoofing algorithm ($\gamma \rightarrow \infty$, light blue) whose distribution is dominated by low scores $z < z^*$. The difference $\Delta(\log z)$ in these two distributions, plotted versus the random variable $\log z$ (b.i.), more clearly demonstrates the relative heaviness of the clean distribution toward scores above threshold $z > z^*$, with the difference in distributions diminishing as the noise rate is increased (black to light blue). Plotting this probability gap $\Delta(\log z)$ for different circuit depths $a = n^{-c}$ with $c = 0.25, 0.3, \ldots, 1$ (b.ii., yellow to dark green) allows us to extract the threshold score $z^*$ as a function of circuit depth (inset). The probability gap $\Delta(z > z^*)$ of obtaining a score above threshold (c) exhibits different scalings with system size $n$ depending on the circuit depth. At shallow depths $c < 1/2$ (yellow to green) the probability gap is finite but diminishes with system size as an inverse polynomial. By contrast, at larger depths $c \geq 1/2$ (green to black) the probability gap is constant or grows monotonically with system size. At the transition point $c = 1/2$ between these two regimes (dotted red) the threshold score $z^* \approx 1.29045$ and the probability gap $\Delta(z > z^*) \approx 0.26263$ are both constant as a function of system size.}
    \label{fig:SSHOG}
\end{figure*}

To make these ideas more concrete, consider generalizing Eqs. \eqref{eq:singlestringporterthomas} and \eqref{eq:porterthomasdeptha} to find the distribution $S(z)$ of scores $z$ obtained from a noisy sampler. To do so, we follow the same procedure as in Section \ref{sec:distributions}: the distribution $S(z)$ is completely determined by its moments
\begin{equation}
    \langle z^k \rangle = \int_0^d dz \ z^k S(z)
\end{equation}
so if we can find expressions for the moments we can construct the moment-generating function and apply the inverse Laplace transform as before to obtain the score distribution $S(z)$. (We emphasize that the score distribution $S(z)$ is distinct from the circuit-averaged probability distribution $P_a(q)$.)
The moment-generating function is given by the sum
\begin{equation}
    N(t) = \sum_k \frac{\langle z^k \rangle}{k!} t^k = \sum_k \frac{\langle q^k \rangle}{k!} (dt)^k
\end{equation}
where $\langle q^k \rangle$ are the moments we already computed in Eq. \eqref{eq:qkmomentsnoisy}. Unfortunately we are not aware of an analytic expression for this sum. Instead, we return to Eq. \eqref{eq:qkmomentsnoisy} and pull the sum over bitstrings outside, yielding
\begin{align}
    N(t) &= \sum_{\mathbf{x}} \sum_k \frac{1}{(d \bx)^{k+1} } \left[ 1 + k \ e^{- n \beta \gamma} \right] (dt)^k \nonumber \\
    &= \frac{1}{d} \sum_{x} \binom{n}{x} \frac{\bx - t(1 - e^{-n \beta \gamma})}{(\bx - t)^2}
\end{align}
Finally we apply the inverse Laplace transform to obtain
\begin{align}
    S(z) &= \mathcal{L}^{-1} \left[ N (-t) \right] \nonumber \\
    &= d^{-1} \sum_{x} \binom{n}{x} \left[1 + e^{-n \beta \gamma} (\bx z - 1) \right] e^{- \bx z}.
\end{align}
where the sum over bitstrings can be pulled outside because the inverse Laplace transform is a linear operation. As before, the sum over bitstrings can be performed efficiently because the factors $\bx$ depend only on the Hamming weight of the string.

Performing this sum numerically yields plots such as those shown in Fig. \ref{fig:SSHOG}.a. for $n = 100$ qubits and circuit depth $a = 1/5$. In the absence of noise $\gamma = 0$ (`Clean,' black) the score distribution $S(z)$ exhibits `score repulsion,' with small scores suppressed relative to the most likely score (black dot). Score repulsion disappears as we crank up the noise rate $\gamma$ (purple to light blue), with extremely small scores dominating the distribution at large noise rates. The formal limit $\gamma \rightarrow \infty$ yields the score distribution for the classical Harvard spoofing algorithm (`Spoofer,' light blue) where score repulsion is completely absent and small scores dominate. As we tune the noise rate $\gamma$ up and down we see that the weight of the distribution $S(z)$ shifts about a central `pivot' point $z^*$ marked with a vertical dotted line, with the clean sampler distribution exhibiting higher weight at large scores $z > z^*$ relative to the classical spoofer. As a result, the pivot point $z^*$ acts as a classification boundary: scores larger than $z^*$ (gray region) are statistically more likely to have been generated by a clean quantum sampler than by a noisy sampler or the classical Harvard spoofer. By the same token, scores smaller than the threshold $z^*$ are more likely to have been generated by the spoofer than by the clean quantum sampler.

We can illustrate this point even more clearly by numerically computing the difference in distributions
\begin{equation}
    \Delta(z) := \left. S(z) \right\rvert_{\gamma = 0} - \left. S(z) \right\rvert_{\gamma = \infty}
\end{equation}
between the ideal clean sampler and the Harvard sampler (this is the largest possible difference since the distribution $S(z) \rvert_{\gamma}$ for a noisy sampler with finite $\gamma > 0$ always lies somewhere in between the clean and spoofer distributions). It is convenient to plot the difference as a function of $\log z$ (Fig. \ref{fig:SSHOG}.b.i.). The distribution difference is positive for scores above the threshold $z^*$, indicating a bias toward higher scores $z > z^*$ (gray region) relative to the classical spoofer.

This rightward bias in the score distribution $S(z)$ about the pivot point $z^*$ suggests a procedure for distinguishing a clean sampler from a noisy one: we sample a bitstring $\mathbf{x}$ and compute its score $z$; if this score is above threshold $z > z^*$ then we are likely to have a clean sampler, otherwise it is likely a noisy sampler. This procedure does not guarantee successful identification every time (since it is always possible to get unlucky and sample a below-threshold bitstring from the clean sampler or an above-threshold bitstring from the noisy sampler) but a straightforward application of game theory yields a strategy that successfully identifies which sampler we have more often than not.

To make this more concrete, we can phrase the procedure in the language of a betting game: every `round' we are given one of two possible samplers $A$ or $B$ (say a quantum sampler versus a classical spoofer) to examine. We aim to make money by attempting to correctly guess which sampler, $A$ or $B$, is in front of us. If we are correct, we win one dollar; if we are wrong we lose one dollar. Without further information we might as well toss a coin, and our expected value in the long run is zero. Of course we can improve our odds by acquiring information about the sampler. To inform our decision we may select one circuit $U$ from the ensemble $\mathcal{B}$ and extract one bitstring $\mathbf{x}$ from the sampler. We send that bitstring to a classical supercomputer to assign a score $z = d q(\mathbf{x})$. If this score is larger than the threshold (decision boundary) $z > z^*$ then the sampler in front of us is more likely to be a noisy quantum sampler as opposed to a classical spoofer.

This betting game is equivalent to differentiating between two biased coins $A,B$, where coin $A$ has probability $p_A$ of coming up heads -- corresponding to the quantum sampler $A$ giving an above threshold score ($z > z^*$) -- while coin $B$ has probability $p_B$ of coming up heads -- corresponding to the classical spoofer $B$ giving an above threshold score ($z > z^*$) where $p_A > p_B$. A sensible strategy is to guess coin $A$ whenever we observe `heads' and guess coin $B$ whenever we observe `tails.' This \emph{pure strategy} yields an expected value of $2 p_A-1$ if we are given coin $A$ and $1-2 p_B$ if we are given coin $B$, which are both positive when $p_A > 1/2 > p_B$. However, if $p_A \leq 1/2$ or $p_B \geq 1/2$ we must consider a more general strategy. Assuming that we know the probabilities $p_A,p_B$ ahead of time (without loss of generality, we take $p_A > p_B \geq 1/2$), a straightforward exercise in game theory (see Appendix \ref{app:sshogbetting}) demonstrates that the optimal approach is a \emph{mixed strategy} in which we apply the `sensible' pure strategy above with probability $x = 1/(p_A+p_B)$, and we blindly guess coin $B$ with probability $1-x$. This mixed strategy yields an expected value of $(p_A - p_B) / (p_A + p_B)$ which is always positive.

We can do even better by sampling multiple times per round. For ease of discussion, we continue to use the language of biased coins $A,B$ as outlined above, where we may now flip each coin $m$ times per round. Our strategy in this case is to count the total number of heads $h$ we obtain, and guess coin $A$ if the number of heads $h > h^*$ is above a threshold $h^*$ and guess coin $B$ if it is below threshold $h \leq h^*$. We emphasize that the threshold $h^*$ for the requisite number of heads is distinct from the HOG score threshold $z^*$. To pick the optimal threshold $h^*$ we demand that the probability of guessing correctly is the same regardless of which coin we are given:
\begin{equation}
    P_{\mathrm{correct}} = P(h > h^* | A) = P(h \leq h^* | B).
    \label{eq:mflipscorrectprob}
\end{equation}
In the limit of large $m$ it is straightforward to show (see Appendix \ref{app:sshogbetting}) that this condition is equivalent to
\begin{equation}
    \Phi \left(\frac{\mu_A - h^*}{\sigma_A} \right) = \Phi \left(\frac{h^* - \mu_B}{\sigma_B} \right)
    \label{eq:cdfequality}
\end{equation}
where $\mu_{A,B} := m p_{A,B}$ and $\sigma_{A,B} = m \ p_{A,B} (1- p_{A,B})$ are the mean and variance of the number of heads we expect from coin $A$ or $B$ respectively, and $\Phi$ is the cumulative distribution function (CDF) of the standard normal distribution. Solving for the optimal threshold yields
\begin{equation}
    h^* = m \frac{p_A \sigma_B + p_B \sigma_A}{\sigma_A + \sigma_B}
    \label{eq:headsthresh}
\end{equation}
which gives a success probability
\begin{equation}
    P_{\mathrm{correct}} = \Phi \left( \sqrt{m} \ x \right) \approx 1 - \frac{1}{\sqrt{2 \pi m} \ x} e^{-m x^2 / 2}
\end{equation}
in the limit of large $m$, where
\begin{equation}
    x := \frac{p_A - p_B}{\sqrt{p_A(1-p_A)} + \sqrt{p_B(1-p_B)}}.
    \label{eq:xdef}
\end{equation}
Therefore we require only $m \sim \log n$ samples in order to achieve a success probability of $1 - 1/\mathrm{poly}(n)$ or greater.

To examine how this procedure scales with system size $n$ and circuit depth $a$ we compute the threshold score $z^*$ by numerically finding the root of the probability difference $\Delta(z)$ as illustrated in Fig. \ref{fig:SSHOG}.b.ii. The threshold score $z^*$ is large at shallow depths (yellow) and rapidly decreases to $z^* = 1$ at large circuit depths (dark green), where the circuit depth is parameterized by $a = n^{-c}$ for $c = 0.25, 0.3, \ldots, 1$ (yellow to dark green). Practically speaking, this means that it is `obvious' to identify a genuine quantum sampler in the shallow depth regime because it tends to generate bitstrings with very large scores $z > z^* \gg 1$; this stands in stark contrast to a trivial uniform sampler which tends to produce bitstrings with low scores $z \sim 1$. With the threshold score $z^*$ in hand, we then numerically compute the probability difference for obtaining a score above threshold:
\begin{equation}
    \Delta(z > z^*) := \int_{z^*}^{\infty} dz \ \Delta(z).
\end{equation}
which we refer to as the \emph{probability gap}. A nonzero gap means that there is a larger probability for the clean quantum sampler to return a score above threshold relative to the classical Harvard spoofer, indicating that we can reliably distinguish these two samplers.

In order for our procedure to be useful at arbitrary system sizes we must ensure that this probability gap remains finite even in the asymptotic limit $n \rightarrow \infty$. In the shallow regime $c = 0.25, \ldots, 0.45$ (Fig. \ref{fig:SSHOG}.c, yellow to green) the probability gap decreases monotonically with $n$ as an inverse polynomial, indicating that our procedure fails asymptotically in this regime. Despite scaling poorly with system size in this regime, the gap is still substantial at small systems sizes and could still be used in principle to reliably distinguish samplers at small system sizes and shallow circuit depths. By contrast, at larger circuit depths $c = 0.5, \ldots, 1$ (Fig. \ref{fig:SSHOG}.c, green to dark green) the probability gap is constant or grows monotonically with system size, indicating a scalable metric in this depth regime capable of reliably distinguishing between different samplers. In other words, the HOG classifier is reliable asymptotically in the log-depth regime $12 \beta J > 1/2 \log n$ but not at shallower depths. The crossover point $c = 1/2$ (dotted red) between these two depth regimes is the same crossover point identified in Fig. \ref{fig:PTBrownianSetup}.b.

\section{Outlook}
\label{sec:discussion}

While experimental milestones achieved with superconducting qubits led by groups from, for example, Google \cite{Arute2019-zu,google2025quantum} and IBM \cite{kim2023evidence,abughanem2025ibm}, have attracted great interest in understanding the power and limitations of geometrically local quantum circuits, especially in 2D, quantum circuit models with an all-to-all interaction graph have received comparatively less attention. More recently, neutral atom quantum computing with reconfigurable atomic qubits has emerged as another strong and promising candidate for realizing fault-tolerant quantum computation in the near future \cite{bluvstein2024logical,manetsch2025tweezer,sales2025experimental,holman2026trapping,cain2026shor}. The same is true for trapped ion platforms, where trapped-ion systems developed by groups such as Quantinuum/Honeywell, IonQ, and Oxford Ionics have demonstrated high-fidelity gates, long-range or effectively all-to-all connectivity, and increasingly sophisticated error-correction primitives. Recent milestones include random-circuit-sampling experiments on Quantinuum's 56-qubit H2 trapped-ion processor with arbitrary connectivity and other interesting features \cite{quantinuum2024rcs,paetznick2024logical,loschnauer2024electronic,perlin2026faulttolerant}.

Compared with superconducting nearest-neighbor architectures, neutral atom and ionic platforms allow for long-range interactions and dynamically reconfigurable interaction graphs at the level of physical qubits. Furthermore, neutral atom architectures may support fault-tolerant schemes with significantly more reconfigurable and lower-overhead logical connectivity than fixed nearest-neighbor superconducting architectures; for example, when combining LDPC-style error-correcting codes \cite{pecorari2025high,guo2026toward} with erasure-aware decoding \cite{wu2022erasure,chow2024circuit,pecorari2025quantum}. As a result, quantum circuit models that allow long-range or $2$-qubit gates between any pair of qubits in a given circuit layer more accurately capture the computational capability of neutral atom quantum hardware. In the near future, logical random circuit sampling evaluated on the nonlinear XEB benchmark may become a valuable tool for assessing quantum advantage and the system-wide logical error rate of error-corrected neutral atom quantum computers with a few dozen logical qubits. We believe the nonlinear XEB benchmark could be especially well-suited for this parameter regime because (1) the number of logical qubits is still in the dozens so that classically computing ideal output probabilities remains feasible, (2) the logical error rate is low enough to be well-modeled by $\propto 1/n$, and (3) long-range logical $2$-qubit gates are available, at least to some extent.

There are also several promising directions for future theoretical work based on our Brownian random circuit methods. Inspired by the binary classifier introduced in Section \ref{sec:sshog}, one might hope to develop a similar binary classifier based on the nonlinear cross-entropy, where samplers generating scores above a certain threshold are classified as genuine quantum samplers and those generating scores below threshold are classified as classical spoofers. In the best case scenario this classifier would also feature logarithmic sample complexity similar to the HOG classifier. Another promising direction would be to apply Brownian circuit models to Bell sampling \cite{hangleiter2024bell} where our methods would allow for direct calculation of higher-moment quantities which could extract additional information about state fidelity, circuit errors and T-gate count. Further, whereas the methods employed here become unreliable at extremely short circuit depths $12 \beta J \leq \log 4$, one can employ relatively simple numerical methods at large-$n$ to capture the time-dependent dynamics occurring at extremely short depths \cite{bentsen2021measurement}. Finally, we anticipate adding different types of noise, including non-unital channels and imperfect circuit implementation, to the Brownian circuit models to further refine our results to realistic experimental scenarios.

\section{Acknowledgments}
\label{sec:ack}

B.F. and S.G. ~acknowledge support from AFOSR
(FA9550-21-1-0008).  This material is based upon work partially
supported by the National Science Foundation under Grant CCF-2044923
(CAREER), by the U.S. Department of Energy, Office of Science,
National Quantum Information Science Research Centers (Q-NEXT) and by
the DOE QuantISED grant DE-SC0020360.  This work was done in part while a subset of the authors were visiting the Simons Institute for the Theory of Computing, supported by NSF QLCI Grant No. 2016245.

\bibliography{References}

@article{gray2018quimb,
    title={quimb: a python library for quantum information and many-body calculations},
    author={Gray, Johnnie},
    journal={Journal of Open Source Software},
    year = {2018},
    volume={3}, number={29}, pages={819},
    doi={10.21105/joss.00819},
}

@misc{quantinuum2024rcs,
  title         = {The computational power of random quantum circuits in arbitrary geometries},
  author        = {{Quantinuum Collaboration}},
  year          = {2024},
  eprint        = {2406.02501},
  archivePrefix = {arXiv},
  primaryClass  = {quant-ph}
}

@misc{paetznick2024logical,
  title         = {Demonstration of logical qubits and repeated error correction with better-than-physical error rates},
  author        = {Paetznick, A. and da Silva, M. P. and Ryan-Anderson, C. and Bello-Rivas, J. M. and Campora III, J. P. and Chernoguzov, A. and Dreiling, J. M. and Foltz, C. and Frachon, F. and Gaebler, J. P. and others},
  year          = {2024},
  eprint        = {2404.02280},
  archivePrefix = {arXiv},
  primaryClass  = {quant-ph}
}

@misc{loschnauer2024electronic,
  title         = {Scalable, high-fidelity all-electronic control of trapped-ion qubits},
  author        = {L{\"o}schnauer, C. M. and Mosca Toba, J. and Hughes, A. C. and King, S. A. and Weber, M. A. and Srinivas, R. and Matt, R. and Nourshargh, R. and Allcock, D. T. C. and Ballance, C. J. and Matthiesen, C. and Malinowski, M. and Harty, T. P.},
  year          = {2024},
  eprint        = {2407.07694},
  archivePrefix = {arXiv},
  primaryClass  = {quant-ph}
}

@misc{perlin2026faulttolerant,
  title         = {Fault-tolerant execution of error-corrected quantum algorithms},
  author        = {Perlin, Michael and He, Zichang and Armenakas, Anthony Alexiades and Andres-Martinez, Pablo and Hao, Tianyi and Herman, Dylan and Jin, Yuwei and Mayer, Karl and Self, Chris and Amaro, David and Ryan-Anderson, Ciaran and Shaydulin, Ruslan},
  year          = {2026},
  eprint        = {2603.04584},
  archivePrefix = {arXiv},
  primaryClass  = {quant-ph}
}

@misc{bouland2025exponentialimprovementsaveragecasehardness,
      title={Exponential improvements to the average-case hardness of BosonSampling}, 
      author={Adam Bouland and Ishaun Datta and Bill Fefferman and Felipe Hernandez},
      year={2025},
      eprint={2411.04566},
      archivePrefix={arXiv},
      primaryClass={quant-ph},
      url={https://arxiv.org/abs/2411.04566}, 
}

@article{Deshpande_2022,
   title={Tight Bounds on the Convergence of Noisy Random Circuits to the Uniform Distribution},
   volume={3},
   ISSN={2691-3399},
   url={http://dx.doi.org/10.1103/PRXQuantum.3.040329},
   DOI={10.1103/prxquantum.3.040329},
   number={4},
   journal={PRX Quantum},
   publisher={American Physical Society (APS)},
   author={Deshpande, Abhinav and Niroula, Pradeep and Shtanko, Oles and Gorshkov, Alexey V. and Fefferman, Bill and Gullans, Michael J.},
   year={2022},
   month=Dec }

@article{bravyi2022simulate,
  title={How to simulate quantum measurement without computing marginals},
  author={Bravyi, Sergey and Gosset, David and Liu, Yinchen},
  journal={Physical Review Letters},
  volume={128},
  number={22},
  pages={220503},
  year={2022},
  publisher={APS}
}

@article{mcginley2025measurement,
  title={Measurement-induced entanglement and complexity in random constant-depth 2D quantum circuits},
  author={McGinley, Max and Ho, Wen Wei and Malz, Daniel},
  journal={Physical Review X},
  volume={15},
  number={2},
  pages={021059},
  year={2025},
  publisher={APS}
}

@article{bene2025quantum,
  title={Quantum advantage from measurement-induced entanglement in random shallow circuits},
  author={Bene Watts, Adam and Gosset, David and Liu, Yinchen and Soleimanifar, Mehdi},
  journal={PRX Quantum},
  volume={6},
  number={1},
  pages={010356},
  year={2025},
  publisher={APS}
}

@InProceedings{barak2020spoofing,
  author =	{Barak, Boaz and Chou, Chi-Ning and Gao, Xun},
  title =	{{Spoofing Linear Cross-Entropy Benchmarking in Shallow Quantum Circuits}},
  booktitle =	{12th Innovations in Theoretical Computer Science Conference (ITCS 2021)},
  pages =	{30:1--30:20},
  series =	{Leibniz International Proceedings in Informatics (LIPIcs)},
  ISBN =	{978-3-95977-177-1},
  ISSN =	{1868-8969},
  year =	{2021},
  volume =	{185},
  editor =	{Lee, James R.},
  publisher =	{Schloss Dagstuhl -- Leibniz-Zentrum f{\"u}r Informatik},
  address =	{Dagstuhl, Germany},
  URL =		{https://drops.dagstuhl.de/entities/document/10.4230/LIPIcs.ITCS.2021.30},
  URN =		{urn:nbn:de:0030-drops-135699},
  doi =		{10.4230/LIPIcs.ITCS.2021.30}
}

@misc{movassagh2020quantumsupremacyrandomcircuits,
      title={Quantum supremacy and random circuits}, 
      author={Ramis Movassagh},
      year={2020},
      eprint={1909.06210},
      archivePrefix={arXiv},
      primaryClass={quant-ph},
      url={https://arxiv.org/abs/1909.06210}, 
}

@misc{zhu2021quantumcomputationaladvantage60qubit,
      title={Quantum Computational Advantage via 60-Qubit 24-Cycle Random Circuit Sampling}, 
      author={Qingling Zhu and Sirui Cao and Fusheng Chen and Ming-Cheng Chen and Xiawei Chen and Tung-Hsun Chung and Hui Deng and Yajie Du and Daojin Fan and Ming Gong and Cheng Guo and Chu Guo and Shaojun Guo and Lianchen Han and Linyin Hong and He-Liang Huang and Yong-Heng Huo and Liping Li and Na Li and Shaowei Li and Yuan Li and Futian Liang and Chun Lin and Jin Lin and Haoran Qian and Dan Qiao and Hao Rong and Hong Su and Lihua Sun and Liangyuan Wang and Shiyu Wang and Dachao Wu and Yulin Wu and Yu Xu and Kai Yan and Weifeng Yang and Yang Yang and Yangsen Ye and Jianghan Yin and Chong Ying and Jiale Yu and Chen Zha and Cha Zhang and Haibin Zhang and Kaili Zhang and Yiming Zhang and Han Zhao and Youwei Zhao and Liang Zhou and Chao-Yang Lu and Cheng-Zhi Peng and Xiaobo Zhu and Jian-Wei Pan},
      year={2021},
      eprint={2109.03494},
      archivePrefix={arXiv},
      primaryClass={quant-ph},
      url={https://arxiv.org/abs/2109.03494}, 
}

@misc{ware2023sharp,
      title={A sharp phase transition in linear cross-entropy benchmarking}, 
      author={Brayden Ware and Abhinav Deshpande and Dominik Hangleiter and Pradeep Niroula and Bill Fefferman and Alexey V. Gorshkov and Michael J. Gullans},
      year={2023},
      eprint={2305.04954},
      archivePrefix={arXiv},
      primaryClass={quant-ph},
      url={https://arxiv.org/abs/2305.04954}, 
}

@article{Dalzell_2022,
   title={Random Quantum Circuits Anticoncentrate in Log Depth},
   volume={3},
   ISSN={2691-3399},
   url={http://dx.doi.org/10.1103/PRXQuantum.3.010333},
   DOI={10.1103/prxquantum.3.010333},
   number={1},
   journal={PRX Quantum},
   publisher={American Physical Society (APS)},
   author={Dalzell, Alexander M. and Hunter-Jones, Nicholas and Brandão, Fernando G. S. L.},
   year={2022},
   month=mar }

@misc{aaronson2020classicalhardnessspoofinglinear,
      title={On the Classical Hardness of Spoofing Linear Cross-Entropy Benchmarking}, 
      author={Scott Aaronson and Sam Gunn},
      year={2020},
      eprint={1910.12085},
      archivePrefix={arXiv},
      primaryClass={quant-ph},
      url={https://arxiv.org/abs/1910.12085}, 
}

@inproceedings{Bouland_2022,
   title={Noise and the Frontier of Quantum Supremacy},
   url={http://dx.doi.org/10.1109/FOCS52979.2021.00127},
   DOI={10.1109/focs52979.2021.00127},
   booktitle={2021 IEEE 62nd Annual Symposium on Foundations of Computer Science (FOCS)},
   publisher={IEEE},
   author={Bouland, Adam and Fefferman, Bill and Landau, Zeph and Liu, Yunchao},
   year={2022},
   month=feb }

@article{Bouland_2018,
   title={On the complexity and verification of quantum random circuit sampling},
   volume={15},
   ISSN={1745-2481},
   url={http://dx.doi.org/10.1038/s41567-018-0318-2},
   DOI={10.1038/s41567-018-0318-2},
   number={2},
   journal={Nature Physics},
   publisher={Springer Science and Business Media LLC},
   author={Bouland, Adam and Fefferman, Bill and Nirkhe, Chinmay and Vazirani, Umesh},
   year={2018},
   month=oct, pages={159–163} }

@misc{morvan2023phasetransitionrandomcircuit,
      title={Phase transition in Random Circuit Sampling}, 
      author={A. Morvan and B. Villalonga and X. Mi and S. Mandrà and A. Bengtsson and P. V. Klimov and Z. Chen and S. Hong and C. Erickson and I. K. Drozdov and J. Chau and G. Laun and R. Movassagh and A. Asfaw and L. T. A. N. Brandão and R. Peralta and D. Abanin and R. Acharya and R. Allen and T. I. Andersen and K. Anderson and M. Ansmann and F. Arute and K. Arya and J. Atalaya and J. C. Bardin and A. Bilmes and G. Bortoli and A. Bourassa and J. Bovaird and L. Brill and M. Broughton and B. B. Buckley and D. A. Buell and T. Burger and B. Burkett and N. Bushnell and J. Campero and H. S. Chang and B. Chiaro and D. Chik and C. Chou and J. Cogan and R. Collins and P. Conner and W. Courtney and A. L. Crook and B. Curtin and D. M. Debroy and A. Del Toro Barba and S. Demura and A. Di Paolo and A. Dunsworth and L. Faoro and E. Farhi and R. Fatemi and V. S. Ferreira and L. Flores Burgos and E. Forati and A. G. Fowler and B. Foxen and G. Garcia and E. Genois and W. Giang and C. Gidney and D. Gilboa and M. Giustina and R. Gosula and A. Grajales Dau and J. A. Gross and S. Habegger and M. C. Hamilton and M. Hansen and M. P. Harrigan and S. D. Harrington and P. Heu and M. R. Hoffmann and T. Huang and A. Huff and W. J. Huggins and L. B. Ioffe and S. V. Isakov and J. Iveland and E. Jeffrey and Z. Jiang and C. Jones and P. Juhas and D. Kafri and T. Khattar and M. Khezri and M. Kieferová and S. Kim and A. Kitaev and A. R. Klots and A. N. Korotkov and F. Kostritsa and J. M. Kreikebaum and D. Landhuis and P. Laptev and K. -M. Lau and L. Laws and J. Lee and K. W. Lee and Y. D. Lensky and B. J. Lester and A. T. Lill and W. Liu and W. P. Livingston and A. Locharla and F. D. Malone and O. Martin and S. Martin and J. R. McClean and M. McEwen and K. C. Miao and A. Mieszala and S. Montazeri and W. Mruczkiewicz and O. Naaman and M. Neeley and C. Neill and A. Nersisyan and M. Newman and J. H. Ng and A. Nguyen and M. Nguyen and M. Yuezhen Niu and T. E. O'Brien and S. Omonije and A. Opremcak and A. Petukhov and R. Potter and L. P. Pryadko and C. Quintana and D. M. Rhodes and E. Rosenberg and C. Rocque and P. Roushan and N. C. Rubin and N. Saei and D. Sank and K. Sankaragomathi and K. J. Satzinger and H. F. Schurkus and C. Schuster and M. J. Shearn and A. Shorter and N. Shutty and V. Shvarts and V. Sivak and J. Skruzny and W. C. Smith and R. D. Somma and G. Sterling and D. Strain and M. Szalay and D. Thor and A. Torres and G. Vidal and C. Vollgraff Heidweiller and T. White and B. W. K. Woo and C. Xing and Z. J. Yao and P. Yeh and J. Yoo and G. Young and A. Zalcman and Y. Zhang and N. Zhu and N. Zobrist and E. G. Rieffel and R. Biswas and R. Babbush and D. Bacon and J. Hilton and E. Lucero and H. Neven and A. Megrant and J. Kelly and I. Aleiner and V. Smelyanskiy and K. Kechedzhi and Y. Chen and S. Boixo},
      year={2023},
      eprint={2304.11119},
      archivePrefix={arXiv},
      primaryClass={quant-ph},
      url={https://arxiv.org/abs/2304.11119}, 
}

@inproceedings{Aharonov_2023, series={STOC ’23},
   title={A Polynomial-Time Classical Algorithm for Noisy Random Circuit Sampling},
   url={http://dx.doi.org/10.1145/3564246.3585234},
   DOI={10.1145/3564246.3585234},
   booktitle={Proceedings of the 55th Annual ACM Symposium on Theory of Computing},
   publisher={ACM},
   author={Aharonov, Dorit and Gao, Xun and Landau, Zeph and Liu, Yunchao and Vazirani, Umesh},
   year={2023},
   month=jun, collection={STOC ’23} }

@ARTICLE{Arute2019-zu,
  title     = "Quantum supremacy using a programmable superconducting processor",
  author    = "Arute, Frank and Arya, Kunal and Babbush, Ryan and Bacon, Dave
               and Bardin, Joseph C and Barends, Rami and Biswas, Rupak and
               Boixo, Sergio and Brandao, Fernando G S L and Buell, David A and
               Burkett, Brian and Chen, Yu and Chen, Zijun and Chiaro, Ben and
               Collins, Roberto and Courtney, William and Dunsworth, Andrew and
               Farhi, Edward and Foxen, Brooks and Fowler, Austin and Gidney,
               Craig and Giustina, Marissa and Graff, Rob and Guerin, Keith and
               Habegger, Steve and Harrigan, Matthew P and Hartmann, Michael J
               and Ho, Alan and Hoffmann, Markus and Huang, Trent and Humble,
               Travis S and Isakov, Sergei V and Jeffrey, Evan and Jiang, Zhang
               and Kafri, Dvir and Kechedzhi, Kostyantyn and Kelly, Julian and
               Klimov, Paul V and Knysh, Sergey and Korotkov, Alexander and
               Kostritsa, Fedor and Landhuis, David and Lindmark, Mike and
               Lucero, Erik and Lyakh, Dmitry and Mandr{\`a}, Salvatore and
               McClean, Jarrod R and McEwen, Matthew and Megrant, Anthony and
               Mi, Xiao and Michielsen, Kristel and Mohseni, Masoud and Mutus,
               Josh and Naaman, Ofer and Neeley, Matthew and Neill, Charles and
               Niu, Murphy Yuezhen and Ostby, Eric and Petukhov, Andre and
               Platt, John C and Quintana, Chris and Rieffel, Eleanor G and
               Roushan, Pedram and Rubin, Nicholas C and Sank, Daniel and
               Satzinger, Kevin J and Smelyanskiy, Vadim and Sung, Kevin J and
               Trevithick, Matthew D and Vainsencher, Amit and Villalonga,
               Benjamin and White, Theodore and Yao, Z Jamie and Yeh, Ping and
               Zalcman, Adam and Neven, Hartmut and Martinis, John M",
  journal   = "Nature",
  publisher = "Springer Science and Business Media LLC",
  volume    =  574,
  number    =  7779,
  pages     = "505--510",
  month     =  oct,
  year      =  2019,
  language  = "en"
}

@article{maldacena2016remarks,
  title = {Remarks on the Sachdev-Ye-Kitaev model},
  author = {Maldacena, Juan and Stanford, Douglas},
  journal = {Phys. Rev. D},
  volume = {94},
  issue = {10},
  pages = {106002},
  numpages = {43},
  year = {2016},
  month = {Nov},
  publisher = {American Physical Society},
  doi = {10.1103/PhysRevD.94.106002},
  url = {https://link.aps.org/doi/10.1103/PhysRevD.94.106002}
}

@article{gao2024limitations,
  title = {Limitations of Linear Cross-Entropy as a Measure for Quantum Advantage},
  author = {Gao, Xun and Kalinowski, Marcin and Chou, Chi-Ning and Lukin, Mikhail D. and Barak, Boaz and Choi, Soonwon},
  journal = {PRX Quantum},
  volume = {5},
  issue = {1},
  pages = {010334},
  numpages = {27},
  year = {2024},
  month = {Feb},
  publisher = {American Physical Society},
  doi = {10.1103/PRXQuantum.5.010334},
  url = {https://link.aps.org/doi/10.1103/PRXQuantum.5.010334}
}

@article{edwards1975theory,
doi = {10.1088/0305-4608/5/5/017},
url = {https://dx.doi.org/10.1088/0305-4608/5/5/017},
year = {1975},
month = {may},
publisher = {},
volume = {5},
number = {5},
pages = {965},
author = {S F Edwards and  P W Anderson},
title = {Theory of spin glasses},
journal = {Journal of Physics F: Metal Physics}
}

@article{sherrington1975solvable,
  title = {Solvable Model of a Spin-Glass},
  author = {Sherrington, David and Kirkpatrick, Scott},
  journal = {Phys. Rev. Lett.},
  volume = {35},
  issue = {26},
  pages = {1792--1796},
  numpages = {0},
  year = {1975},
  month = {Dec},
  publisher = {American Physical Society},
  doi = {10.1103/PhysRevLett.35.1792},
  url = {https://link.aps.org/doi/10.1103/PhysRevLett.35.1792}
}

@article{almeida1978stability,
doi = {10.1088/0305-4470/11/5/028},
url = {https://dx.doi.org/10.1088/0305-4470/11/5/028},
year = {1978},
month = {may},
publisher = {},
volume = {11},
number = {5},
pages = {983},
author = {J R L de Almeida and  D J Thouless},
title = {Stability of the Sherrington-Kirkpatrick solution of a spin glass model},
journal = {Journal of Physics A: Mathematical and General}
}

@article{thouless1977solution,
author = {D. J. Thouless, P. W. Anderson and R. G. Palmer},
title = {Solution of 'Solvable model of a spin glass'},
journal = {The Philosophical Magazine: A Journal of Theoretical Experimental and Applied Physics},
volume = {35},
number = {3},
pages = {593--601},
year = {1977},
publisher = {Taylor \& Francis},
doi = {10.1080/14786437708235992},
URL = {https://doi.org/10.1080/14786437708235992},
eprint = {https://doi.org/10.1080/14786437708235992}
}

@article{parisi1979toward,
title = {Toward a mean field theory for spin glasses},
journal = {Physics Letters A},
volume = {73},
number = {3},
pages = {203-205},
year = {1979},
issn = {0375-9601},
doi = {https://doi.org/10.1016/0375-9601(79)90708-4},
url = {https://www.sciencedirect.com/science/article/pii/0375960179907084},
author = {G. Parisi}
}

@article{gross1984simplest,
title = {The simplest spin glass},
journal = {Nuclear Physics B},
volume = {240},
number = {4},
pages = {431-452},
year = {1984},
issn = {0550-3213},
doi = {https://doi.org/10.1016/0550-3213(84)90237-2},
url = {https://www.sciencedirect.com/science/article/pii/0550321384902372},
author = {D.J. Gross and M. Mezard}
}

@book{mezard1986spin,
author = {Mezard, M and Parisi, G and Virasoro, M},
title = {Spin Glass Theory and Beyond},
publisher = {WORLD SCIENTIFIC},
year = {1986},
doi = {10.1142/0271},
address = {},
edition   = {},
URL = {https://www.worldscientific.com/doi/abs/10.1142/0271},
eprint = {https://www.worldscientific.com/doi/pdf/10.1142/0271}
}

@book{young1997spin,
author = {Young, A P},
title = {Spin Glasses and Random Fields},
publisher = {WORLD SCIENTIFIC},
year = {1997},
doi = {10.1142/3517},
address = {},
edition   = {},
URL = {https://www.worldscientific.com/doi/abs/10.1142/3517},
eprint = {https://www.worldscientific.com/doi/pdf/10.1142/3517}
}

@article{bray1980replica,
doi = {10.1088/0022-3719/13/24/005},
url = {https://dx.doi.org/10.1088/0022-3719/13/24/005},
year = {1980},
month = {aug},
publisher = {},
volume = {13},
number = {24},
pages = {L655},
author = {A J Bray and  M A Moore},
title = {Replica theory of quantum spin glasses},
journal = {Journal of Physics C: Solid State Physics}
}

@article{sachdev1993gapless,
  title = {Gapless spin-fluid ground state in a random quantum Heisenberg magnet},
  author = {Sachdev, Subir and Ye, Jinwu},
  journal = {Phys. Rev. Lett.},
  volume = {70},
  issue = {21},
  pages = {3339--3342},
  numpages = {0},
  year = {1993},
  month = {May},
  publisher = {American Physical Society},
  doi = {10.1103/PhysRevLett.70.3339},
  url = {https://link.aps.org/doi/10.1103/PhysRevLett.70.3339}
}

@article{georges2001quantum,
  title = {Quantum fluctuations of a nearly critical Heisenberg spin glass},
  author = {Georges, A. and Parcollet, O. and Sachdev, S.},
  journal = {Phys. Rev. B},
  volume = {63},
  issue = {13},
  pages = {134406},
  numpages = {17},
  year = {2001},
  month = {Mar},
  publisher = {American Physical Society},
  doi = {10.1103/PhysRevB.63.134406},
  url = {https://link.aps.org/doi/10.1103/PhysRevB.63.134406}
}

@article{miller1993zerotemperature,
  title = {Zero-temperature critical behavior of the infinite-range quantum Ising spin glass},
  author = {Miller, Jonathan and Huse, David A.},
  journal = {Phys. Rev. Lett.},
  volume = {70},
  issue = {20},
  pages = {3147--3150},
  numpages = {0},
  year = {1993},
  month = {May},
  publisher = {American Physical Society},
  doi = {10.1103/PhysRevLett.70.3147},
  url = {https://link.aps.org/doi/10.1103/PhysRevLett.70.3147}
}

@article{read1995landau,
  title = {Landau theory of quantum spin glasses of rotors and Ising spins},
  author = {Read, N. and Sachdev, Subir and Ye, J.},
  journal = {Phys. Rev. B},
  volume = {52},
  issue = {1},
  pages = {384--410},
  numpages = {0},
  year = {1995},
  month = {Jul},
  publisher = {American Physical Society},
  doi = {10.1103/PhysRevB.52.384},
  url = {https://link.aps.org/doi/10.1103/PhysRevB.52.384}
}

@article{georges2000mean,
  title = {Mean Field Theory of a Quantum Heisenberg Spin Glass},
  author = {Georges, Antoine and Parcollet, Olivier and Sachdev, Subir},
  journal = {Phys. Rev. Lett.},
  volume = {85},
  issue = {4},
  pages = {840--843},
  numpages = {0},
  year = {2000},
  month = {Jul},
  publisher = {American Physical Society},
  doi = {10.1103/PhysRevLett.85.840},
  url = {https://link.aps.org/doi/10.1103/PhysRevLett.85.840}
}

@article{kopec1995continuous,
  title = {Discontinuous spin-glass transition in a random quantum Heisenberg magnet},
  author = {Kope\ifmmode \acute{c}\else \'{c}\fi{}, T. K.},
  journal = {Phys. Rev. B},
  volume = {52},
  issue = {13},
  pages = {9590--9594},
  numpages = {0},
  year = {1995},
  month = {Oct},
  publisher = {American Physical Society},
  doi = {10.1103/PhysRevB.52.9590},
  url = {https://link.aps.org/doi/10.1103/PhysRevB.52.9590}
}

@article{parcollet1999nonfermi,
  title = {Non-Fermi-liquid regime of a doped Mott insulator},
  author = {Parcollet, Olivier and Georges, Antoine},
  journal = {Phys. Rev. B},
  volume = {59},
  issue = {8},
  pages = {5341--5360},
  numpages = {0},
  year = {1999},
  month = {Feb},
  publisher = {American Physical Society},
  doi = {10.1103/PhysRevB.59.5341},
  url = {https://link.aps.org/doi/10.1103/PhysRevB.59.5341}
}

@article{fitzpatrick2014nonfermi,
  title = {Non-Fermi-liquid behavior of large-${N}_{B}$ quantum critical metals},
  author = {Fitzpatrick, A. Liam and Kachru, Shamit and Kaplan, Jared and Raghu, S.},
  journal = {Phys. Rev. B},
  volume = {89},
  issue = {16},
  pages = {165114},
  numpages = {9},
  year = {2014},
  month = {Apr},
  publisher = {American Physical Society},
  doi = {10.1103/PhysRevB.89.165114},
  url = {https://link.aps.org/doi/10.1103/PhysRevB.89.165114}
}

@Article{werman2017nonquasiparticle,
author={Werman, Yochai
and Kivelson, Steven A.
and Berg, Erez},
title={Non-quasiparticle transport and resistivity saturation: a view from the large-N limit},
journal={npj Quantum Materials},
year={2017},
month={Feb},
day={08},
volume={2},
number={1},
pages={7},
issn={2397-4648},
doi={10.1038/s41535-017-0009-8},
url={https://doi.org/10.1038/s41535-017-0009-8}
}

@article{chowdhury2022sachdev,
  title = {Sachdev-Ye-Kitaev models and beyond: Window into non-Fermi liquids},
  author = {Chowdhury, Debanjan and Georges, Antoine and Parcollet, Olivier and Sachdev, Subir},
  journal = {Rev. Mod. Phys.},
  volume = {94},
  issue = {3},
  pages = {035004},
  numpages = {78},
  year = {2022},
  month = {Sep},
  publisher = {American Physical Society},
  doi = {10.1103/RevModPhys.94.035004},
  url = {https://link.aps.org/doi/10.1103/RevModPhys.94.035004}
}

@article{sachdev2015bekenstein,
  title = {Bekenstein-Hawking Entropy and Strange Metals},
  author = {Sachdev, Subir},
  journal = {Phys. Rev. X},
  volume = {5},
  issue = {4},
  pages = {041025},
  numpages = {13},
  year = {2015},
  month = {Nov},
  publisher = {American Physical Society},
  doi = {10.1103/PhysRevX.5.041025},
  url = {https://link.aps.org/doi/10.1103/PhysRevX.5.041025}
}

@Article{gu2020notes,
author={Gu, Yingfei
and Kitaev, Alexei
and Sachdev, Subir
and Tarnopolsky, Grigory},
title={Notes on the complex Sachdev-Ye-Kitaev model},
journal={Journal of High Energy Physics},
year={2020},
month={Feb},
day={25},
volume={2020},
number={2},
pages={157},
issn={1029-8479},
doi={10.1007/JHEP02(2020)157},
url={https://doi.org/10.1007/JHEP02(2020)157}
}

@article{davison2017thermoelectric,
  title = {Thermoelectric transport in disordered metals without quasiparticles: The Sachdev-Ye-Kitaev models and holography},
  author = {Davison, Richard A. and Fu, Wenbo and Georges, Antoine and Gu, Yingfei and Jensen, Kristan and Sachdev, Subir},
  journal = {Phys. Rev. B},
  volume = {95},
  issue = {15},
  pages = {155131},
  numpages = {27},
  year = {2017},
  month = {Apr},
  publisher = {American Physical Society},
  doi = {10.1103/PhysRevB.95.155131},
  url = {https://link.aps.org/doi/10.1103/PhysRevB.95.155131}
}

@incollection{thooft1993planar,
  title={A planar diagram theory for strong interactions},
  author={'t Hooft, Gerard},
  booktitle={The Large N Expansion In Quantum Field Theory And Statistical Physics: From Spin Systems to 2-Dimensional Gravity},
  pages={80--92},
  year={1993},
  publisher={World Scientific}
}

@inbook{coleman1985aspects,
place={Cambridge},
title={1/N},
booktitle={Aspects of Symmetry: Selected Erice Lectures},
publisher={Cambridge University Press},
author={Coleman, Sidney},
year={1985},
pages={351–402}
}

@Inbook{witten1980expansion,
author="Witten, Edward",
editor="Hooft, G.'t
and Itzykson, C.
and Jaffe, A.
and Lehmann, H.
and Mitter, P. K.
and Singer, I. M.
and Stora, R.",
title="The 1/N Expansion in Atomic and Particle Physics",
bookTitle="Recent Developments in Gauge Theories",
year="1980",
publisher="Springer US",
address="Boston, MA",
pages="403--419",
isbn="978-1-4684-7571-5",
doi="10.1007/978-1-4684-7571-5_21",
url="https://doi.org/10.1007/978-1-4684-7571-5_21"
}

@article{moshe2003quantum,
title = {Quantum field theory in the large N limit: a review},
journal = {Physics Reports},
volume = {385},
number = {3},
pages = {69-228},
year = {2003},
issn = {0370-1573},
doi = {https://doi.org/10.1016/S0370-1573(03)00263-1},
url = {https://www.sciencedirect.com/science/article/pii/S0370157303002631},
author = {Moshe Moshe and Jean Zinn-Justin}
}

@inproceedings{kitaev2015simple,
  title={A simple model of quantum holography},
  author={Kitaev, Alexei},
  booktitle={KITP strings seminar and Entanglement},
  volume={12},
  year={2015}
}

@book{hartnoll2018holographic,
  title={Holographic quantum matter},
  author={Hartnoll, Sean A and Lucas, Andrew and Sachdev, Subir},
  year={2018},
  publisher={MIT press}
}

@Article{gu2017local,
author={Gu, Yingfei
and Qi, Xiao-Liang
and Stanford, Douglas},
title={Local criticality, diffusion and chaos in generalized Sachdev-Ye-Kitaev models},
journal={Journal of High Energy Physics},
year={2017},
month={May},
day={23},
volume={2017},
number={5},
pages={125},
issn={1029-8479},
doi={10.1007/JHEP05(2017)125},
url={https://doi.org/10.1007/JHEP05(2017)125}
}

@article{fu2016numerical,
  title = {Numerical study of fermion and boson models with infinite-range random interactions},
  author = {Fu, Wenbo and Sachdev, Subir},
  journal = {Phys. Rev. B},
  volume = {94},
  issue = {3},
  pages = {035135},
  numpages = {9},
  year = {2016},
  month = {Jul},
  publisher = {American Physical Society},
  doi = {10.1103/PhysRevB.94.035135},
  url = {https://link.aps.org/doi/10.1103/PhysRevB.94.035135}
}

@Article{berkooz2021complex,
author={Berkooz, Micha
and Narovlansky, Vladimir
and Raj, Himanshu},
title={Complex Sachdev-Ye-Kitaev model in the double scaling limit},
journal={Journal of High Energy Physics},
year={2021},
month={Feb},
day={15},
volume={2021},
number={2},
pages={113},
issn={1029-8479},
doi={10.1007/JHEP02(2021)113},
url={https://doi.org/10.1007/JHEP02(2021)113}
}

@Article{kitaev2018soft,
author={Kitaev, Alexei
and Suh, S. Josephine},
title={The soft mode in the Sachdev-Ye-Kitaev model and its gravity dual},
journal={Journal of High Energy Physics},
year={2018},
month={May},
day={29},
volume={2018},
number={5},
pages={183},
issn={1029-8479},
doi={10.1007/JHEP05(2018)183},
url={https://doi.org/10.1007/JHEP05(2018)183}
}

@article{rosenhaus2019introduction,
doi = {10.1088/1751-8121/ab2ce1},
url = {https://dx.doi.org/10.1088/1751-8121/ab2ce1},
year = {2019},
month = {jul},
publisher = {IOP Publishing},
volume = {52},
number = {32},
pages = {323001},
author = {Vladimir Rosenhaus},
title = {An introduction to the SYK model},
journal = {Journal of Physics A: Mathematical and Theoretical}
}

@Article{berkooz2017higher,
author={Berkooz, Micha
and Narayan, Prithvi
and Rozali, Moshe
and Sim{\'o}n, Joan},
title={Higher dimensional generalizations of the SYK model},
journal={Journal of High Energy Physics},
year={2017},
month={Jan},
day={31},
volume={2017},
number={1},
pages={138},
issn={1029-8479},
doi={10.1007/JHEP01(2017)138},
url={https://doi.org/10.1007/JHEP01(2017)138}
}

@article{sarosi2018ads2,
  author = "Sarosi, Gabor",
  title = "{AdS$_{2}$ holography and the SYK model}",
  doi = "10.22323/1.323.0001",
  journal = "PoS",
  year = 2018,
  volume = "Modave2017",
  pages = "001"
}

@article{fu2017supersymmetric,
  title = {Supersymmetric Sachdev-Ye-Kitaev models},
  author = {Fu, Wenbo and Gaiotto, Davide and Maldacena, Juan and Sachdev, Subir},
  journal = {Phys. Rev. D},
  volume = {95},
  issue = {2},
  pages = {026009},
  numpages = {20},
  year = {2017},
  month = {Jan},
  publisher = {American Physical Society},
  doi = {10.1103/PhysRevD.95.026009},
  url = {https://link.aps.org/doi/10.1103/PhysRevD.95.026009}
}

@Article{brezin1978planar,
author={Br{\'e}zin, E.
and Itzykson, C.
and Parisi, G.
and Zuber, J. B.},
title={Planar diagrams},
journal={Communications in Mathematical Physics},
year={1978},
month={Feb},
day={01},
volume={59},
number={1},
pages={35-51},
issn={1432-0916},
doi={10.1007/BF01614153},
url={https://doi.org/10.1007/BF01614153}
}

@article{bentsen2021measurement,
  title = {Measurement-induced purification in large-$N$ hybrid Brownian circuits},
  author = {Bentsen, Gregory S. and Sahu, Subhayan and Swingle, Brian},
  journal = {Phys. Rev. B},
  volume = {104},
  issue = {9},
  pages = {094304},
  numpages = {32},
  year = {2021},
  month = {Sep},
  publisher = {American Physical Society},
  doi = {10.1103/PhysRevB.104.094304},
  url = {https://link.aps.org/doi/10.1103/PhysRevB.104.094304}
}

@article{sahu2022entanglement,
  title = {Entanglement phases in large-$N$ hybrid Brownian circuits with long-range couplings},
  author = {Sahu, Subhayan and Jian, Shao-Kai and Bentsen, Gregory and Swingle, Brian},
  journal = {Phys. Rev. B},
  volume = {106},
  issue = {22},
  pages = {224305},
  numpages = {16},
  year = {2022},
  month = {Dec},
  publisher = {American Physical Society},
  doi = {10.1103/PhysRevB.106.224305},
  url = {https://link.aps.org/doi/10.1103/PhysRevB.106.224305}
}

@Article{jian2023linear,
author={Jian, Shao-Kai
and Bentsen, Gregory
and Swingle, Brian},
title={Linear growth of circuit complexity from Brownian dynamics},
journal={Journal of High Energy Physics},
year={2023},
month={Aug},
day={28},
volume={2023},
number={8},
pages={190},
issn={1029-8479},
doi={10.1007/JHEP08(2023)190},
url={https://doi.org/10.1007/JHEP08(2023)190}
}

@misc{saad2019semiclassicalrampsykgravity,
      title={A semiclassical ramp in SYK and in gravity}, 
      author={Phil Saad and Stephen H. Shenker and Douglas Stanford},
      year={2019},
      eprint={1806.06840},
      archivePrefix={arXiv},
      primaryClass={hep-th},
      url={https://arxiv.org/abs/1806.06840}, 
}

@article{NappShallow2022,
  title = {Efficient Classical Simulation of Random Shallow 2D Quantum Circuits},
  author = {Napp, John C. and La Placa, Rolando L. and Dalzell, Alexander M. and Brand\~ao, Fernando G. S. L. and Harrow, Aram W.},
  journal = {Phys. Rev. X},
  volume = {12},
  issue = {2},
  pages = {021021},
  numpages = {32},
  year = {2022},
  month = {Apr},
  publisher = {American Physical Society},
  doi = {10.1103/PhysRevX.12.021021},
  url = {https://link.aps.org/doi/10.1103/PhysRevX.12.021021}
}

@Article{Lashkari2013towards,
author={Lashkari, Nima
and Stanford, Douglas
and Hastings, Matthew
and Osborne, Tobias
and Hayden, Patrick},
title={Towards the fast scrambling conjecture},
journal={Journal of High Energy Physics},
year={2013},
month={Apr},
day={03},
volume={2013},
number={4},
pages={22},
issn={1029-8479},
doi={10.1007/JHEP04(2013)022},
url={https://doi.org/10.1007/JHEP04(2013)022}
}

@article{bluvstein2024logical,
  title={Logical quantum processor based on reconfigurable atom arrays},
  author={Bluvstein, Dolev and Evered, Simon J and Geim, Alexandra A and Li, Sophie H and Zhou, Hengyun and Manovitz, Tom and Ebadi, Sepehr and Cain, Madelyn and Kalinowski, Marcin and Hangleiter, Dominik and others},
  journal={Nature},
  volume={626},
  number={7997},
  pages={58--65},
  year={2024},
  publisher={Nature Publishing Group UK London}
}

@article{manetsch2025tweezer,
  title={A tweezer array with 6,100 highly coherent atomic qubits},
  author={Manetsch, Hannah J and Nomura, Gyohei and Bataille, Elie and Lv, Xudong and Leung, Kon H and Endres, Manuel},
  journal={Nature},
  volume={647},
  number={8088},
  pages={60--67},
  year={2025},
  publisher={Nature Publishing Group UK London}
}

@article{sales2025experimental,
  title={Experimental demonstration of logical magic state distillation},
  author={Sales Rodriguez, Pedro and Robinson, John M and Jepsen, Paul Niklas and He, Zhiyang and Duckering, Casey and Zhao, Chen and Wu, Kai-Hsin and Campo, Joseph and Bagnall, Kevin and Kwon, Minho and others},
  journal={Nature},
  volume={645},
  number={8081},
  pages={620--625},
  year={2025},
  publisher={Nature Publishing Group UK London}
}

@article{holman2026trapping,
  title={Trapping of single atoms in metasurface optical tweezer arrays},
  author={Holman, Aaron and Xu, Yuan and Sun, Ximo and Wu, Jiahao and Wang, Mingxuan and Zhu, Zezheng and Seo, Bojeong and Yu, Nanfang and Will, Sebastian},
  journal={Nature},
  pages={1--7},
  year={2026},
  publisher={Nature Publishing Group UK London}
}

@article{kim2023evidence,
  title={Evidence for the utility of quantum computing before fault tolerance},
  author={Kim, Youngseok and Eddins, Andrew and Anand, Sajant and Wei, Ken Xuan and Van Den Berg, Ewout and Rosenblatt, Sami and Nayfeh, Hasan and Wu, Yantao and Zaletel, Michael and Temme, Kristan and others},
  journal={Nature},
  volume={618},
  number={7965},
  pages={500--505},
  year={2023},
  publisher={Nature Publishing Group UK London}
}

@article{google2025quantum,
  title={Quantum error correction below the surface code threshold},
  author={Google Quantum AI and Collaborators},
  journal={Nature},
  volume={638},
  number={8052},
  pages={920--926},
  year={2025},
  publisher={Nature Publishing Group UK London}
}

@article{abughanem2025ibm,
  title={IBM quantum computers: evolution, performance, and future directions: M. AbuGhanem},
  author={AbuGhanem, Muhammad},
  journal={The Journal of Supercomputing},
  volume={81},
  number={5},
  pages={687},
  year={2025},
  publisher={Springer}
}

@article{cain2026shor,
  title={Shor's algorithm is possible with as few as 10,000 reconfigurable atomic qubits},
  author={Cain, Madelyn and Xu, Qian and King, Robbie and Picard, Lewis RB and Levine, Harry and Endres, Manuel and Preskill, John and Huang, Hsin-Yuan and Bluvstein, Dolev},
  journal={arXiv preprint arXiv:2603.28627},
  year={2026}
}

@article{pecorari2025high,
  title={High-rate quantum LDPC codes for long-range-connected neutral atom registers},
  author={Pecorari, Laura and Jandura, Sven and Brennen, Gavin K and Pupillo, Guido},
  journal={Nature Communications},
  volume={16},
  number={1},
  pages={1111},
  year={2025},
  publisher={Nature Publishing Group UK London}
}

@article{guo2026toward,
  title={Toward Self-Correcting Quantum Codes for Neutral Atom Arrays},
  author={Guo, Jinkang and Hong, Yifan and Kaufman, Adam and Lucas, Andrew},
  journal={PRX Quantum},
  volume={7},
  number={1},
  pages={010301},
  year={2026},
  publisher={APS}
}

@article{pecorari2025quantum,
  title={Quantum low-density parity-check codes for erasure-biased atomic quantum processors},
  author={Pecorari, Laura and Pupillo, Guido},
  journal={Physical Review A},
  volume={112},
  number={5},
  pages={052417},
  year={2025},
  publisher={APS}
}

@article{wu2022erasure,
  title={Erasure conversion for fault-tolerant quantum computing in alkaline earth Rydberg atom arrays},
  author={Wu, Yue and Kolkowitz, Shimon and Puri, Shruti and Thompson, Jeff D},
  journal={Nature communications},
  volume={13},
  number={1},
  pages={4657},
  year={2022},
  publisher={Nature Publishing Group UK London}
}

@article{chow2024circuit,
  title={Circuit-based leakage-to-erasure conversion in a neutral-atom quantum processor},
  author={Chow, Matthew NH and Buchemmavari, Vikas and Omanakuttan, Sivaprasad and Little, Bethany J and Pandey, Saurabh and Deutsch, Ivan H and Jau, Yuan-Yu},
  journal={PRX Quantum},
  volume={5},
  number={4},
  pages={040343},
  year={2024},
  publisher={APS}
}

@article{boixo2018characterizing,
  title={Characterizing quantum supremacy in near-term devices},
  author={Boixo, Sergio and Isakov, Sergei V and Smelyanskiy, Vadim N and Babbush, Ryan and Ding, Nan and Jiang, Zhang and Bremner, Michael J and Martinis, John M and Neven, Hartmut},
  journal={Nature Physics},
  volume={14},
  number={6},
  pages={595--600},
  year={2018},
  publisher={Nature Publishing Group UK London}
}

@article{zabalo2020critical,
  title={Critical properties of the measurement-induced transition in random quantum circuits},
  author={Zabalo, Aidan and Gullans, Michael J and Wilson, Justin H and Gopalakrishnan, Sarang and Huse, David A and Pixley, JH},
  journal={Physical Review B},
  volume={101},
  number={6},
  pages={060301},
  year={2020},
  publisher={APS}
}

@article{bao2020theory,
  title = {Theory of the phase transition in random unitary circuits with measurements},
  author = {Bao, Yimu and Choi, Soonwon and Altman, Ehud},
  journal = {Phys. Rev. B},
  volume = {101},
  issue = {10},
  pages = {104301},
  numpages = {26},
  year = {2020},
  month = {Mar},
  publisher = {American Physical Society},
  doi = {10.1103/PhysRevB.101.104301},
  url = {https://link.aps.org/doi/10.1103/PhysRevB.101.104301}
}

@article{choi2020quantum,
  title = {Quantum Error Correction in Scrambling Dynamics and Measurement-Induced Phase Transition},
  author = {Choi, Soonwon and Bao, Yimu and Qi, Xiao-Liang and Altman, Ehud},
  journal = {Phys. Rev. Lett.},
  volume = {125},
  issue = {3},
  pages = {030505},
  numpages = {6},
  year = {2020},
  month = {Jul},
  publisher = {American Physical Society},
  doi = {10.1103/PhysRevLett.125.030505},
  url = {https://link.aps.org/doi/10.1103/PhysRevLett.125.030505}
}

@article{li2019measurement,
  title={Measurement-driven entanglement transition in hybrid quantum circuits},
  author={Li, Yaodong and Chen, Xiao and Fisher, Matthew PA},
  journal={Physical Review B},
  volume={100},
  number={13},
  pages={134306},
  year={2019},
  publisher={APS}
}

@article{skinner2019measurement,
  title = {Measurement-Induced Phase Transitions in the Dynamics of Entanglement},
  author = {Skinner, Brian and Ruhman, Jonathan and Nahum, Adam},
  journal = {Phys. Rev. X},
  volume = {9},
  issue = {3},
  pages = {031009},
  numpages = {21},
  year = {2019},
  month = {Jul},
  publisher = {American Physical Society},
  doi = {10.1103/PhysRevX.9.031009},
  url = {https://link.aps.org/doi/10.1103/PhysRevX.9.031009}
}

@article{heinrich2022randomized,
  title={Randomized benchmarking with random quantum circuits},
  author={Heinrich, Markus and Kliesch, Martin and Roth, Ingo},
  journal={arXiv preprint arXiv:2212.06181},
  year={2022}
}

@article{brandao2021models,
  title={Models of quantum complexity growth},
  author={Brand{\~a}o, Fernando GSL and Chemissany, Wissam and Hunter-Jones, Nicholas and Kueng, Richard and Preskill, John},
  journal={PRX Quantum},
  volume={2},
  number={3},
  pages={030316},
  year={2021},
  publisher={APS}
}

@article{choi1975completely,
  title={Completely positive linear maps on complex matrices},
  author={Choi, Man-Duen},
  journal={Linear algebra and its applications},
  volume={10},
  number={3},
  pages={285--290},
  year={1975},
  publisher={Elsevier}
}

@article{jamiolkowski1972linear,
  title={Linear transformations which preserve trace and positive semidefiniteness of operators},
  author={Jamio{\l}kowski, Andrzej},
  journal={Reports on mathematical physics},
  volume={3},
  number={4},
  pages={275--278},
  year={1972},
  publisher={Elsevier}
}

@article{pednault2019leveraging,
  title={Leveraging secondary storage to simulate deep 54-qubit sycamore circuits},
  author={Pednault, Edwin and Gunnels, John A and Nannicini, Giacomo and Horesh, Lior and Wisnieff, Robert},
  journal={arXiv preprint arXiv:1910.09534},
  year={2019}
}

@article{bentsen2024complexity,
  title={On the complexity of sampling from shallow Brownian circuits},
  author={Bentsen, Gregory and Fefferman, Bill and Ghosh, Soumik and Gullans, Michael J and Liu, Yinchen},
  journal={arXiv preprint arXiv:2411.04169},
  year={2024}
}

@article{aaronson2016complexity,
  title={Complexity-theoretic foundations of quantum supremacy experiments},
  author={Aaronson, Scott and Chen, Lijie},
  journal={arXiv preprint arXiv:1612.05903},
  year={2016}
}

@article{hangleiter2024bell,
  title={Bell sampling from quantum circuits},
  author={Hangleiter, Dominik and Gullans, Michael J},
  journal={Physical Review Letters},
  volume={133},
  number={2},
  pages={020601},
  year={2024},
  publisher={APS}
}

\clearpage
\pagebreak
\newpage

\onecolumngrid

\appendix

\section{Numerical Simulations of Discrete All-to-All Random Quantum Circuits}
\label{app:numerics}
In this appendix, we perform numerical simulations to investigate whether our Brownian analytical results generalize to predict the behaviors of all-to-all random quantum circuits composed of Haar-random $2$-qubit gates. First, we recall the relevant context and state some necessary definitions.

In the following, we will need to numerically simulate sampling from the output distribution of noisy random quantum circuits. Since we only consider independent single-qubit depolarizing noise, the quantum channels induced by such noisy circuits are manifestly mixed unitary channels, which allow for the stochastic simulation of sampling from the noisy output distribution without storing density matrices. Concretely, let $D$ be an $n$-qubit mixed unitary channel defined by $D(\rho)=\sum_{i=1}^r \lambda_i V_i\rho V_i^\dagger$ with $\lambda_i\geq 0$, $V_i$ unitary for every $i\in\{1,\ldots,r\}$, and $\sum_{i=1}^r\lambda_i=1$. Let $\tilde{q}(x)=\bra{x}D(\rho)\ket{x}, x\in\{0,1\}^n$ denote the output distribution induced by measuring all qubits of $D(\rho)$ in the computational basis. Since the initial state is $\rho=\ket{0^n}\bra{0^n}$ in our case, for every $x\in\{0,1\}^n$, we have
\begin{equation}
\tilde{q}(x)=\bra{x}D(\rho)\ket{x}=\sum_{i=1}^r \lambda_i\bra{x}V_i\ket{0^n}\bra{0^n} V_i^\dagger\ket{x}=\sum_{i=1}^r\lambda_i|\bra{x}V_i\ket{0^n}|^2.
\end{equation}
Thus, to sample $x\sim\tilde{q}$, it suffices to first sample a $V_i$ with probability $\lambda_i$, and then sample a string $x$ with probability $|\bra{x}V_i\ket{0^n}|^2$, which only requires the classical simulation of the noiseless unitary circuit $V_i$. Furthermore, note that the tensor products of single-qubit depolarizing channels are mixed unitary channels, a unitary gate is a mixed unitary channel with one term in it (e.g, $\lambda_1=1$), and the composition of mixed unitary channels is again a mixed unitary channel.

We will numerically simulate $n$-qubit depth-$d$ (in this section, $d$ denotes circuit depth as opposed to Hilbert space dimension) all-to-all random quantum circuits composed of Haar-random $2$-qubit gates. To draw a circuit instance, for each layer of gates from $1$ to $d$, we draw a random permutation $\sigma$ of $\{1,2,\ldots,n\}$ and apply a Haar-random $2$-qubit gate to qubits $\sigma(i)$ and $\sigma(i+1)$ for every $i\in\{1,3,5,\ldots,n-1\}$, so every layer always consists of exactly $\frac{n}{2}$ gates. Let $\mathcal{D}$ denote the distribution of $n$-qubit depth-$d$ all-to-all random quantum circuits formed in this way, and for every $U\sim\mathcal{D}$, we use $q_U$ to denote its output distribution; namely, $q_U(x)=|\bra{x}U\ket{0^n}|^2$ for every $x\in\{0,1\}^n$.


We further consider noisy all-to-all random quantum circuits with single-qubit depolarizing noise, with noise strength controlled by a parameter $\gamma\geq 0$. Define a stochastic single-qubit gate $N_\gamma$ as follows. With $\gamma'=\frac{\gamma}{n}$, $N_\gamma$ applies the single-qubit identity gate with probability $1-\gamma'$ and $X$, $Y$, or $Z$ gate with probability $\frac{\gamma'}{3}$ each. Note that under this parameterization, the completely depolarizing channel corresponds to $\gamma'=0.75$. To obtain a noisy circuit $D_\gamma$ from a noiseless circuit $U$, for every $2$-qubit gate in $U$ acting on qubits $j$ and $j'$, we include the same gate in $D_\gamma$ but with an $N_\gamma$ gate appended to qubits $j$ and $j'$. For a noisy circuit $D_\gamma$, we use $q_{D_\gamma}$ to denote its output distribution.

\begin{figure}[t]
    \centering
    \includegraphics[width=0.9\linewidth]{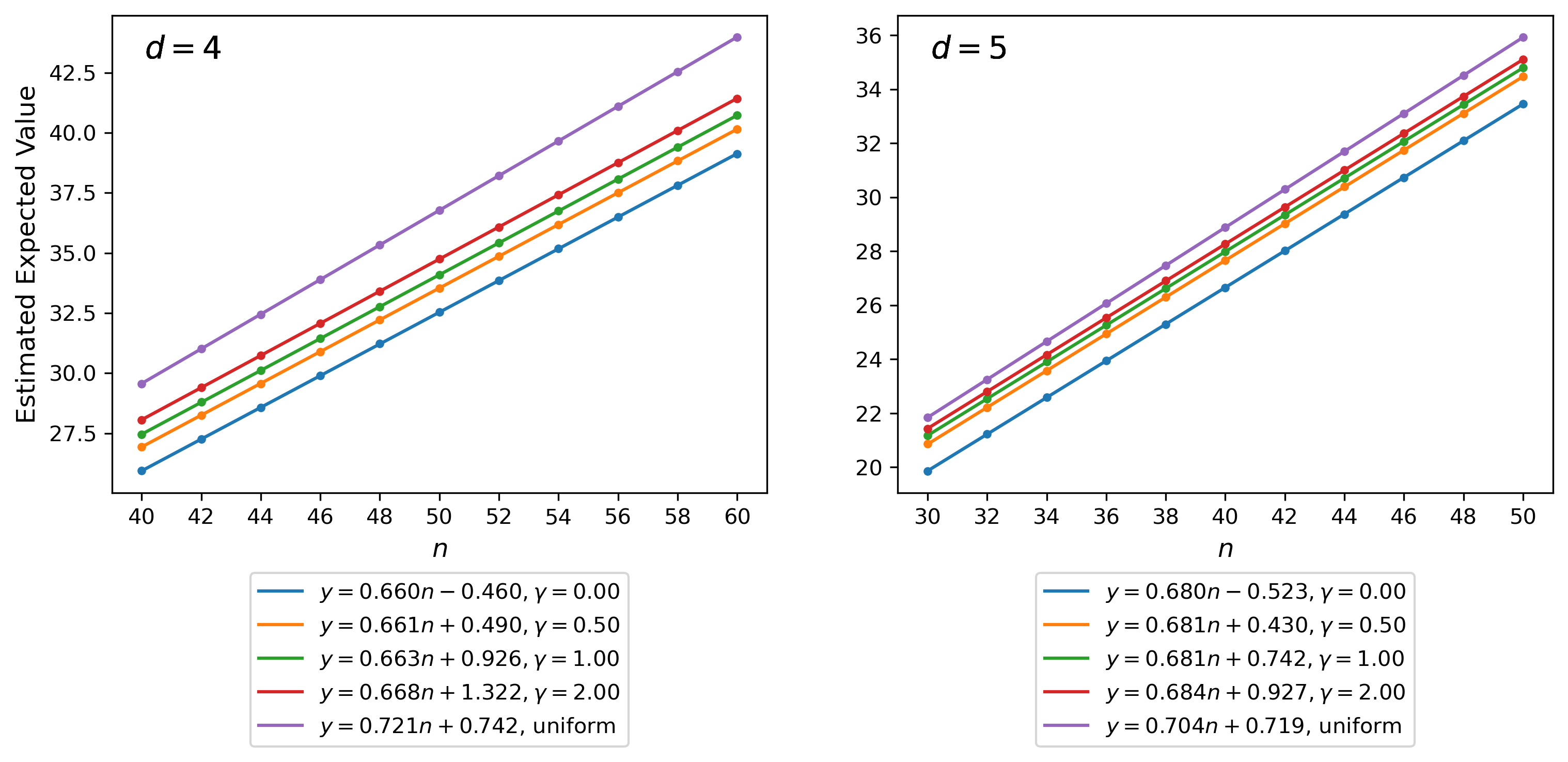}
    \caption{Numerical estimation of $\Exp[X_\gamma]$ as functions of $n$ for all-to-all random quantum circuits composed of Haar-random $2$-qubit gates. Each data point is estimated from $10^5$ independent draws of the random variable $X_\gamma$. Uniform means the string $x\in\{0,1\}^n$ is sampled from the uniform distribution. At $d=4$ and $d=5$, the estimated slopes appear to be increasing mildly in $\gamma$, showing behavior inconsistent with the Brownian prediction that the slopes should be independent of $\gamma$.}
    \label{fig:expected_value_d=45}
\end{figure}

Fix some $\gamma\geq 0$. We consider a continuous random variable $X_\gamma$ defined by the following procedure. To draw a sample from $X_\gamma$, we first draw a noiseless circuit $U\sim\mathcal{D}$, then stochastically simulate the induced noisy circuit $D_\gamma$ once to draw a string $x\sim q_{D_\gamma}$, and finally output the negative log probability $-\log(q_U(x))$. The last step requires calculating the noiseless output probability $q_U(x)$, and to repeatedly draw from $X_\gamma$ independently using the above procedure, a fresh noiseless circuit $U$ is drawn every time. Note that $\Exp[X_\gamma]$ is precisely the XEB score of $\mathcal{D}$ with depolarizing noise strength $\gamma$. 

Our work has characterized the XEB score and its variance for noisy Brownian circuits in the regime where $n\rightarrow\infty$, $\beta J$ is at least some sufficiently large constant, and $\gamma$ is at most some sufficiently small constant (so $\gamma'=\frac{O(1)}{n}$). In this regime, our Brownian results state that
\begin{equation}
\Exp[X_\gamma]=H\left(\frac{1+a}{2}\right)n+\gamma_e-e^{-\gamma\beta}
\label{eq:brownian_EX}
\end{equation}
where $H$ is the binary entropy function defined with $\log$, $a=e^{-12\beta J}$, and $\gamma_e$ is Euler's constant. The variance of $X_\gamma$ has the closed-form formula
\begin{equation}
\Var(X_\gamma)=\frac{(1-a^2)}{4} \mathrm{logit}^2 \left( \frac{1+a}{2} \right)n+\frac{\pi^2}{6}-e^{-2\gamma\beta}.
\label{eq:brownian_Var}
\end{equation}
The main qualitative predictions about $\Exp[X_\gamma]$ and $\Var(X_\gamma)$ coming from our Brownian calculations \eqref{eq:brownian_EX} and \eqref{eq:brownian_Var} are as follows.
\begin{enumerate}
\item (Noise sensitivity) The expected score $\Exp[X_\gamma]$ scales linearly in $n$, with the slope depending only on the circuit depth $\beta J$ and being independent of the noise strength $\gamma$. The $y$-intercept increases in $\gamma$.

\item (Sample efficiency) The variance $\Var(X_\gamma)$ scales linearly in $n$, with the slope of $\Var(X_\gamma)$ also depending only on the circuit depth $\beta J$ and being independent of the noise strength $\gamma$.
\end{enumerate}
Therefore, the goal of our numerics is to check whether the noise sensitivity and sample efficiency properties translate from Brownian to the setting of discrete all-to-all random quantum circuits.

Next, we describe the experimental setup. In our numerics, we estimate $\Exp[X_\gamma]$ and $\Var(X_\gamma)$ for discrete all-to-all random quantum circuits for circuit depths $d$ ranging from $4$ to $7$ and various noise strengths $\gamma$. To estimate $\Exp[X_\gamma]$ and $\Var(X_\gamma)$ as functions of $n$ for each choice of $d$ and $\gamma$, we consider a list of values for $n$, estimate one data point per value of $n$ by taking empirical averages over $10^5$ independent draws of the random variable $X_\gamma$, and finally perform linear regression to solve for the best linear fit over the data points. The simulations are performed using the tensor network library \cite{gray2018quimb}, and the sampling step uses quimb's implementation of the gate-by-gate algorithm \cite{bravyi2022simulate}, which we empirically observe to be faster than the alternative qubit-by-qubit algorithm for the specific circuits we are trying to simulate in this experiment.

\begin{figure}[t]
    \centering

    \includegraphics[width=0.9\linewidth]{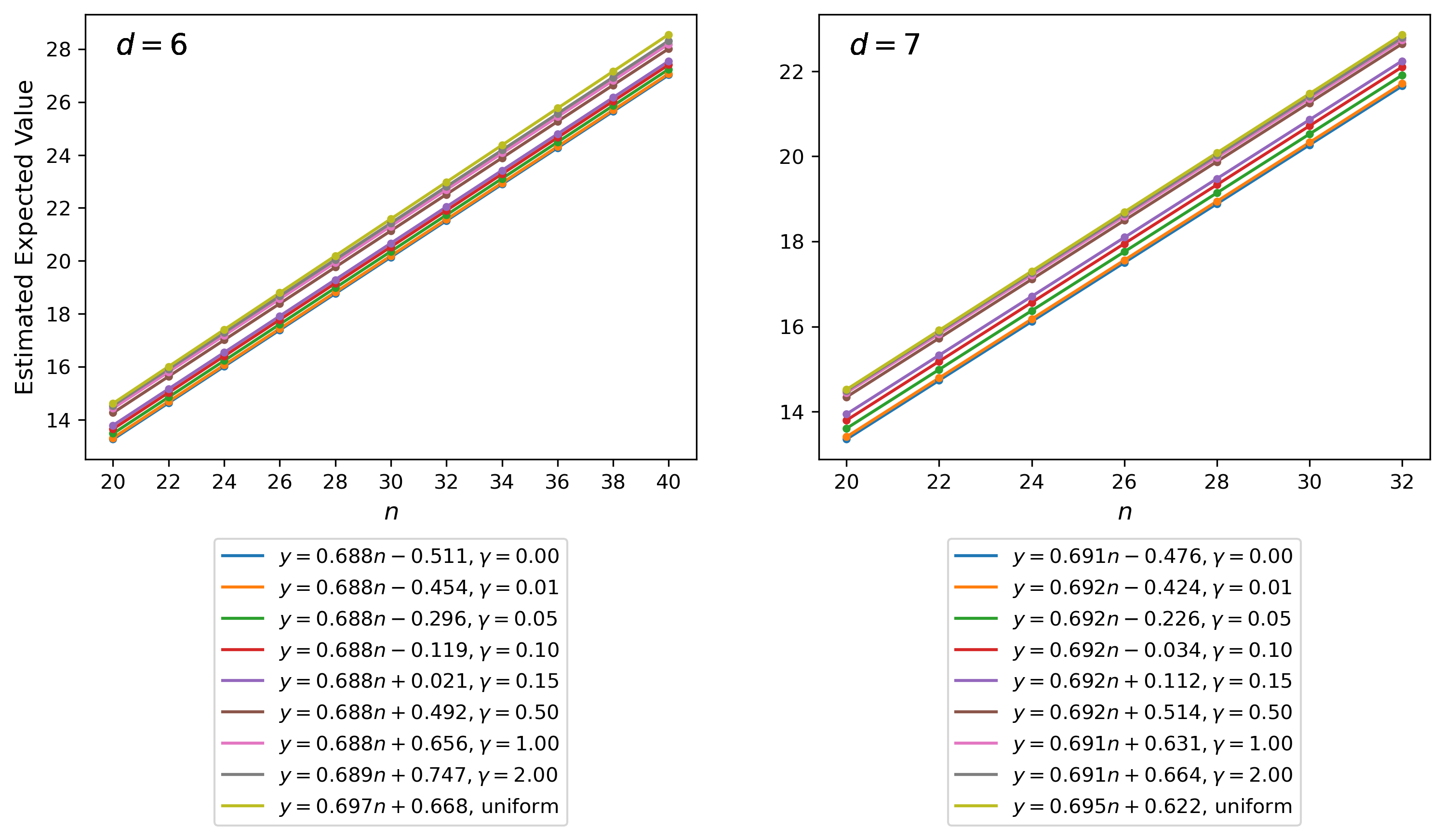}
    \caption{Numerical estimation of $\Exp[X_\gamma]$ as functions of $n$ for all-to-all random quantum circuits composed of Haar-random $2$-qubit gates. Each data point is estimated from $10^5$ independent draws of the random variable $X_\gamma$. Uniform means the string $x\in\{0,1\}^n$ is sampled from the uniform distribution. At $d=6$ and $d=7$, the estimated slopes come out to be independent of $\gamma$ up to statistical uncertainties, which is in agreement with the Brownian prediction. At the same time, the estimated $y$-intercepts are monotonically increasing in $\gamma$, showing sensitivity to the noise strength.}
    \label{fig:expected_value_d=67}
\end{figure}

We first discuss our findings on the behaviors of the expected score $\Exp[X_\gamma]$. From Fig. \ref{fig:expected_value_d=45}, we see that for $d=4$ and $d=5$, the estimated slopes appear to be increasing mildly in $\gamma$, showing behaviors inconsistent with the fine-grained Brownian prediction that the slopes should be independent of $\gamma$. We believe that this is because $d=4$ and $d=5$ are not quite large enough to match the assumption of $\beta J>>1$ required by our analytical calculations. In contrast, from Fig. \ref{fig:expected_value_d=67}, we see that for $d=6$ and $d=7$, the estimated slopes come out to be independent of $\gamma$ up to statistical uncertainties, which is in agreement with the Brownian prediction. At the same time, the estimated $y$-intercepts increase monotonically in $\gamma$, showing sensitivity to the noise strength. Note that \eqref{eq:brownian_EX} predicts that as the circuit gets deeper, which corresponds to $\beta J\rightarrow\infty$ and $a\rightarrow 0$ in the Brownian setting, the slope of $\Exp[X_\gamma]$ should approach $\log(2)\approx 0.693147$. At $d=6$, the estimated slope appears to be around $0.688$, which is close to but distinguishably different from $\log(2)$. At $d=7$, the estimated slope of around $0.692$ is inching towards the limiting value $\log(2)\approx 0.693147$. In conclusion, for the XEB score $\Exp[X_\gamma]$, we observed fine-grained qualitative behaviors consistent with those predicted by our analytical Brownian calculations in our experiments involving depths $6$ and $7$ discrete all-to-all random quantum circuits.

\begin{figure}[t]
    \centering
    \includegraphics[width=0.9\linewidth]{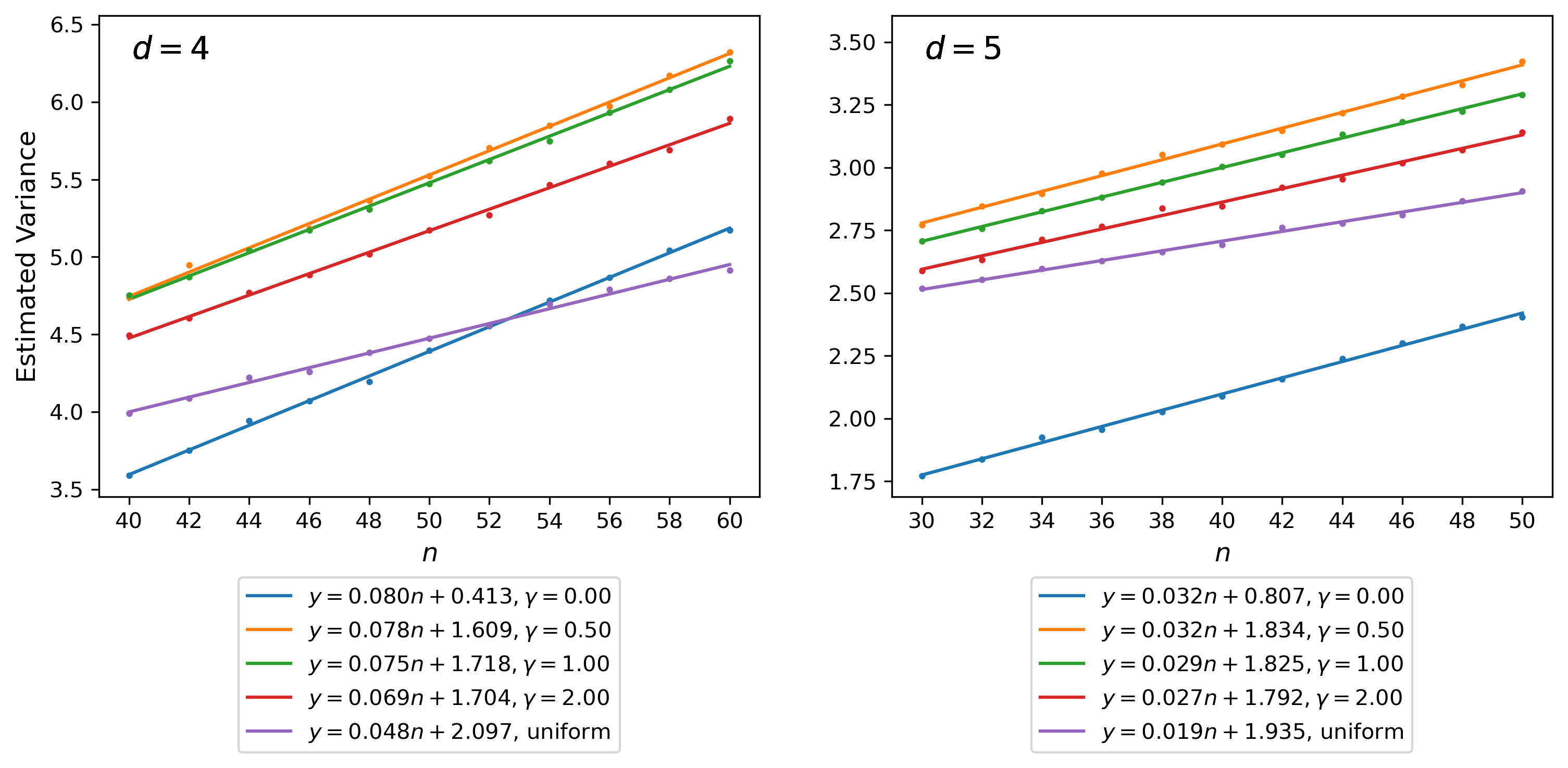}
    \caption{Numerical estimation of $\Var[X_\gamma]$ as functions of $n$ for all-to-all random quantum circuits composed of Haar-random $2$-qubit gates. Each data point is estimated from $10^5$ independent draws of the random variable $X_\gamma$. Uniform means the string $x\in\{0,1\}^n$ is sampled from the uniform distribution. For $d=4$ and $d=5$, the estimated slopes are visibly decreasing in $\gamma$, while \eqref{eq:brownian_Var} predicts that they should be independent of $\gamma$.}
    \label{fig:variance_d=45}
\end{figure}

\begin{figure}[t]
    \centering
    \includegraphics[width=0.9\linewidth]{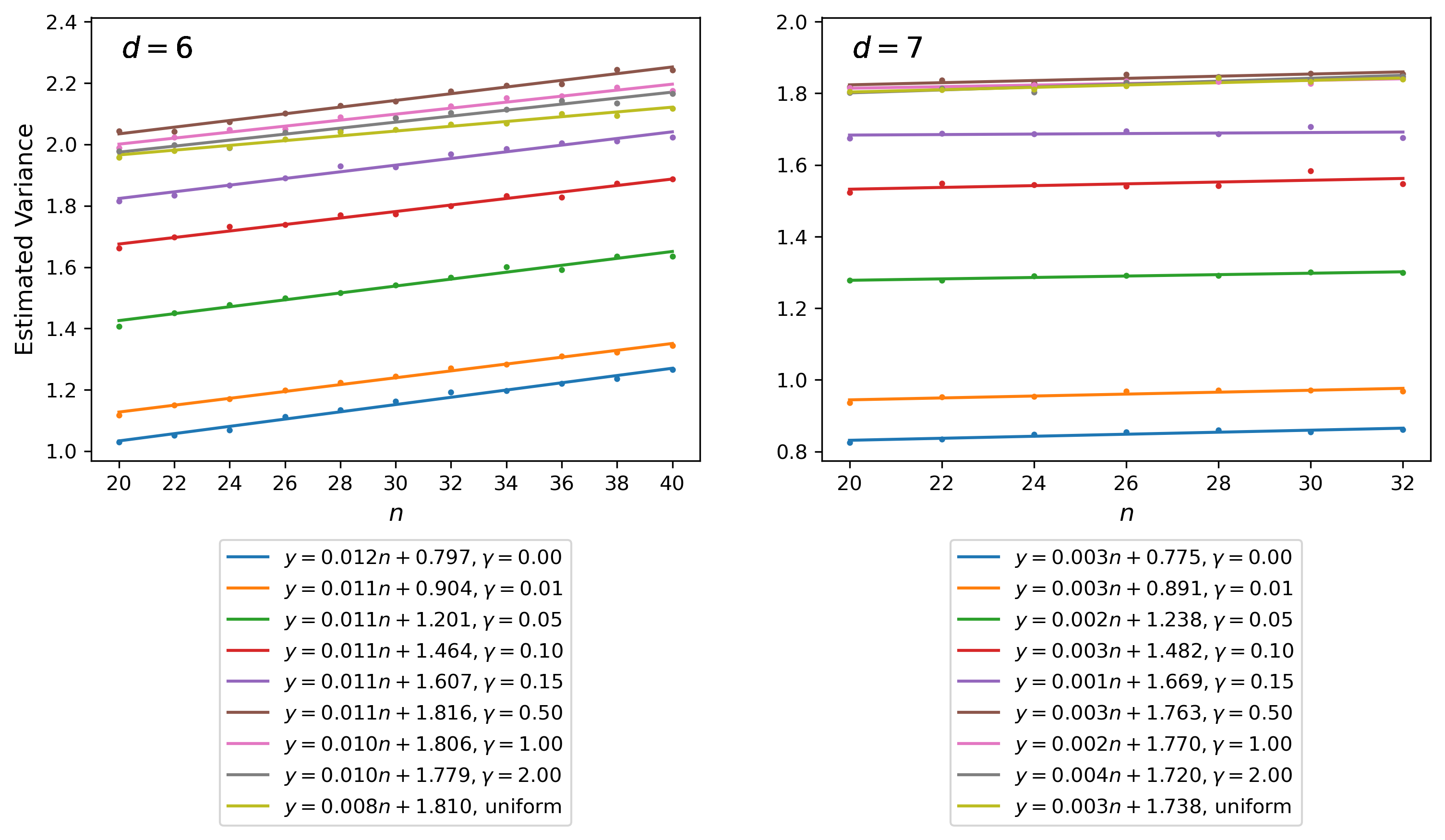}
    \caption{Numerical estimation of $\Var[X_\gamma]$ as functions of $n$ for all-to-all random quantum circuits composed of Haar-random $2$-qubit gates. Each data point is estimated from $10^5$ independent draws of the random variable $X_\gamma$. Uniform means the string $x\in\{0,1\}^n$ is sampled from the uniform distribution. At $d=6$, it is unclear whether the slight negative correlation between the slopes and $\gamma$ is statistically significant. For $d=7$, we find that $10^5$ samples are no longer sufficient to meaningfully estimate the growth rate of $\Var[X_\gamma]$.}
    \label{fig:variance_d=67}
\end{figure}

Next, we discuss our numerically observed behaviors for the variance $\Var(X_\gamma)$. From Fig. \ref{fig:variance_d=45}, we see that for $d=4$ and $d=5$, the estimated slopes are visibly decreasing in $\gamma$, while \eqref{eq:brownian_Var} predicts that they should be independent of $\gamma$. In particular, for $d=4$, the slope of $\Var(X_\gamma)$ for the noiseless $\gamma=0$ case is set to overtake the noisy variance curves, which is a phenomenon unexplained by our Brownian analysis. Thus, for $d=4$ and $d=5$, certain specific properties of $\Var(X_\gamma)$ for all-to-all random quantum circuits deviate from our Brownian predictions, and again, we believe this is the case because $d=4$ and $d=5$ do not yet correspond to the $\beta J>>1$ regime. Looking at Fig. \ref{fig:variance_d=67}, it is unclear whether the slight negative correlation between the slopes and $\gamma$ is statistically significant for $d=6$. For $d=7$, we find that $10^5$ samples are no longer sufficient to meaningfully estimate the growth rate of $\Var[X_\gamma]$, which is consistent with the prediction that the slope of $\Var[X_\gamma]$ should approach $0$ as the circuit depth increases. Overall, the findings in Figs. \ref{fig:variance_d=45} and \ref{fig:variance_d=67} support our central claim that $\Exp[X_\gamma]$ should be a sample-efficient quantity to estimate for all-to-all random quantum circuits.

\begin{figure}[ht]
    \centering
    \includegraphics[width=0.9\linewidth]{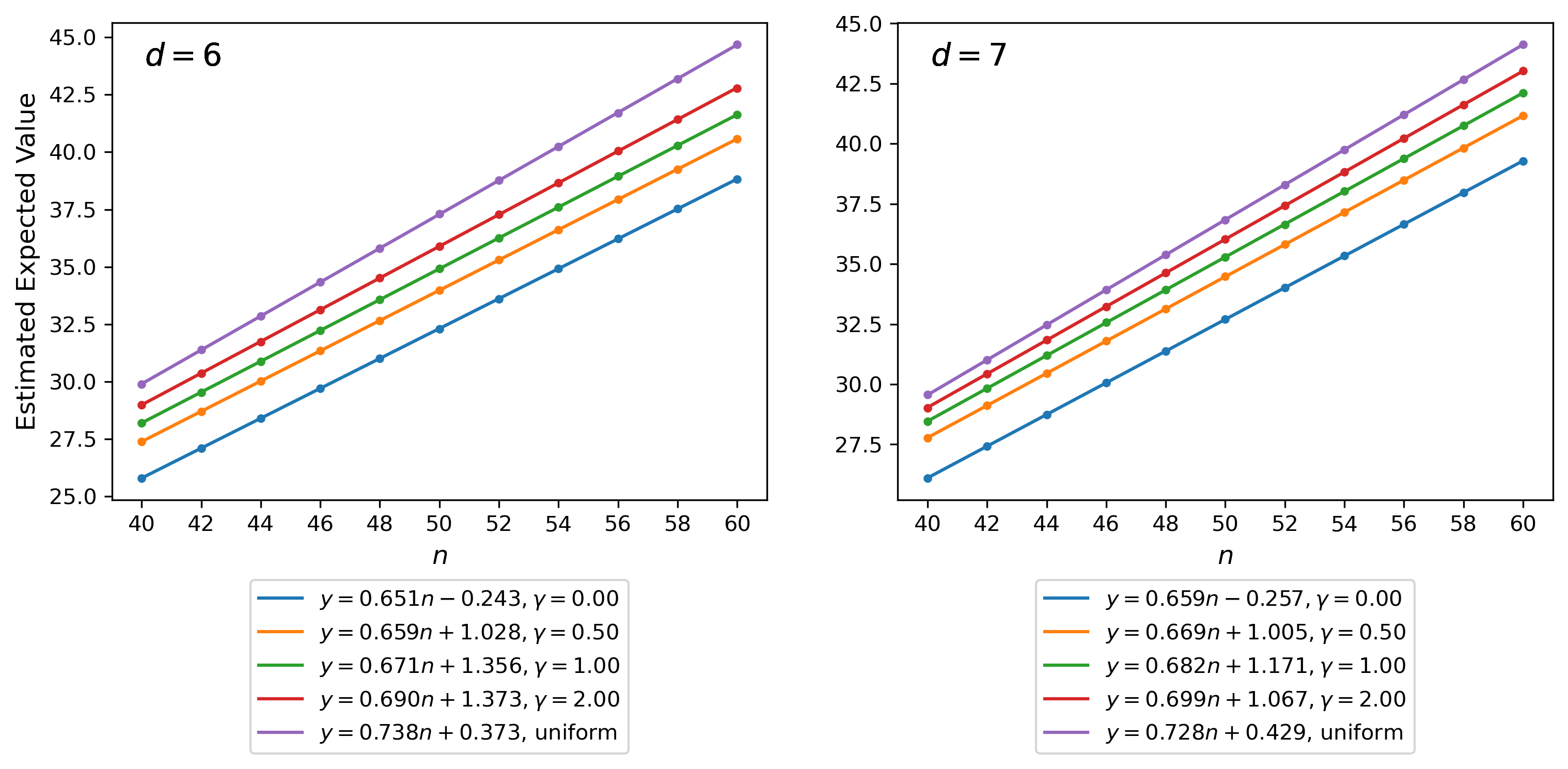}
    \caption{Numerical estimation of $\Exp[X_\gamma]$ as functions of $n$ for 1D brickwork random quantum circuits composed of Haar-random $2$-qubit gates. Each data point is estimated from $10^5$ independent draws of the random variable $X_\gamma$. Uniform means the string $x\in\{0,1\}^n$ is sampled from the uniform distribution. In the 1D case, even for $d=6$ and $d=7$, the estimated slopes are visibly increasing in $\gamma$, deviating from the Brownian prediction.}
    \label{fig:expected_value_1D_d=67}
\end{figure}

As a final test to probe the generality of our Brownian analysis, in Fig. \ref{fig:expected_value_1D_d=67}, we repeat the same numerical experiments to investigate the behavior of $\Exp[X_\gamma]$ at $d=6$ and $d=7$ for 1D brickwork random quantum circuits. We observe that for the 1D architecture, even for $d=6$ and $d=7$, the estimated slopes are visibly increasing in $\gamma$, deviating from the Brownian prediction. Although this does not rule out the possibility that $\Exp[X_\gamma]$ could be an informative and sample-efficient quantity to estimate for 1D brickwork random quantum circuits, at least the Brownian analysis may no longer accurately predict certain fine-grained features of $\Exp[X_\gamma]$ in this setting.  

\FloatBarrier

\section{Incorporation of Single-Qubit Depolarizing Noise}
\label{app:noise}

In this appendix we incorporate single-qubit depolarizing noise into the Brownian circuit model and study the resulting spectrum of $\Heff$ using perturbation theory. To model a noisy quantum computer, we consider interleaving unitary Brownian dynamics with single-qubit depolarizing noise at a rate $\gamma$ described by the channel
\begin{equation}
    \mathcal{E}(\rho) = (1-n\gamma dt) \rho + \frac{1}{3} \gamma dt\sum_{i,\alpha} \sigma_i^{\alpha} \rho \sigma_i^{\alpha}
\end{equation}
which can equivalently be expressed in terms of $3n+1$ Kraus operators as
\begin{equation}
    \mathcal{E}(\rho) = \sum_{\mu} E_{\mu} \rho \adj{E}_{\mu} \quad \quad \quad E_0 = \sqrt{1- n \gamma dt} \quad \quad \quad \quad E_{i \alpha} = \sqrt{\frac{\gamma dt}{3}} \sigma_i^{\alpha}
\end{equation}
and which manifestly satisfy the completeness relation
\begin{equation}
    \sum_{\mu} \adj{E}_{\mu} E_{\mu} = \adj{E}_{0} E_{0} + \sum_{i, \alpha} \adj{E}_{i \alpha} E_{i \alpha } = 1.
\end{equation}
Interleaving these depolarizing channels with the unitary Brownian channel
\begin{equation}
    \mathcal{U}_t(\rho) = U_t \rho \adj{U}_t
\end{equation}
 (see Eq. \eqref{eq:brownianunitary} of the main text) gives the total iterated quantum channel
\begin{equation}
    \Phi = \mathcal{E}_K \circ \mathcal{U}_K \circ \mathcal{E}_{K-1} \circ \mathcal{U}_{K-1} \circ \cdots \circ \mathcal{E}_1 \circ \mathcal{U}_1
\end{equation}
where we use the index $t = 1,2,\ldots,K$ to label discrete timesteps, and $\beta = K dt$. As usual, at the end of the calculation we will take a continuum limit $dt \rightarrow 0$ and $K \rightarrow \infty$ while keeping the total circuit depth $\beta$ fixed.

Consider first the transition probability $\ket{\mathbf{0}} \rightarrow \ket{\mathbf{x}}$ for a particular realization of a noisy Brownian circuit:
\begin{align}
    q_{\gamma}(\mathbf{x}) &= \bra{\mathbf{x}} \Phi \left( \ket{\mathbf{0}} \! \bra{\mathbf{0}} \right) \ket{\mathbf{x}} \nonumber \\
    &= \bra{\mathbf{x}} \mathcal{E}_K \circ \mathcal{U}_K \circ \cdots \circ \mathcal{E}_1 \circ \mathcal{U}_1 \left( \ket{\mathbf{0}} \! \bra{\mathbf{0}} \right) \ket{\mathbf{x}} \nonumber \\
    &= \sum_{\mu_K} \cdots \sum_{\mu_1} \bra{\mathbf{x}} E_{\mu_K} U_K \cdots E_{\mu_1} U_1 \ket{\mathbf{0}} \! \bra{\mathbf{0}} \adj{U}_1 \adj{E}_{\mu_1} \cdots \adj{U}_K \adj{E}_{\mu_K} \ket{\mathbf{x}} \nonumber
\end{align}
where in the last line we have decomposed the noisy channels $\mathcal{E}_t$ into their respective Kraus operators $E_{\mu_t}$. As in prior calculations, it is convenient to express this as a doubled system with two replicas $L,R$:
\begin{align}
    q_{\gamma}(\mathbf{x}) &= \sum_{\mu_K\cdots\mu_1} \bra{\mathbf{x}} E_{\mu_K} U_K \cdots E_{\mu_1} U_1 \ket{\mathbf{0}} \left( \bra{\mathbf{x}} E_{\mu_K} U_K \cdots E_{\mu_1} U_1 \ket{\mathbf{0}} \right)^* \nonumber \\
    &= \sum_{\mu_K\cdots\mu_1} \bra{\mathbf{x} \mathbf{x}} \left(E_{\mu_K} \otimes E_{\mu_K}^* \right) \left( U_K \otimes U_K^* \right) \cdots \left(E_{\mu_1} \otimes E_{\mu_1}^* \right) \left( U_1 \otimes U_1^* \right) \ket{\mathbf{0} \mathbf{0}} \nonumber \\
    &= \sum_{\mu_K\cdots\mu_1} (-1)^{\mathbf{x}} \bra{\mathbf{x\overline{x}}} \left(E_{\mu_K} \otimes E_{\mu_K}^{\mathcal{T}} \right) \left( U_K \otimes U_K^{\mathcal{T}} \right) \cdots \left(E_{\mu_1} \otimes E_{\mu_1}^{\mathcal{T}} \right) \left( U_1 \otimes U_1^{\mathcal{T}} \right) \ket{\mathbf{0} \mathbf{1}} \nonumber \\
    &= (-1)^{\mathbf{x}} \bra{\mathbf{x\overline{x}}} \prod_t \left( \sum_{\mu_t} E_{\mu_t} \otimes E_{\mu_t}^{\mathcal{T}} \right) \left( U_t \otimes U_t^{\mathcal{T}} \right) \ket{\mathbf{0} \mathbf{1}} 
\end{align}
where in the second-to-last last line we have inserted factors of $(\pm i Y)$ as usual to give properly time-reversed operators. Note that the Kraus operators for the depolarizing channel behave differently under time-reversal:
\begin{equation}
    E_0^{\mathcal{T}} = E_0^{\mathcal{T}} \quad \quad \quad \quad E_{i \alpha}^{\mathcal{T}} = -E_{i \alpha}^{\mathcal{T}}
\end{equation}
Next, assuming that $dt$ is infinitesimally small, we can express each depolarizing step in terms of an effective Hamiltonian acting on the replicas $L,R$:
\begin{align}
    \sum_{\mu} E_{\mu} \otimes E_{\mu}^{\mathcal{T}} &= \left(1 - n \gamma dt \right) \ \mathbb{I} \otimes \mathbb{I} - \frac{1}{3} \gamma dt \sum_{i, \alpha} \sigma_i^{\alpha} \otimes \sigma_i^{\alpha} \nonumber \\
    &\approx \exp \left[ - n \gamma dt - \frac{1}{3} \gamma dt \sum_i \vec{\sigma}_{iL} \cdot \vec{\sigma}_{iR} \right] + \mathcal{O}( dt^2) \nonumber \\
    &\approx \exp(- H_{\gamma} dt)
\end{align}
where the effective depolarizing Hamiltonian is
\begin{equation}
    H_{\gamma} = \gamma \sum_i \left( 1 + \frac{1}{3} \vec{\sigma}_{iL} \cdot \vec{\sigma}_{iR} \right).
\end{equation}
Finally, we take the expectation value over Brownian circuits to obtain an effective Hamiltonian for the Brownian timesteps:
\begin{align}
    \mathbb{E}_U \left[ q_{\gamma}(\mathbf{x}) \right] &= (-1)^{\mathbf{x}} \bra{\mathbf{x\overline{x}}} \prod_t \left( \sum_{\mu_t} E_{\mu_t} \otimes E_{\mu_t}^{\mathcal{T}} \right) \left( \mathbb{E}_{U_t} \left[ U_t \otimes U_t^{\mathcal{T}} \right] \right) \ket{\mathbf{0} \mathbf{1}} \nonumber \\
    &= (-1)^{\mathbf{x}} \bra{\mathbf{x\overline{x}}} \prod_t e^{- H_{\gamma}  dt} e^{- H_{\mathrm{eff}} dt} \ket{\mathbf{0} \mathbf{1}} \nonumber \\
    &\approx (-1)^{\mathbf{x}} \bra{\mathbf{x\overline{x}}} e^{- \beta H'_{\mathrm{eff}}} \ket{\mathbf{0} \mathbf{1}}
\end{align}
where in the last step we have used the lowest-order term of the Baker-Campbell-Hausdorff formula to give
\begin{equation}
    e^{- H_{\gamma} dt} e^{- H_{\mathrm{eff}}  dt} \approx e^{- \left( H_{\gamma} + H_{\mathrm{eff}} \right) dt}
\end{equation}
assuming $\Delta t \rightarrow 0$. Hence the final effective Hamiltonian in this case is
\begin{equation}
    H'_{\mathrm{eff}} = H_{\mathrm{eff}} + H_{\gamma}.
\end{equation}
where $H_{\mathrm{eff}}$ is the usual $k=1$ effective Hamiltonian for Brownian dynamics. This derivation is straightforward to generalize to higher moments $k > 1$, leading to an effective Hamiltonian consisting of the usual Brownian term $\Heff$ plus a contribution $H_{\gamma}$ from depolarizing noise, where the depolarizing Hamiltonian acts non-trivially on replicas $ra = 1L,1R$, but acts trivially on all other replicas.

Finally, we apply degenerate perturbation theory to study the spectrum of $H_{\mathrm{eff}}'$ in the case where $\gamma \ll J$. We consider the moments $m^{(2)}$ first to illustrate the basic ideas. The ground space has dimension $2! = 2$ and is spanned by the states $\ket{\Omega},\ket{\Omega'}$, where
\begin{align}
    \ket{\Omega} &= \bigotimes_{i} \frac{1}{2} \left( \ket{01} - \ket{10} \right)_{i,1L1R} \left( \ket{01} - \ket{10} \right)_{i,2L2R} \nonumber \\
    &= \bigotimes_{i} \frac{1}{2} \left( \ket{0101} - \ket{1001} - \ket{0110} + \ket{1010} \right)_i
\end{align}
is the `ladder' state consisting of singlets pairing spins $i1L,i1R$ and $i2L,i2R$ (see Eq. \eqref{eq:ladderground}) and
\begin{equation}
    \ket{\Omega'} := \bigotimes_{i} \frac{1}{2} \left( \ket{0101} - \ket{1100} - \ket{0011} + \ket{1010} \right)_i
\end{equation}
is the `crossed' state consisting of singlets pairing spins $i1L,i2R$ and $i2L,i1R$.
Although these states are not strictly orthogonal, they are nearly orthogonal in the large-$n$ limit
\begin{equation}
    \bracket{\Omega}{\Omega'} = 1/2^n \rightarrow 0 
\end{equation}
as $n \rightarrow \infty$, which allows us to apply standard degenerate perturbation theory to the problem.
The lowest-order energy shifts are given by the matrix elements of the perturbing Hamiltonian, giving
\begin{align}
    &\Delta E = \bra{\Omega} H_{\gamma} \ket{\Omega} = 0 \nonumber \\
    & \Delta E' = \bra{\Omega'} H_{\gamma} \ket{\Omega'} = n \gamma
\end{align}
where the off-diagonal matrix elements vanish $\bra{\Omega'} H_{\gamma} \ket{\Omega} = 0$ because $\ket{\Omega}$ is a zero-energy eigenstate of $H_{\gamma}$. Therefore the `ladder' state does not acquire an energy shift, while the `crossed' state acquires an energy shift $\Delta E' = n \gamma$.

\begin{figure}[h!]
    \centering
    \includegraphics{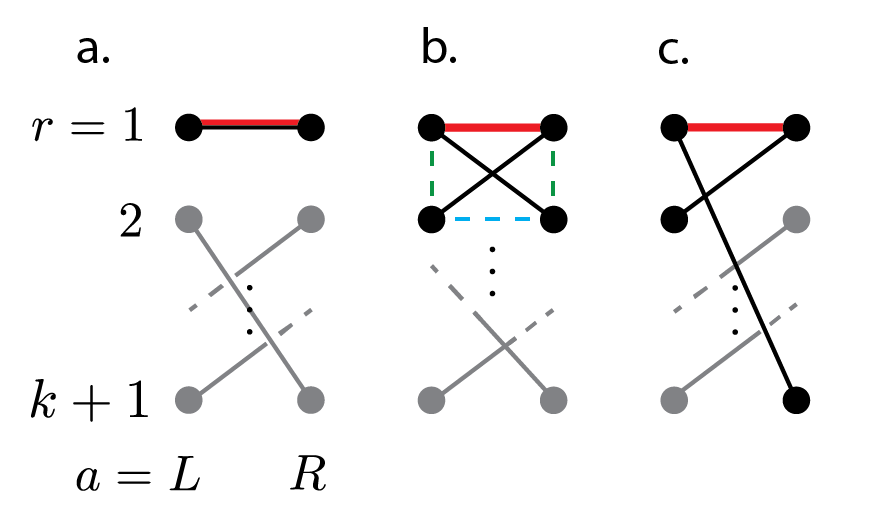}
    \caption{Ground-state pairings relevant to the spectrum of $H_{\mathrm{eff}}'$ including perturbatively small depolarizing noise. The red line represents the noisy Heisenberg coupling term $H_{\mathcal{\gamma}}$ acting on replicas $1L$ and $1R$. (a) `Trivial' pairings, in which the system has a singlet state pairing replicas $1L$ and $1R$ (overlapping with the noisy Heisenberg term), account for $k!$ pairing possibilities and have vanishing energy shift $\Delta E = 0$ at lowest order in $\gamma$. (b) The simplest `non-trivial' pairings, in which the singlet pairings do not overlap with the noisy Heisenberg coupling, account for $(k-1)!$ possible pairings. (c) Another set of `non-trivial' pairings account for another $(k-1)!$ possibilities. Continuing in this fashion leads to a total of $k^2 (k-1)! = k \cdot k!$ `non-trivial' pairing possibilities with energy shift $\Delta E' = n \gamma$ at lowest order in $\gamma$.}
    \label{fig:ladderdiagrams2}
\end{figure}

These arguments generalize straightforwardly to higher moments. In particular, to compute the expectation value $\langle q^k \rangle_{\gamma}$ appearing in Eq. \eqref{eq:qkmomentsnoisy} of the main text, we must compute the moments $m^{(k+1)}$. Similar to the `ladder' and `crossed' states, the calculation in this case separates into diagrams with `trivial' and `non-trivial' spin-singlet pairings as illustrated in Fig. \ref{fig:ladderdiagrams2}. There are $k!$ `trivial' pairings (analogous to the `ladder' state) with vanishing energy shift $\Delta E = 0$. In addition, there are $k \cdot k!$ `non-trivial' pairings (analogous to the `crossed' state) with energy shift $\Delta E' = n \gamma$. Incorporating these energy shifts into the spectrum of $H_{\mathrm{eff}}^{(k+1)}$ yields the noisy expression Eq. \eqref{eq:qkmomentsnoisy} in the main text.

\section{Harvard Spoofing Algorithm in All-to-All Brownian Circuits}
\label{app:harvard}

\begin{figure}
    \centering
    \includegraphics[width=0.4\linewidth]{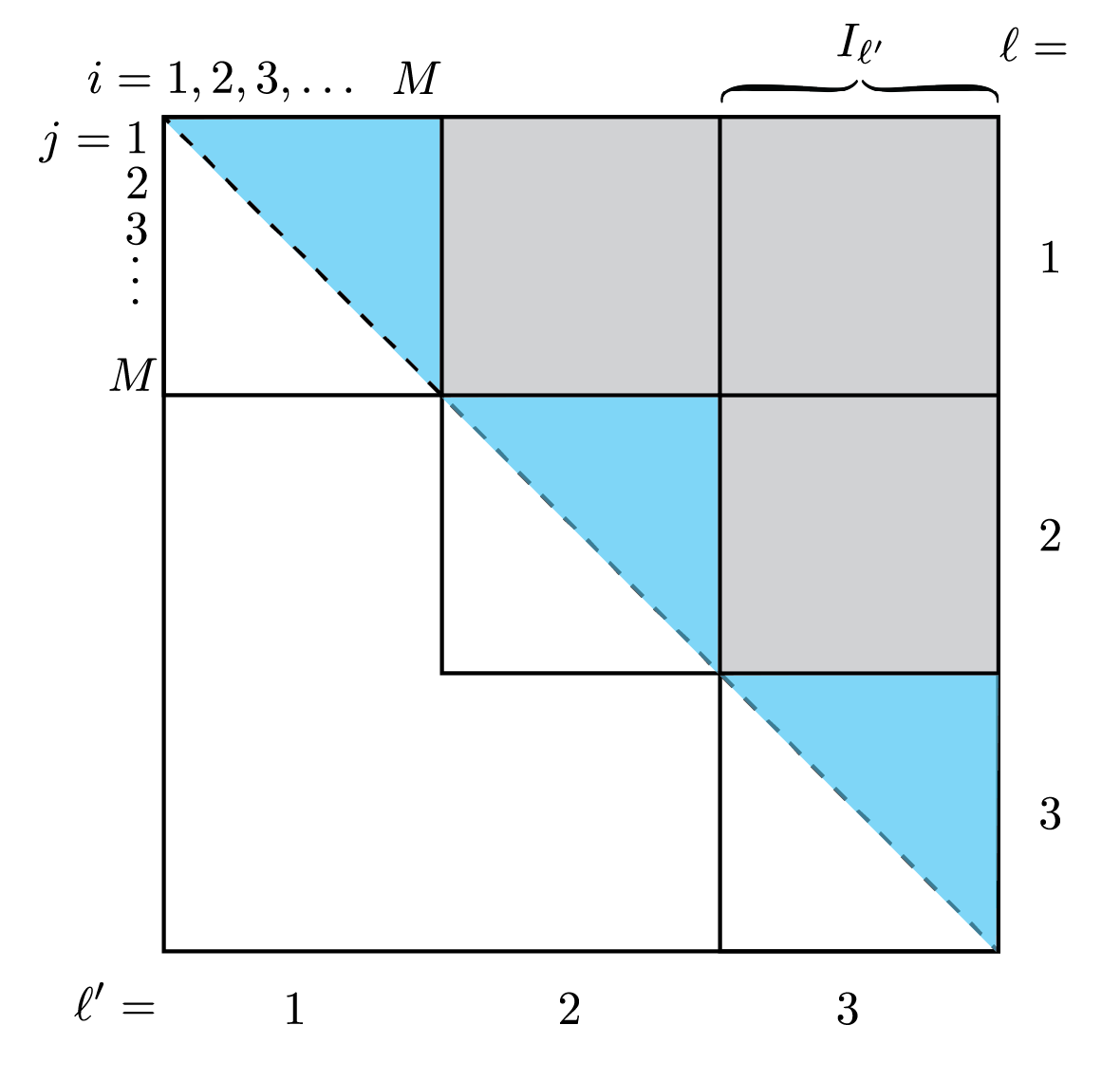}
    \caption{Two categories of coupling coefficients $J_{ij}^{\alpha \beta}(t)$ in the disjoint Brownian circuit model. Blue coefficients couple subsystems to themselves while gray coefficients couple two different subsystems. Gray coefficients are set to zero in the classical spoofer.}
    \label{fig:disjointcouplings}
\end{figure}

In this appendix we demonstrate that the spoofing algorithm of \cite{gao2024limitations} is formally equivalent to a noisy Brownian circuit ensemble where the noise rate is very large $\gamma \rightarrow \infty$. Consider computing the moments
\begin{equation}
    \sum_{\mathbf{x}} \mathbb{E}_U \left[ p(\mathbf{x}) q^k(\mathbf{x}) \right]
\end{equation}
where $q(\mathbf{x})$ is the ideal transition probability calculated by a classical supercomputer, whereas
\begin{equation}
    p(\mathbf{x}) = \magn{\bra{\mathbf{x}}V\ket{\mathbf{0}}}^2
\end{equation}
is the probability of sampling $\mathbf{x}$ from a classical spoofer implementing a unitary $V$ where the system has been subdivided into small decoupled subsystems to render its simulation classically efficient. In particular, given an all-to-all random circuit $U$ governed by Brownian couplings $J_{ij}^{\alpha \beta}(t)$ as in Eq. \eqref{eq:brownianunitary}, we construct a classical spoofer $V$ by subdividing the $n$-qubit system into $L$ subsystems each consisting of $M$ qubits ($ML = n$) and deleting all couplings spanning between different subsystems as shown in Fig. \ref{fig:disjointcouplings}. To ensure that the spoofer is classically tractable, we anticipate setting the subsystem size to be $M \sim \log n$ and $L \sim n / \log n$. In the all-to-all case, the deleted couplings account for the vast majority of couplings in the system; to compensate for the commensurate reduction in scrambling rate we must artificially boost the remaining couplings so that the timescales in the ideal circuit $U$ and the spoofing circuit $V$ remain comparable. We refer to the circuit $V$ as a `disjoint' Brownian circuit model \cite{bentsen2024complexity}.

The original circuit $U$ features $\frac{9}{2} n(n-1)$ Brownian coefficients $J_{ij}^{\alpha \beta}(t)$ for $i < j$. The disjoint spoofer circuit $V = \otimes_{\ell} V_{\ell}$ is split into $L$ disjoint subsystems labeled by $\ell = 1,2,\ldots,L$. Each subsystem $V_{\ell}$ consists of $M$ qubits indexed by $i \in I_{\ell}$, where the interval $I_{\ell}$ consists of exactly $M$ indices. The $\frac{9}{2} n(n-1)$ coefficients $J_{ij}^{\alpha \beta}(t)$ in the original circuit can then be divided into categories depending on whether they couple a subsystem $\ell$ to itself or whether they couple a subsystem $\ell$ to a different subsystem $\ell'$ as illustrated in Fig. \ref{fig:disjointcouplings}. There are $\frac{9}{2} L M (M-1)$ coefficients that couple subsystems to themselves (blue in Fig. \ref{fig:disjointcouplings}), where $i < j \in I_{\ell}$ for any $\ell$. In addition, there are $\frac{9}{2} M^2 L(L-1)$ coefficients that couple different subsystems (gray in Fig. \ref{fig:disjointcouplings}), where $i \in I_{\ell}$ and $j \in I_{\ell'}$ for $\ell \neq \ell'$. In the disjoint circuit $V$, this second category of coefficients are all set to zero. Together, these two categories add up to the total $\frac{9}{2} n (n-1)$ for $n = M L$.

Introducing replicas and applying the time-reversal operation as usual, we find
\begin{align}
    \mathbb{E}[p(\mathbf{x}) q^k(\mathbf{x})] &= 
    (-1)^{x(k+1)}\bra{\mathbf{x} \overline{\mathbf{x}} \ldots } \mathbb{E}_U \left[ V \otimes V_{\mathcal{T}} \otimes U \otimes U_{\mathcal{T}} \otimes \cdots \right] \ket{\mathbf{0} \mathbf{1} \ldots}
\end{align}
where replicas $1L,1R$ refer to the disjoint Brownian circuit $V$ and the remaining replicas refer to the ideal Brownian circuit $U$. Performing the ensemble average over the Gaussian distributed couplings $J_{ij}^{\alpha \beta}(t)$ at each time $t$ yields:
\begin{align}
    &\mathbb{E}_U[ V_t \otimes V_t^{\mathcal{T}} \otimes U_t \otimes U_t^{\mathcal{T}} \otimes \cdots] = \nonumber \\
    & \int \prod_{\substack{i<j \\ \alpha \beta}} dJ_{ij}^{\alpha \beta}(t) \exp \left[ - \sum_{\substack{i<j \\ \alpha\beta}} \frac{\left( J_{ij}^{\alpha \beta}(t)\right)^2}{2J / n dt} \right] \nonumber \\
    &\times \exp \left[ -i \sum_{\ell} \sum_{i < j \in I_{\ell}} \sum_{\alpha \beta} J_{ij}^{\alpha \beta}(t) \left( A \sigma_{i1L}^{\alpha} \sigma_{j1L}^{\beta} - A \sigma_{i1R}^{\alpha} \sigma_{j1R}^{\beta} + \sum_{r=2}^{k+1} \left[ \sigma_{irL}^{\alpha} \sigma_{jrL}^{\beta} - \sigma_{irR}^{\alpha} \sigma_{jrR}^{\beta} \right] \right) dt \right] \nonumber \\
    &\times \exp \left[ -i \sum_{\ell < m} \sum_{\substack{i \in I_{\ell} \\ j \in I_{m}}} \sum_{\alpha \beta} J_{ij}^{\alpha \beta}(t) \left( \sum_{r = 2}^{k+1} \sigma_{irL}^{\alpha} \sigma_{jrL}^{\beta} - \sigma_{irR}^{\alpha} \sigma_{jrR}^{\beta} \right) dt \right]
\end{align}
where the third line accounts for the couplings within each subsystem (blue in Fig. \ref{fig:disjointcouplings}) that are present in both circuits, while the fourth line accounts for the couplings between different subsystems (gray in Fig. \ref{fig:disjointcouplings}) that are present in the full Brownian circuit $U$ but not the disjoint circuit $V$. Notice that we have included an extra factor of $A$ in replicas $1L,1R$ to increase the strength of these couplings so that the disjoint system reaches its Haar value on the same timescale as the full circuit. We will find that the choice $A = \sqrt{L}$ gives the right timescale, but we leave this parameter undetermined for the moment. Performing the Gaussian integrals, we find that the effective Hamiltonian is
\begin{align}
    H_{\mathrm{eff}} &= \frac{J}{2n} \sum_{\ell} \sum_{i < j \in I_{\ell}} \sum_{\alpha \beta} \left( A \sigma_{i1L}^{\alpha} \sigma_{j1L}^{\beta} - A \sigma_{i1R}^{\alpha} \sigma_{j1R}^{\beta} + \sum_{r=2}^{k+1} \left[ \sigma_{irL}^{\alpha} \sigma_{jrL}^{\beta} - \sigma_{irR}^{\alpha} \sigma_{jrR}^{\beta} \right] \right)^2 \nonumber \\
    &+  \frac{J}{2 n} \sum_{\ell < m} \sum_{\substack{i \in I_{\ell} \\ j \in I_{m}}} \sum_{\alpha \beta} \left( \sum_{r=2}^{k+1} \sigma_{irL}^{\alpha} \sigma_{jrL}^{\beta} - \sigma_{irR}^{\alpha} \sigma_{jrR}^{\beta} \right)^2.
    \label{eq:harvardalgham}
\end{align}
After some manipulation and keeping only leading-order terms in $n$ and $L \sim n / \log n$ we find that the Hamiltonian is approximately
\begin{equation}
    H_{\mathrm{eff}} \approx H_0 + \sum_{\ell} H_{\ell}
\end{equation}
where
\begin{align}
    H_0 &= \frac{J n}{2} \sum_{\substack{ra<sb \\ 2 \leq r,s \leq k+1}} (-1)^{a+b} \left( \frac{1}{n} \sum_{i=1}^n \heis{ira}{isb} \right)^2 + \frac{9 J}{2} k n \nonumber \\
    H_{\ell} &= - \frac{J}{2} \frac{A^2 M}{L} \left( \frac{1}{M} \sum_{i \in I_{\ell}} \heis{i1L}{i1R} \right)^2 + \frac{9 J}{2} \frac{A^2 M}{L}
\end{align}
where we have dropped terms scaling like $O(A)$ and $O(1)$. We see that $A^2 = L$ naturally ensures that the timescales for $H_0$ and $\sum_{\ell} H_{\ell}$ are the same. In this form we see that the effective Hamiltonian splits into $\ell+1$ disjoint pieces: $H_0$ acts only on spins in replicas $2 \leq r,s \leq k+1$ whereas each $H_{\ell}$ only acts on spins inside the subsystem $I_{\ell}$ and replica indices $ra = 1L,1R$. Because these disjoint pieces do not interact with one another, we may solve each of them separately, and their contributions to the expectation value factorize. Using prior results, we have
\begin{align}
    \mathbb{E}_U \left[ p(\mathbf{x}) q^k(\mathbf{x}) \right] &= \left[ \frac{k!}{2^{nk}} \left(1 - a \right)^{k\magn{\mathbf{x}}} \left(1 + a \right)^{k(n-\magn{\mathbf{x}})} \right] \prod_{\ell=1}^L \left[ \frac{1}{2^{M}} \left(1 - a \right)^{\magn{\mathbf{x}_{\ell}}} \left(1 + a \right)^{(M-\magn{\mathbf{x}_{\ell}})} \right] \nonumber \\
     &= \frac{k!}{2^{n(k+1)}} \left[ \left(1 - a \right)^{(k+1)\magn{\mathbf{x}}} \left(1 + a \right)^{(k+1)(n-\magn{\mathbf{x}})} \right]
     \label{eq:harvardmoments}
\end{align}
where $\mathbf{x}_{\ell}$ is the portion of the bitstring $\mathbf{x}$ restricted to the subregion $I_{\ell}$. Finally, performing the sum over bitstrings we obtain
\begin{equation}
    \sum_{\mathbf{x}} \mathbb{E}_U \left[ p(\mathbf{x}) q^k(\mathbf{x}) \right] = \frac{k!}{2^{n(k+1)}} \left[ \left(1 - a \right)^{k+1} + \left(1 + a \right)^{k+1} \right]^n
\end{equation}
which is formally equivalent to Eq. \eqref{eq:qkmomentsnoisy} in the limit $\gamma \rightarrow \infty$.

\section{Sample-Inefficiency of Linear XEB at Shallow Circuit Depths}
\label{app:linearxeb}

Here we apply Brownian circuit tools to study the behavior of the linear cross-entropy (XEB) in the shallow-depth regime and demonstrate that this benchmark is not sample efficient in this regime \cite{bentsen2024complexity}. The expectation value for the linear XEB is given by
\begin{equation}
    \XEB = \langle q \rangle = \sum_{\mathbf{x}} \mathbb{E}_U \left[ p(\mathbf{x}) q(\mathbf{x}) \right]
\end{equation}
corresponding to $k=1$ in Eq. \eqref{eq:qkmomentsnoisy}. We may immediately write down the result using our existing calculations:
\begin{equation}
    \XEB = \frac{1}{2^{n}} \left( 1+a^2 \right)^n \left[ 1 + e^{- n \beta \gamma} \right].
\end{equation}
The ensemble fluctuations in the linear XEB are captured by the variance
\begin{equation}
    \mathrm{Var} \XEB = \langle q^2 \rangle - \langle q \rangle^2
\end{equation}
which we may also immediately write down from prior results. We have:
\begin{equation}
    \langle q^2 \rangle = \frac{2}{2^{2n}} \left( 1+3 a^2 \right)^n \left( 1 + 2 e^{- n \beta \gamma} \right)
\end{equation}
for $k=2$, and therefore the signal-to-noise ratio is
\begin{equation}
    \frac{\XEB}{\sqrt{\mathrm{Var} \XEB}} = \frac{1}{\sqrt{\langle q^2 \rangle / \langle q \rangle^2 - 1}}
\end{equation}
where
\begin{equation}
    \langle q^2 \rangle / \langle q \rangle^2 = 2 \frac{1+2 e^{- n \beta \gamma}}{\left(1+ e^{- n \beta \gamma} \right)^2} \left[ \frac{1 + 3 a^2}{(1+a^2)^2} \right]^n
\end{equation}
is exponentially large at constant depth $a \sim O(1)$ because the factor
\begin{equation}
    \frac{1 + 3 a^2}{(1+a^2)^2} > 1
\end{equation}
is larger than 1. Therefore the linear XEB is not sample efficient to estimate at constant depth because the fluctuations are exponentially large compared to the signal. By contrast, at logarithmic depth $a \sim O(1/n)$ we may expand this factor in powers of $n a^2 \ll 1$ to obtain
\begin{equation}
    \langle q^2 \rangle / \langle q \rangle^2 \approx 2 ( 1 + n a^2) + O(n^2 a^4)
\end{equation}
where we have dropped the polynomially-small noise factors $e^{-n \beta \gamma}$. Therefore the linear XEB is sample efficient to estimate at logarithmic depth.

\section{Expected Value for Heavy Output Generation Classifier}
\label{app:sshogbetting}

Here we elaborate on the `betting game' discussed in Section \ref{sec:sshog}. Suppose we are playing against the `house' where there are two biased coins $A,B$ with probabilities $p_A,p_B$ of giving heads, where $p_A > p_B$. Each round we are given one of these two coins and can perform a single toss and attempt to determine the coin's identity. If we guess correctly we win one dollar, otherwise we lose one dollar. In the simplest case where $p_A > 1/2 > p_B$ we may play a `sensible' \emph{pure strategy} $S$ in which we guess coin $A$ when we see heads and we guess coin $B$ when we see tails, leading to a positive-definite expected value $\mathrm{min}[2p_A-1,1-2p_B]$ (for a maximally adversarial house). In the more general case $p_A > p_B \geq 1/2$ we must consider a more sophisticated strategy because the house could always force a negative expected value by simply giving us coin $B$ every time. To hedge against this possibility, it is advantageous to pursue a \emph{mixed strategy} where we sometimes blindly guess coin $B$ regardless of the toss outcome. (The case where $1/2 \geq p_A > p_B$ can be handled by the same methods but working instead with the probabilities $p'_A = 1-p_B$ and $p'_B = 1-p_A$.)

\begin{figure}[h!]
    \centering
    \includegraphics[width=0.3\linewidth]{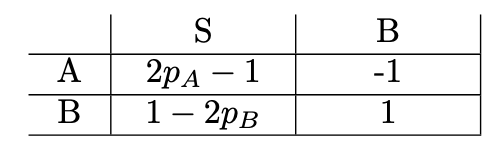}
    \label{fig:payoffmatrix}
\end{figure}

This is a zero-sum game that can be analyzed using standard tools in game theory. The strategies that can be played by the house and the player yield the payoff matrix in Fig. \ref{fig:payoffmatrix}
where the house's strategies appear along the left side and the player's strategies appear along the top. A \emph{mixed strategy} for the player consists of choosing the pure strategy $S$ with probability $x$ and the pure strategy $B$ with probability $1-x$. To solve for $x$ we demand that the expected payoff is the same no matter what the house does. This occurs when
\begin{equation}
    x(2p_A-1) - (1-x) = x(1-2p_B) + (1-x)
\end{equation}
so that the optimal probability is $x = 1/(p_A+p_B)$, yielding an expected value $(p_A-p_B)/(p_A+p_B)$ that is always positive-definite.

The above analysis assumes that we know the probabilities $p_A,p_B$ in advance. This occurs, for example, when we are trying to distinguish a perfectly clean quantum sampler $\gamma = 0$ from the classical Harvard spoofer $\gamma \rightarrow \infty$ where the probabilities $S(z > z^*)$ can be calculated explicitly for both samplers. It also applies when we are dealing with a noisy quantum sampler $\infty > \gamma > 0$ and we know the noise rate $\gamma$ in advance. But what about the situation where $\gamma$ is known to be finite but we do not know its exact value? In this case $p_A - p_B \geq \delta > 0$ where $\delta$ is a minimum gap between probabilities. In this case the optimal strategy is to pick $x = 1/(2p_B + \delta)$, yielding an expected value $(2(p_A - p_B) - \delta) / (2 p_B + \delta)$ if the house chooses coin $A$ and $\delta / (2 p_B + \delta)$ if the house chooses coin $B$. In either case the expected value is lower-bounded by $\delta / (2 p_B + \delta)$ which is positive-definite so long as $\delta > 0$. Further, we see that these results reproduce our previous analysis in the case where $p_A - p_B = \delta$.

To make contact with the HOG classifier problem discussed in the main text, suppose we are trying to distinguish a noisy quantum sampler with noise rate $\gamma < \gamma_{\mathrm{max}}$ from the classical Harvard spoofing algorithm, where we do not know the exact value of $\gamma$ but we are guaranteed that it is no larger than $\gamma_{\mathrm{max}}$. In this case the difference in probabilities is given by
\begin{equation}
    p_A - p_B = e^{-n \beta \gamma} \Delta(z > z^*) \geq e^{-n \beta \gamma_{\mathrm{max}}} \Delta(z > z^*) \equiv \delta
\end{equation}
meaning that we obtain a positive expected value in the betting game despite not knowing the exact value of the noise rate $\gamma$.

Finally, we elaborate on the calculations for multiple samples $m$ per round discussed in the main text. Given a coin $A$, the probability of getting exactly $h$ heads from $m$ flips is given by the binomial distribution
\begin{equation}
    P(h|A) = \binom{m}{h} p_A^h (1-p_A)^{m-h} \approx \frac{1}{\sqrt{2 \pi \sigma_A^2}} e^{-(h-\mu_A)^2 / 2 \sigma_A^2} 
\end{equation}
where in the final expression we have used the normal distribution approximation in the limit of large $m \gg 1$, where $\mu_A = m p_A$ and $\sigma_A^2 = m p_A (1-p_A)$ are the mean and variance in the number of heads obtained from coin $A$. An analogous expression holds for $P(h|B)$. To decide which coin we have, we select a threshold $h^*$; if our $m$ coin flips yield more heads than this threshold $h > h^*$ then we guess coin $A$, otherwise we guess coin $B$. The optimal strategy to beat a maximally adversarial house is to choose a threshold $h^*$ for which the probability of guessing correctly is the same regardless of which coin we are given, leading to Eq. \eqref{eq:mflipscorrectprob} in the main text, where
\begin{equation}
    P(h > h^* | A) \approx \int_{h^*}^{\infty} dh \ P(h | A) = \Phi\left( \frac{\mu_A - h^*}{\sigma_A} \right)
    \label{eq:PAabovethresh}
\end{equation}
in the limit of large $m \gg 1$, where
\begin{equation}
    \Phi(y) := \frac{1}{2} \left[ 1+ \mathrm{erf} \left( \frac{y}{\sqrt{2}} \right) \right]
\end{equation}
is the cumulative distribution function (CDF) for the normal distribution. Similarly, for coin $B$ we have
\begin{equation}
    P(h \leq h^* | B) \approx \int^{h^*}_{-\infty} dh \ P(h | B) = \Phi\left( \frac{h^* - \mu_B}{\sigma_B} \right)
    \label{eq:PBbelowthresh}
\end{equation}
and setting these CDFs equal to each other we arrive at Eq. \eqref{eq:cdfequality} in the main text. Because the CDF $\Phi(y)$ is monotonic we may simply set the arguments of these functions equal:
\begin{equation}
    \frac{\mu_A - h^*}{\sigma_A} = \frac{h^* - \mu_B}{\sigma_B}
\end{equation}
and solving for $h^*$ yields Eq. \eqref{eq:headsthresh} in the main text. Plugging this into Eq. \eqref{eq:PAabovethresh} (or Eq. \eqref{eq:PBbelowthresh}) and simplifying gives the probability of guessing the coin correctly:
\begin{equation}
    P_{\mathrm{correct}} = \Phi \left( \sqrt{m} \ x \right) \approx 1 - \frac{1}{\sqrt{2 \pi m} \ x} e^{-m x^2 / 2}
\end{equation}
where $x$ is defined in Eq. \eqref{eq:xdef} of the main text.

\end{document}